\documentclass[iop]{emulateapj}
\usepackage{color}
\usepackage[encapsulated]{CJK}
\usepackage{ucs}
\usepackage[utf8x]{inputenc}
\usepackage{longtable}
\usepackage{gensymb}
\usepackage{amsmath}

\newenvironment{Contfigure*}{%
\renewcommand{\thefigure}{\arabic{figure} (Cont.)}%
\addtocounter{figure}{-1}%
\begin{figure*}}{%
\renewcommand{\thefigure}{\arabic{figure}}%
\end{figure*}}

\begin{document}
\begin{CJK}{UTF8}{gbsn}

\title{The Kinematics of \textrm{C}~\textsc{iv} in Star-forming Galaxies at $z\sim1.2$}

\author{Xinnan Du (杜辛楠)\altaffilmark{1}, Alice E. Shapley\altaffilmark{1}, Crystal L. Martin\altaffilmark{2}, Alison L. Coil\altaffilmark{3}}

\altaffiltext{1}{Department of Physics and Astronomy, University of California, Los Angeles CA, 90095, USA}

\altaffiltext{2}{Department of Physics, Univerisity of California, Santa Barbara CA, 93106, USA}

\altaffiltext{3}{Department of Physics, Univerisity of California, Santa Diego CA, 92093, USA}

\slugcomment{Draft Version \today}

\shorttitle{Kinematics of \textrm{C}~\textsc{iv}}
\shortauthors{Du}

\begin{abstract}
We present the first statistical sample of rest-frame far-UV spectra 
of star-forming galaxies at $z \sim 1$. These spectra are unique in that they 
cover the high-ionization \textrm{C}~\textsc{iv}$\lambda\lambda$1548, 1550 
doublet. We also detect low-ionization features such as 
\textrm{Si}~\textsc{ii}$\lambda$1527, \textrm{Fe}~\textsc{ii}$\lambda$1608, 
\textrm{Al}~\textsc{ii}$\lambda$1670, \textrm{Ni}~\textsc{ii}$\lambda\lambda$1741, 1751 and
 \textrm{Si}~\textsc{ii}$\lambda$1808, and intermediate-ionization features 
 from \textrm{Al}~\textsc{iii}$\lambda\lambda$1854, 1862. Comparing the 
 properties of absorption lines of lower- and higher- ionization states provides 
 a window into the multi-phase nature of circumgalactic gas. Our sample is drawn 
 from the DEEP2 survey and spans the redshift range 
 1.01 $\leqslant z \leqslant$ 1.35 ($\langle z \rangle$ = 1.25). By isolating the 
 interstellar \textrm{C}~\textsc{iv} absorption from the stellar P-Cygni wind profile 
 we find that 69$\%$ of the \textrm{C}~\textsc{iv} profiles are blueshifted with 
 respect to the systemic velocity. Furthermore, \textrm{C}~\textsc{iv} shows a 
 small but significant blueshift relative to \textrm{Fe}~\textsc{ii} (offset of the 
 best-fit linear regression -76 $\pm$ 26 km $\mbox{s}^{-1}$). At the same time, 
 the \textrm{C}~\textsc{iv} blueshift is on average comparable to that 
 of \textrm{Mg}~\textsc{ii}$\lambda\lambda$2796, 2803. At this point, in explaining 
 the larger blueshift of \textrm{C}~\textsc{iv} absorption at the $\sim$ 3-sigma level, 
 we cannot distinguish between the faster motion of highly-ionized gas relative to 
 gas traced by \textrm{Fe}~\textsc{ii}, and filling in on the red side from 
 resonant \textrm{C}~\textsc{iv} emission. We investigate how far-UV 
 interstellar absorption kinematics correlate with other galaxy properties 
 using stacked spectra. These stacking results show
 a direct link between \textrm{C}~\textsc{iv} absorption and the current SFR, 
 though we only observe small velocity differences among different ionization 
 states tracing the outflowing ISM.

\end{abstract}

\keywords{galaxies: evolution -- ISM: structure -- ultraviolet: galaxies}

\section{INTRODUCTION}
\label{sec:Intro}

Outflows play a key role in the evolution of galaxies. Along with inflows, they regulate the amount of gas available for star formation, and remove metals from galaxies, polluting the intergalactic medium (IGM). These galactic-scale outflows have high velocities, typically several hundreds $\mbox{km s}^{-1}$, and high mass loss rates, commonly ranging from a few tenths up to the same order as the SFR \citep[e.g.,][]{Heckman2000,Pettini2001,Shapley2003,Martin2005,Weiner2009,Steidel2010}. Galactic winds may either be driven by thermal pressure associated with the supernova explosions, or else radiation pressure on dust grains from absorption and scattering of continuum photons emitted by massive stars \citep{Murray2005}.

\begin{figure*}
\includegraphics[width=0.5\linewidth]{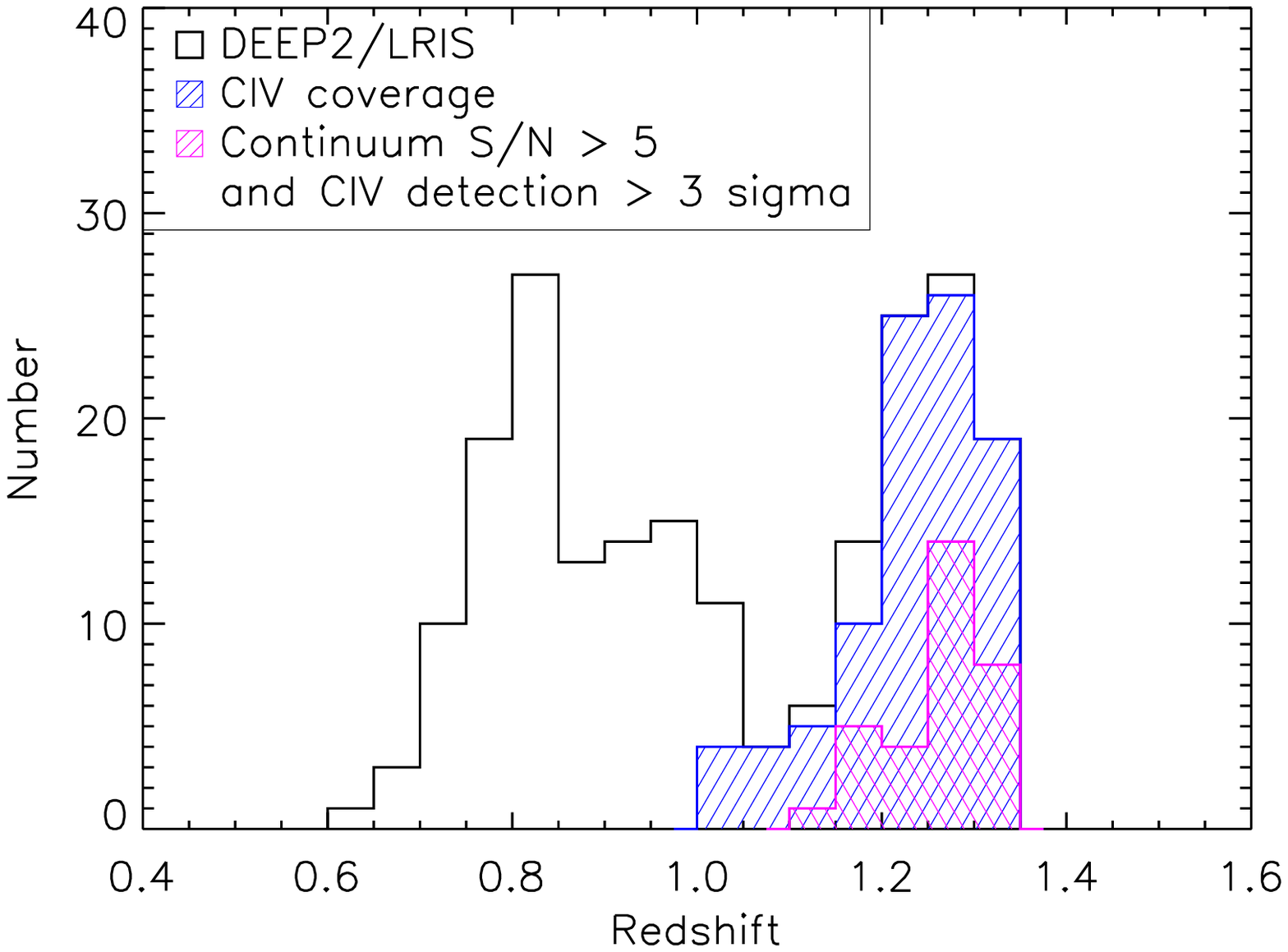}
\includegraphics[width=0.5\linewidth]{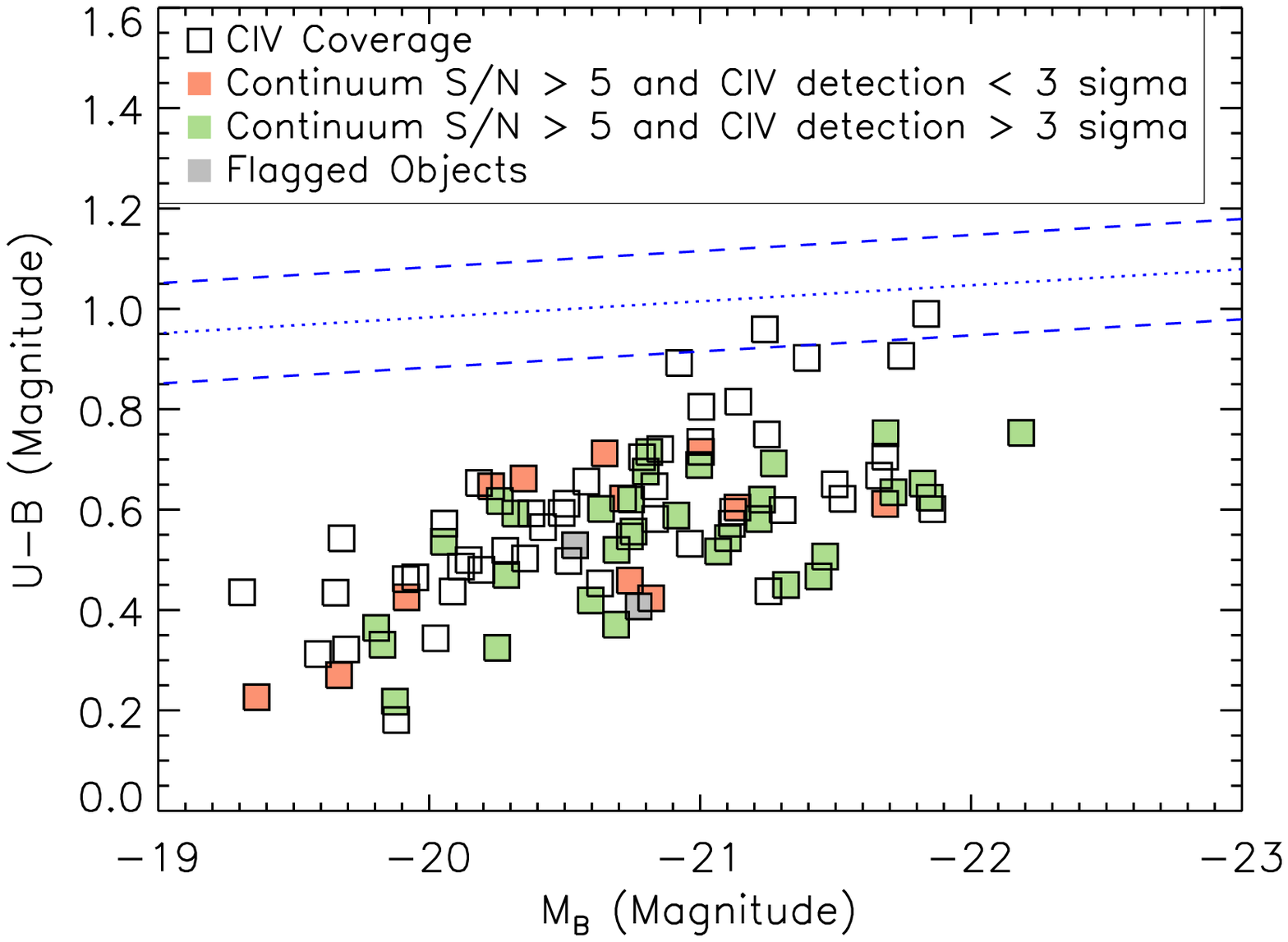}
\label{fig:mbub}
\includegraphics[width=0.5\linewidth]{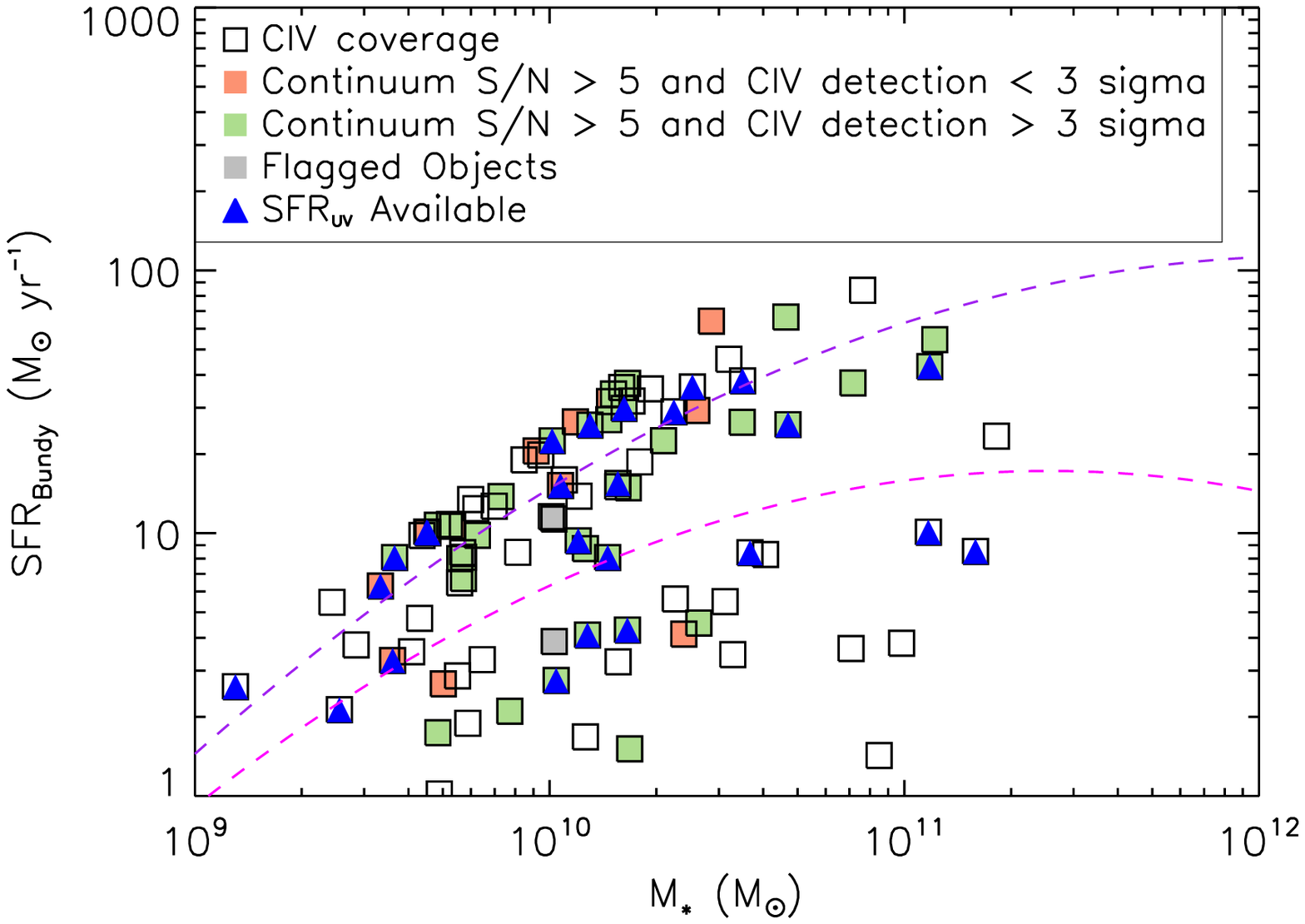}
\label{fig:sfrm}
\includegraphics[width=0.5\linewidth]{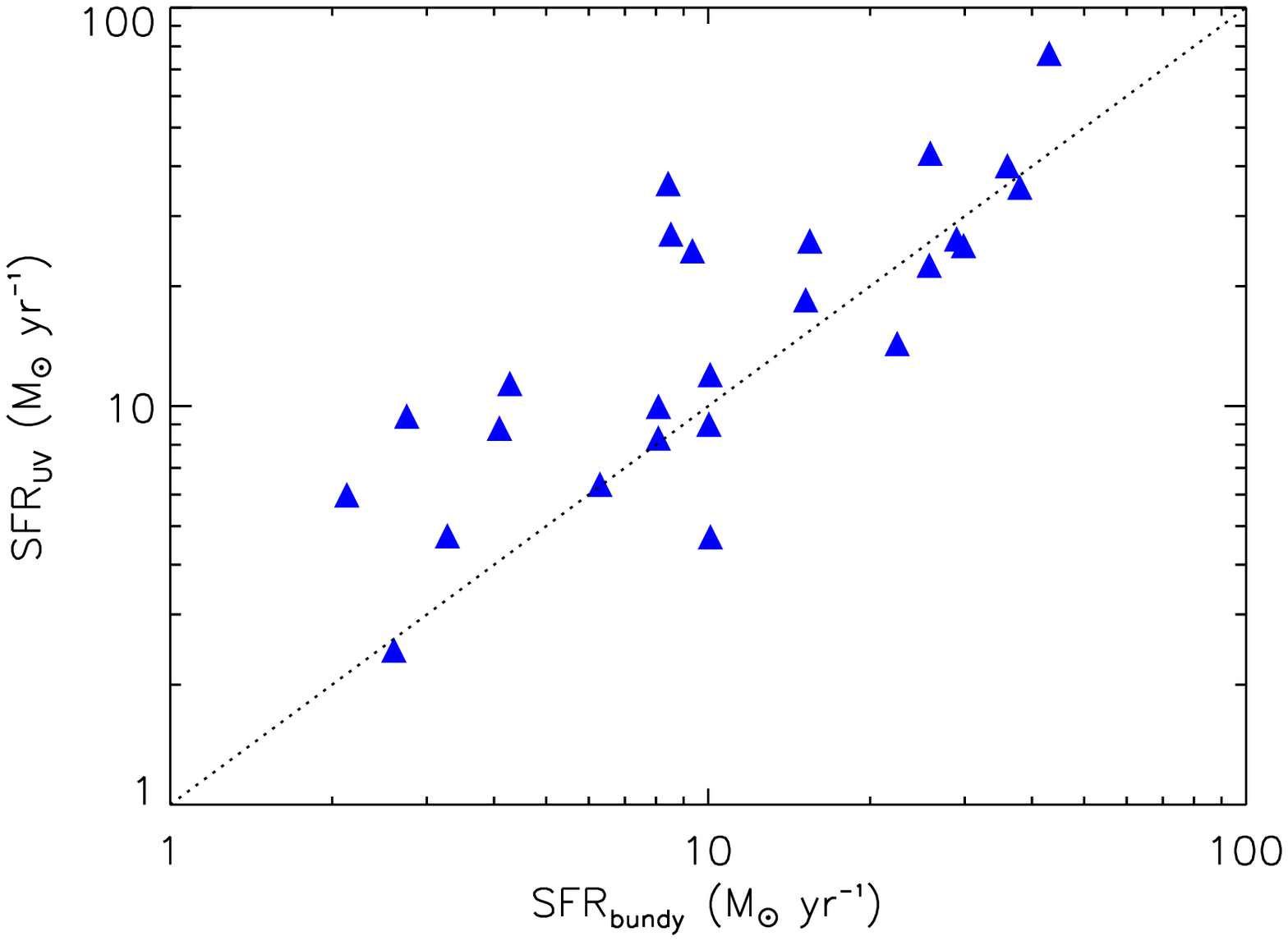}
\label{fig:sfrcom}
\caption{Properties of the galaxies in the DEEP2 \textrm{C}~\textsc{iv} sample. \textbf{Top left:} Redshift distribution. The black open bar represents the parent sample of 208 DEEP2/LRIS galaxies; the dashed blue bar represents 93 galaxies with \textrm{C}~\textsc{iv} coverage; 
The dashed magenta bar represents 32 galaxies meeting the criterion of continuum $S/N>5$ and \textrm{C}~\textsc{iv} 
detection $>$ 3$\sigma$ ($EW/\delta EW >3$). \textbf{Top right:} $\ub$ vs. 
$\mbox{M}_{B}$ color-magnitude diagram. Squares represent 93 objects with \textrm{C}~\textsc{iv} coverage; 
filled squares represent those with continuum $S/N>5$; 
green squares denote galaxies with \textrm{C}~\textsc{iv} $S/N>3$ 
(32 objects) while orange squares denote those with \textrm{C}~\textsc{iv} $S/N<3$ 
(12 objects). gray squares represent galaxies showing asymmetric absorption on the red end of \textrm{C}~\textsc{iv} (2 objects, see Section~\ref{sec:civ}). 
The dotted line marks the division between the ``red sequence" and the ``blue cloud" at $z\sim1$ in the DEEP2 sample \citep{Willmer2006}.
\textbf{Bottom left:} Bundy SFR vs. stellar mass. Both stellar mass and Bundy SFR were derived from SED fitting assuming a Chabrier IMF. Color coding of the symbols is the same as in the top right panel. The blue filled triangles show where both Bundy SFR and $\mbox{SFR}_{UV}$ are available. The pink and purple dashed lines denote the SFR - $\mbox{M}_{*}$ relation for redshift ranges of 0.5 $< z <$1.0 and 1.0 $< z <$1.5, respectively \citep{whitaker2014}. \textbf{Bottom right:} Comparison between Bundy SFR and $\mbox{SFR}_{UV}$ for the subset of 25 objects in the AEGIS field with both types of measurements. $\mbox{SFR}_{UV}$ was measured from the dust-corrected UV luminosity from GALEX (see the text in Section \ref{sec:data} for a full description) and has been converted to a Chabrier IMF. The black dotted line indicates a 1:1 relation.}
\label{fig:galprop}
\end{figure*}

\begin{figure*}
\includegraphics[width=1.0\linewidth]{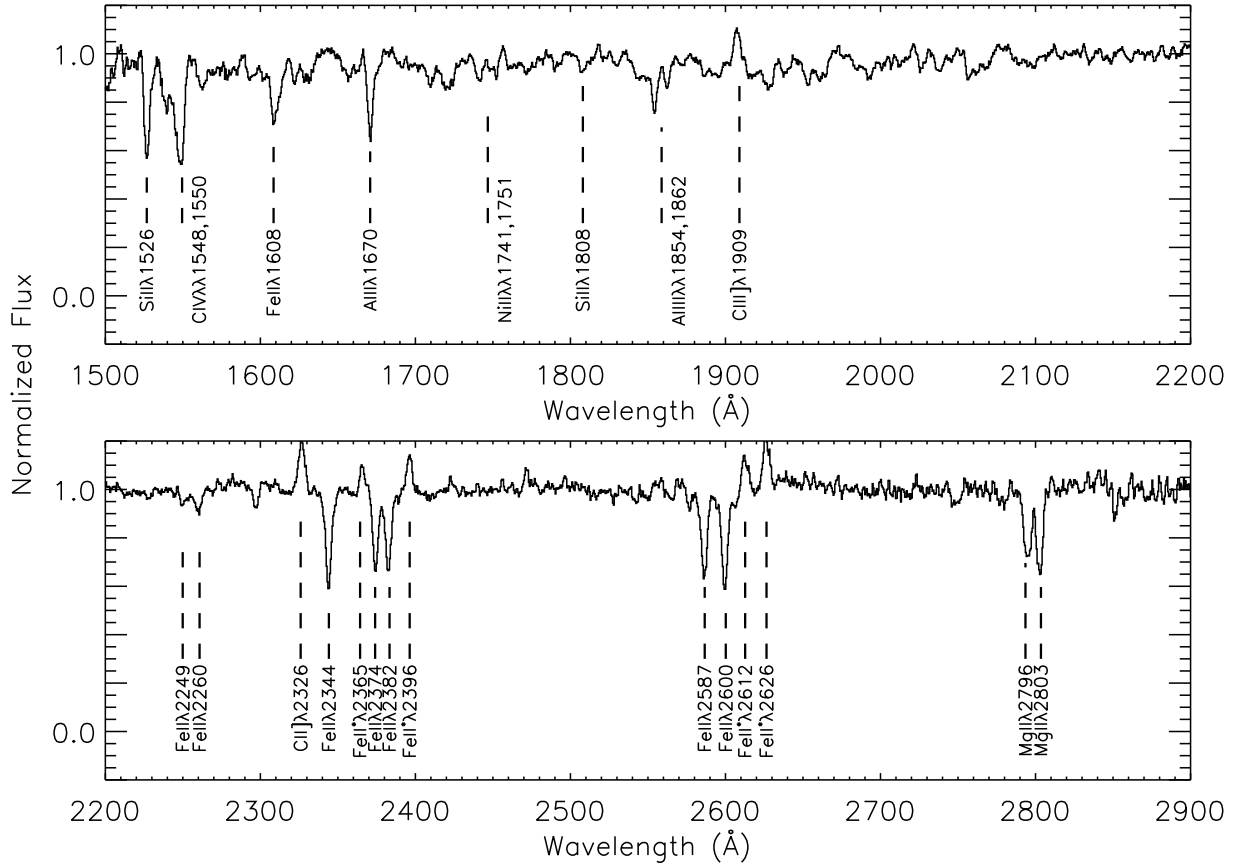}
\caption{Composite continuum-normalized UV spectrum constructed from 93 objects with \textrm{C}~\textsc{iv} coverage from the DEEP2/LRIS sample. Both strong and weak absorption features are identified. At the resolution of the LRIS spectra included in this composite, the \textrm{C}~\textsc{iv} doublet is blended. }
\label{fig:spec}
\end{figure*}

Galactic winds appear to be ubiquitous in galaxies with star-formation surface densities above a given threshold \citep{Heckman2001}, ranging from dwarf galaxies \citep{Martin1999,Heckman2001,Schwartz2004} to Ultra Luminous Infrared Galaxies \citep[ULIRGs;][]{Martin2005,Martin2009}, and are especially 
common at high redshift \citep[e.g.,][]{Pettini2002,Shapley2003,Weiner2009}.
Outflows have typically been probed with narrowband imaging (for example, $\mbox{H}\alpha$) and spectra of metal absorption lines. Unsaturated lines can be used to infer the ionic column density, which can potentially provides estimates of the mass and metallicity of outflowing material. Saturated lines are still useful for probing outflow kinematics.

Outflows have a complex multi-phase structure, which can be probed using neutral, low- and high-ionization absorption features. Most studies of outflows have focused on low-ionization absorption lines, although there are observations of high-ionization lines for a small number of individual galaxies and from specific examples of composite spectra, \citep[e.g.,][]{Heckman2001,Shapley2003,Schwartz2006,Grimes2009}. It is not clear how the kinematics of high-ionization features relate in general to those of low-ionization lines. Some results show that the high-ionization absorption lines have a larger blueshift than the low-ionization lines by studying the coronal-phase \textrm{O}~\textsc{vi} \citep[e.g.][]{Heckman2001,Grimes2009} and \textrm{C}~\textsc{iv} \citep{Wolfe2000}, while others claim that there is no significant difference between the kinematics of high- and low-ionization lines \citep{Pettini2002,Shapley2003,Chisholm2016}. In order to gain a robust handle on the overall mass outflow rate in galactic winds, it is important, however, to understand their detailed multi-phase structure, and compare the properties of different phases of the outflow. Despite the challenges associated with using \textrm{C}~\textsc{iv} as a tracer of galactic outflows, mainly because of the superposition of stellar and interstellar \textrm{C}~\textsc{iv} absorption and emission, this resonant transition potentially provides a very useful probe of the warm and hot phases of outflowing interstellar gas \citep[e.g.][]{Heckman2001,Shapley2003,Schwartz2006,Grimes2009}.

Starburst-driven outflows were first studied in the local universe \citep[e.g.,][]{Heckman1990, Martin1999} and at $z > 2$ \citep{Shapley2003, Pettini2001, Steidel2004}. Relatively recently, $z\sim1$ has become a focus of gas outflow studies. Low-ionization lines (e.g., from \textrm{Mg}~\textsc{ii} and \textrm{Fe}~\textsc{ii}) are commonly used for probing the cool phase of $z\sim1$ outflows \citep[e.g.][]{Weiner2009,Martin2012,Kornei2012,Kornei2013,Rubin2014}. Although outflows are much more commonly detected than inflows at this redshift, rare instances of infalling gas have also been discovered \citep{Coil2011,Martin2012,Kornei2012,Rubin2014}. In terms of outflow demographics, the outflow velocity traced by low-ionization lines is correlated with stellar mass, SFR and SFR surface density \citep{Weiner2009,Kornei2012}, and the strength of these lines increases with stellar mass, $B$-band luminosity, $\ub$ color and SFR \citep{Martin2012,Rubin2014}. Finally, in most star-forming galaxies at $z \leqslant 1.5$ outflows are inferred to have a bipolar geometry, as indicated by the detection fraction of blueshifted gas and extraplanar absorption \citep[e.g.][]{Rupke2005,Martin2012,Bordoloi2014,Rubin2014} and the fact that face-on galaxies tend to have a slightly faster outflow speed \citep{Kornei2012}.

In this study we focus on the high-ionization \textrm{C}~\textsc{iv}$\lambda\lambda$1548,1550 doublet for the first time at $z\sim1$. We also measure the kinematics of other far-UV low-ionization lines (\textrm{Si}~\textsc{ii}$\lambda$1526, \textrm{Fe}~\textsc{ii}$\lambda$1608 and \textrm{Al}~\textsc{ii}$\lambda$1670) and compare them with motions traced by near-UV \textrm{Fe}~\textsc{ii} and \textrm{Mg}~\textsc{ii} lines. Both observations and simulations have shown that \textrm{C}~\textsc{iv} appears to trace two distinct types of gas: the hot (T $> 10^{5}$ K), denser, collisionally ionized gas in the coronal phase also probed by \textrm{O}~\textsc{vi} absorption, and the warm (T $< 10^{5}$ K), less dense, photoionized gas that may be comoving with the gas the low-ionization lines trace \citep{Lehner2011,Shen2013}. Given the intermediate ionization stage of \textrm{C}~\textsc{iv} relative to that of  \textrm{O}~\textsc{vi} and the low-ionization lines, it is essential to examine how its kinematics differ from those inferred from low-ionization lines. Comparing the properties of different ionization phases will provide us a better understanding of the multi-phase structure of gas flows.

We provide a brief overview of the observations and data reduction, and describe the selection criteria of our \textrm{C}~\textsc{iv} sample in Section~\ref{sec:data}. In Section~\ref{sec:meas}, we describe the measurements of gas kinematics traced by \textrm{C}~\textsc{iv} and by other low-ionization lines. We compare the kinematics of \textrm{C}~\textsc{iv} and the far-UV low-ionization lines with those of the near-UV \textrm{Fe}~\textsc{ii} and \textrm{Mg}~\textsc{ii} lines in Section~\ref{sec:kine}, and present scaling relations between outflows and galaxy properties in Section~\ref{sec:galprop}. Finally, we summarize and discuss our results in Section~\ref{sec:sum}. 

Throughout this paper, we use a standard $\Lambda$-CDM model with $\Omega_{m}=0.3$, $\Omega_{\Lambda}=0.7$ and $H_{0}=$70 km $\mbox{s}^{-1}$. All wavelengths are measured in vacuum. Magnitudes and colors are on the Vega system.

\section{Observations, Data Reduction and Sample}
\label{sec:data}

As described in \citet{Martin2012}, our sample was drawn from the Deep Extragalatic Evolutionary Probe 2 \citep[DEEP2;][]{Newman2013} galaxy redshift survey and observed
with the dichroic Low Resolution Imager and Spectrometer \citep[LRIS,][]{Oke1995,Steidel2004} on the Keck I telescope. The LRIS observations were conducted over the course of 4 observing runs from 2007 to 2009. Data were collected for 9 multi-object slitmasks using $1.2''$ slits. 208 galaxies were targeted based on 
 apparent magnitude $B<24.5$ varying in redshift from $z=0.4-1.4$ with $\langle z \rangle=1.01$. Two sets of masks were developed for the DEEP2/LRIS sample: one set (6 masks) was observed with the 400 lines $\mbox{mm}^{-1}$ grism on the blue side
 with the average effective resolution of 435 $\mbox{km s}^{-1}$ full width at half-maximum (FWHM), and the 831 lines $\mbox{mm}^{-1}$ grating on the red side with FWHM of 150 $\mbox{km s}^{-1}$ ($R=700$). The other set (3 masks) was observed with the 600 lines $\mbox{mm}^{-1}$ grism on the blue side with FWHM of 282 $\mbox{km s}^{-1}$, and the 600 lines $\mbox{mm}^{-1}$ red grating with FWHM of 220 $\mbox{km s}^{-1}$ ($R=1100$). The integration time for most 400-line masks ranges from 5 to 9 hours. Galaxies at $1.19\leqslant z \leqslant1.35$ were prioritized on these masks, as the \textrm{C}~\textsc{iv}$\lambda\lambda1548, 1550$ doublet fell in a region of decent sensitivity ($\lambda\geqslant3400\mbox{\AA}$). The 600-line grism masks were mainly designed for less 
 optimal observing conditions in order to obtain the near-UV spectroscopy of brighter galaxies, with typically shorter total exposure time (3-5 hours). For objects observed with the 400-line masks, the blue side of the spectra were designed to contain rest-UV interstellar absorption lines, with the red side covering the [\textrm{O}~\textsc{ii}] emission doublet. The 600-line spectra were typically designed for continuous wavelength coverage between the blue and red sides.

The data were reduced as described in \citet{Martin2012}. In brief, the two-dimensional (2D) spectra were first flat fielded, 
 cleaned of cosmic rays and background-subtracted. Individual 2D exposures were then combined, and extracted into one dimension (1D), and wavelength and flux 
 calibrated. In most cases, the systemic redshift of each galaxy was determined from the [\textrm{O}~\textsc{ii}] doublet, which typically fell on the red side of the LRIS spectrum. For galaxies with no [\textrm{O}~\textsc{ii}] coverage in the LRIS spectra, the DEEP2 Keck/DEIMOS redshift was adopted. As discussed in \citet{Martin2012}, we detect no systematic differences on average between the DEEP2 and LRIS [\textrm{O}~\textsc{ii}] systemic redshifts, for the 167 galaxies with both redshift measurements.
 
Rest-frame $B$-band luminosities and $\ub$ colors were taken from \citet{Willmer2006}, and the stellar mass was derived from SED fitting with $BRIK$ photometry, assuming a \citet{Chabrier2003} initial mass function (IMF) and \citet{Bruzual2003} spectral templates (see \citet{Bundy2006} for a full description). Two types of SFR are shown in Figure~\ref{fig:galprop}: the one from \citet{Bundy2006} was estimated from SED fitting assuming a Chabrier IMF, while $\mbox{SFR}_{UV}$ was derived from $Galaxy$ $Evolution$ $Explorer$ (GALEX) measurements and corrected for dust attenuation. GALEX far-UV and near-UV photometry, along with $B$-band observations were used to determine $\beta$, the spectral slope where the continuum flux is in the form $f_{\lambda} \propto \lambda^{\beta}$ over the rest-frame wavelength range of 1250 - 2500$\mbox{\AA}$. The $\beta/A_{UV}$ relationship from \citet{Seibert2005} was then used to estimate the UV attenuation and thus to correct GALEX luminosities. Given the relation from \citet{Salim2007}, the dust-corrected UV SFR was then calculated assuming a  \citet{Salpeter1955} IMF and converted to a Chabrier IMF by dividing by a factor of 1.8. 25 out of 93 objects in our sample are in the AEGIS field, where both types of SFR measurements are available. Since the SED-based SFR has several limitations and possible biases \citep[e.g., age-dust-metallicity degeneracy, systematic dependence on assumed star-formation history;][]{Papo2001}, we only adopted $\mbox{SFR}_{UV}$ for the analysis presented here, despite its restricted coverage to our sample.

One unique aspect of the DEEP2/LRIS sample is that we optimized our mask design to cover \textrm{C}~\textsc{iv} and the far-UV spectral region at $z\sim1$. With this approach, we obtained \textrm{C}~\textsc{iv} coverage for 93 out of 208 objects. A composite rest-UV spectrum of these 93 objects is shown in Figure~\ref{fig:spec}, with the strongest rest-UV interstellar features marked. Of the 93 galaxies with \textrm{C}~\textsc{iv} coverage, 46 had continuum signal-to-noise ratio ($S/N$) $>5$ in the vicinity of \textrm{C}~\textsc{iv} from $1570\mbox{\AA}$ to $1590\mbox{\AA}$, a region in which the continuum is free of line features. We adopted this $S/N$ threshold to isolate objects with robust \textrm{C}~\textsc{iv} absorption-line measurements or limits. These 46 objects comprise the main sample for analysis, and they are representative of the full sample of galaxies with \textrm{C}~\textsc{iv} coverage in terms of stellar mass, SFR, $B$-band luminosity and $\ub$ color, except being slightly bluer than average at the brightest $B$-band absolute magnitudes (Figure~\ref{fig:galprop}). In the full sample of 93 objects with \textrm{C}~\textsc{iv} coverage, 81 were observed with the 400-line grism, and 12 with the 600-line grism. However, all but one object (ID: 32016683) in the high $S/N$ sample were observed with the 600-line grism.

As illustrated in Figure~\ref{fig:galprop}, our sample with \textrm{C}~\textsc{iv} coverage spans in redshift from 1.01 to 1.35 with a median of 1.25, $B$-band luminosities $-19.31>\mbox{M}_{B} >-22.19$ with a median of -20.74, stellar mass $9.11< \log(\mbox{M}_{*}/\mbox{M}_{\sun}) <11.26$ with a median of 10.10, $\ub$ color from 0.18 to 0.99 with a median of 0.58 and $\mbox{SFR}_{UV}$ from 2 to 77 $\mbox{M}_{\sun}$ $\mbox{yr}^{-1}$ with a median of 14 $\mbox{M}_{\sun}$ $\mbox{yr}^{-1}$. These galaxies primarily fall in the ``blue cloud" \citep{Faber2007} of star-forming galaxies at $z\sim1$.

\section{Measurements}
\label{sec:meas}

\begin{figure}
\includegraphics[width=1.0\linewidth]{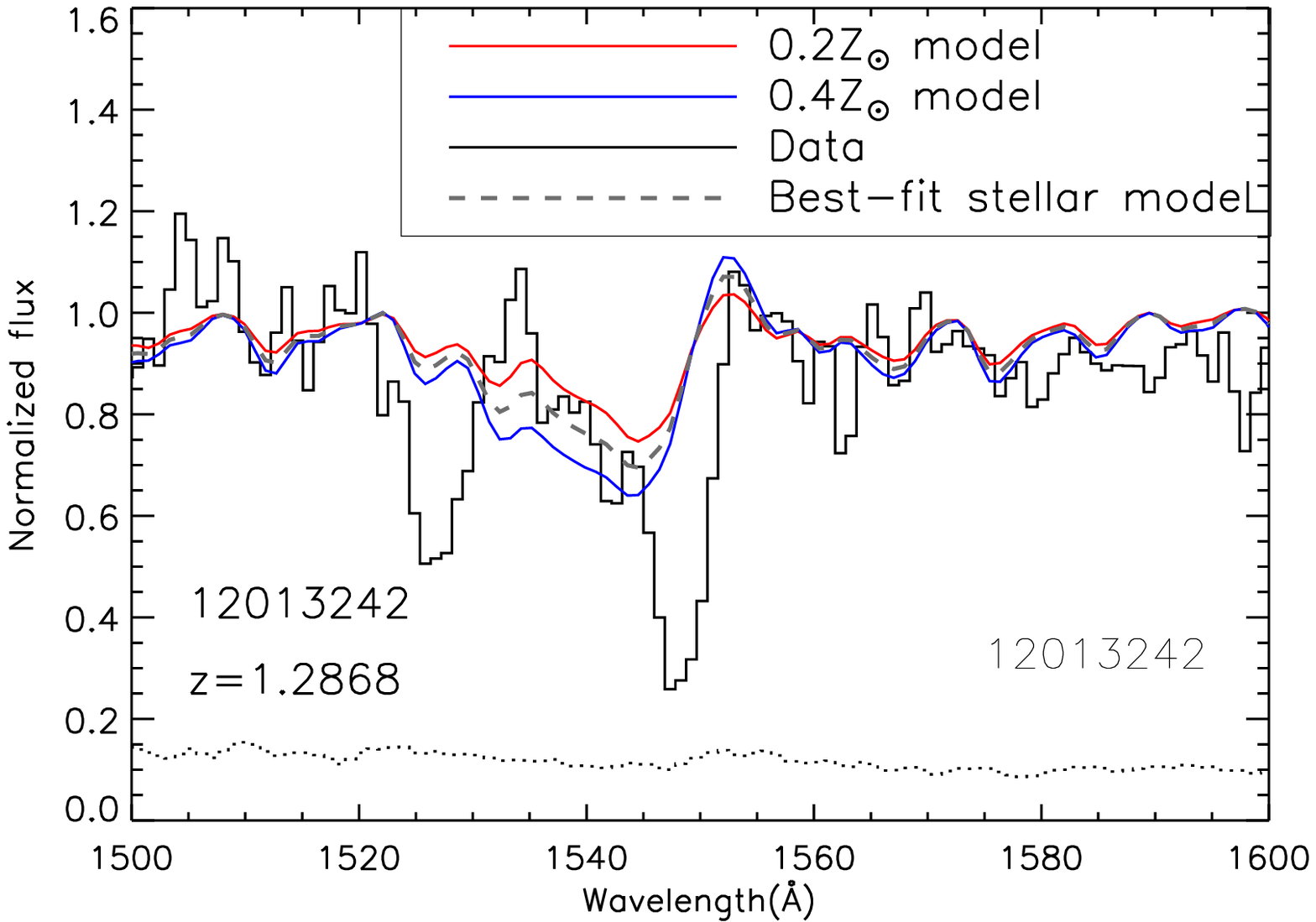}
\includegraphics[width=1.0\linewidth]{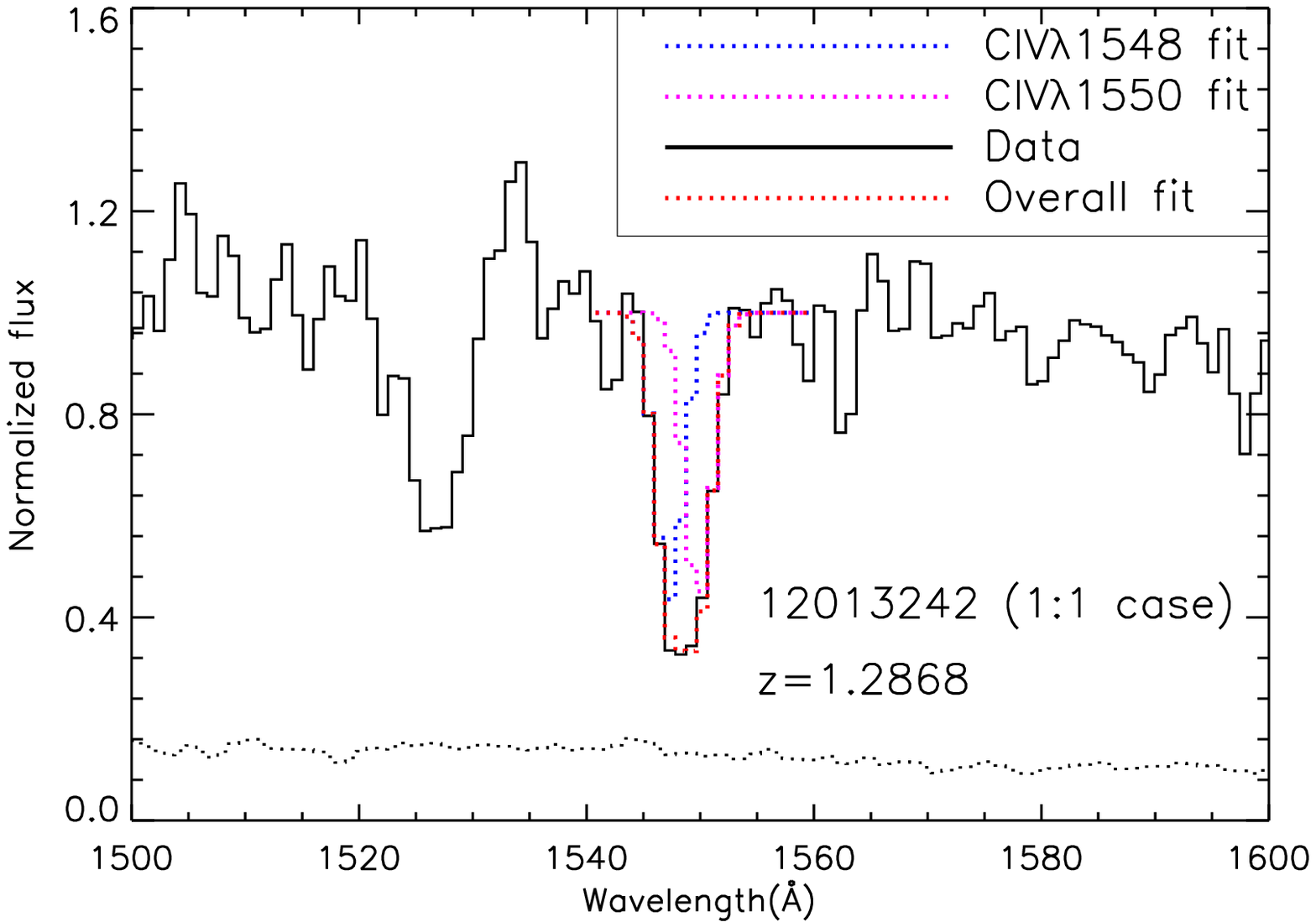}
\caption{\textbf{Top:} Determination of the best-fit stellar model of object 12013242 (continuum $S/N=16.4$). The red and blue lines indicate two stellar models (0.2$Z_{\sun}$ and 0.4$Z_{\sun}$, with residuals 0.0624 and -0.0659, respectively) bracketing the P-Cygni absorption feature of \textrm{C}~\textsc{iv}. The best-fit stellar model was calculated from a linear combination of these two models based on inverse residual weighting. The black dotted line represents the corresponding error spectrum of this object. \textbf{Bottom:} Deblending the interstellar \textrm{C}~\textsc{iv} absorption trough into two Gaussian profiles. The fits for \textrm{C}~\textsc{iv}$\lambda1548$ and \textrm{C}~\textsc{iv}$\lambda1550$, respectively, are shown in blue and magenta dotted lines, and the overall fit is marked as the red dotted line. The centroid of the doublet members was fixed at the rest-wavelength ratio, and the EW of each member was assumed equal.}
\label{fig:meastep}
\end{figure}

In this section, we describe the measurements of strong far-UV absorption features at wavelengths 1500$\mbox{\AA}$ to 1900$\mbox{\AA}$, including the high-ionization doublet \textrm{C}~\textsc{iv}$\lambda\lambda1548,1550$, and low-ionization \textrm{Si}~\textsc{ii}$\lambda1526$, \textrm{Fe}~\textsc{ii}$\lambda1608$ and \textrm{Al}~\textsc{ii}$\lambda1670$ features. We adopted single-component Gaussian fits as the simplest possible functional form to describe our absorption lines, as the data are not of sufficient resolution and $S/N$ to consider more complicated models (especially for \textrm{C}~\textsc{iv}, which already requires both stellar and interstellar model components).

\subsection{Measurement of \textrm{C}~\textsc{iv}}
\label{sec:civ}
 
The \textrm{C}~\textsc{iv} absorption profile is a complex superposition of stellar and interstellar components that needs to be disentangled in order to characterize the interstellar \textrm{C}~\textsc{iv} absorption. The doublet members, \textrm{C}~\textsc{iv}$\lambda$1548 and \textrm{C}~\textsc{iv}$\lambda$1550, have a relative oscillator strength of 2:1. Therefore, the doublet ratio ranges from 2:1 in the optically thin case to 1:1 in the optically thick case. In this paper we adopted the saturated doublet ratio of 1:1 (yielding a rest-frame wavelength of 1549.5 $\mbox{\AA}$ for the blended \textrm{C}~\textsc{iv} profile), based on the typical EW of the \textrm{C}~\textsc{iv} feature in our sample \citep{Pettini2002}. Assuming the doublet ratio of 2:1 results in a velocity shift $\sim$ 84 km $\mbox{s}^{-1}$ less blueshifted than in the 1:1 case, although the optically-thin 2:1 ratio is unlikely to apply given the type of systems we probe.

We fit the rest-frame UV continuum in the 1D calibrated LRIS spectra using spectral regions (`windows') defined by \citet{Rix2004}, which are clean of spectral features. We applied these `windows' for fitting the continuum for all the spectra with the IRAF $continuum$ routine, using a $spline3$ function of order $=8$. In cases where the fitted continuum level does not quite follow the observed spectrum due to the 
limited coverage of windows from \citet{Rix2004}, 
we added additional windows, customized for each object, to keep the fitted continuum 
reasonable and make sure that the continuum scatters around unity. 

The complexity of the observed 
\textrm{C}~\textsc{iv} absorption profile stems from the joint contributions of stellar and interstellar components. The stellar part is the P-Cygni profile mainly produced by O and B stars, where the presence of blue shifted absorption and redshifted emission is the characteristic profile of a fast-moving stellar wind. In order to isolate the interstellar \textrm{C}~\textsc{iv} absorption, we must normalize the continuum by the stellar P-Cygni profile.
The spectral synthesis code of \citet{Leitherer2010} outputs the rest-frame UV spectra for stellar populations based on the assumption of constant star formation, a Salpeter IMF with mass limits of 1 and 100 $M_{\sun}$, and a range in metallicity including $0.05Z_{\sun}$, $0.2Z_{\sun}$, $0.4Z_{\sun}$, $1.0Z_{\sun}$ and $2.0Z_{\sun}$. These spectra include the predicted stellar P-Cygni profiles based on models for the winds of hot stars from WM-Basic, which ranged in wavelength from 900 to 3000$\mbox{\AA}$ at a resolution of 0.4$\mbox{\AA}$. 
 We smoothed all 5 models to the resolution of our LRIS spectra
(282 km $\mbox{s}^{-1}$ FWHM or 1.4$\mbox{\AA}$ at 1550$\mbox{\AA}$ for 600-line masks, and 435 km $\mbox{s}^{-1}$ FWHM or 2.2$\mbox{\AA}$ at 1550$\mbox{\AA}$ for 400-line masks) and continuum normalized them using the windows from \citet{Rix2004}.
For each stellar model we calculated the residual to the data based on inverse square weighting over the blue wing of \textrm{C}~\textsc{iv} from $1535\mbox{\AA}$ to $1544\mbox{\AA}$.

\begin{figure*}
\includegraphics[width=0.5\linewidth]{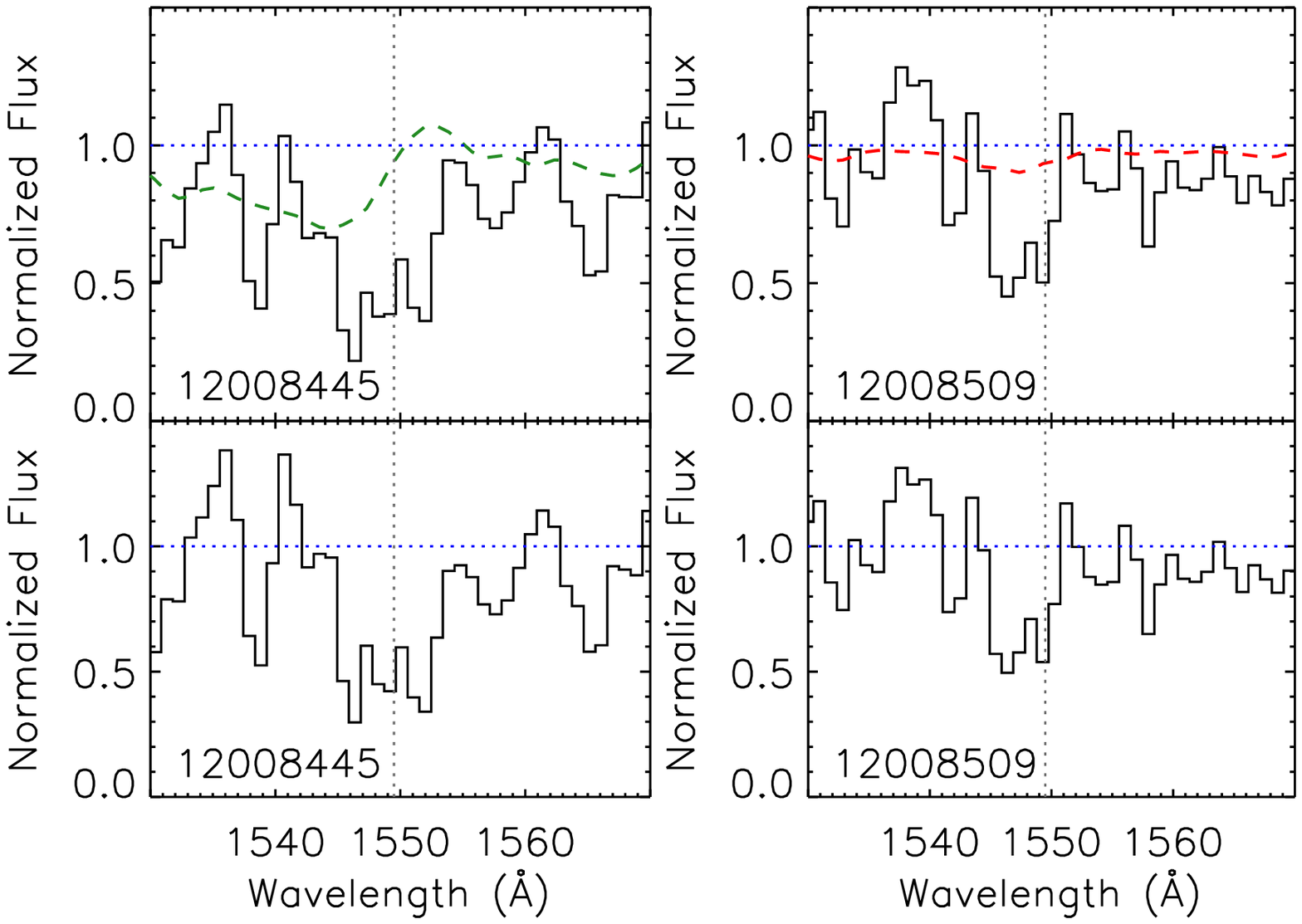}
\includegraphics[width=0.5\linewidth]{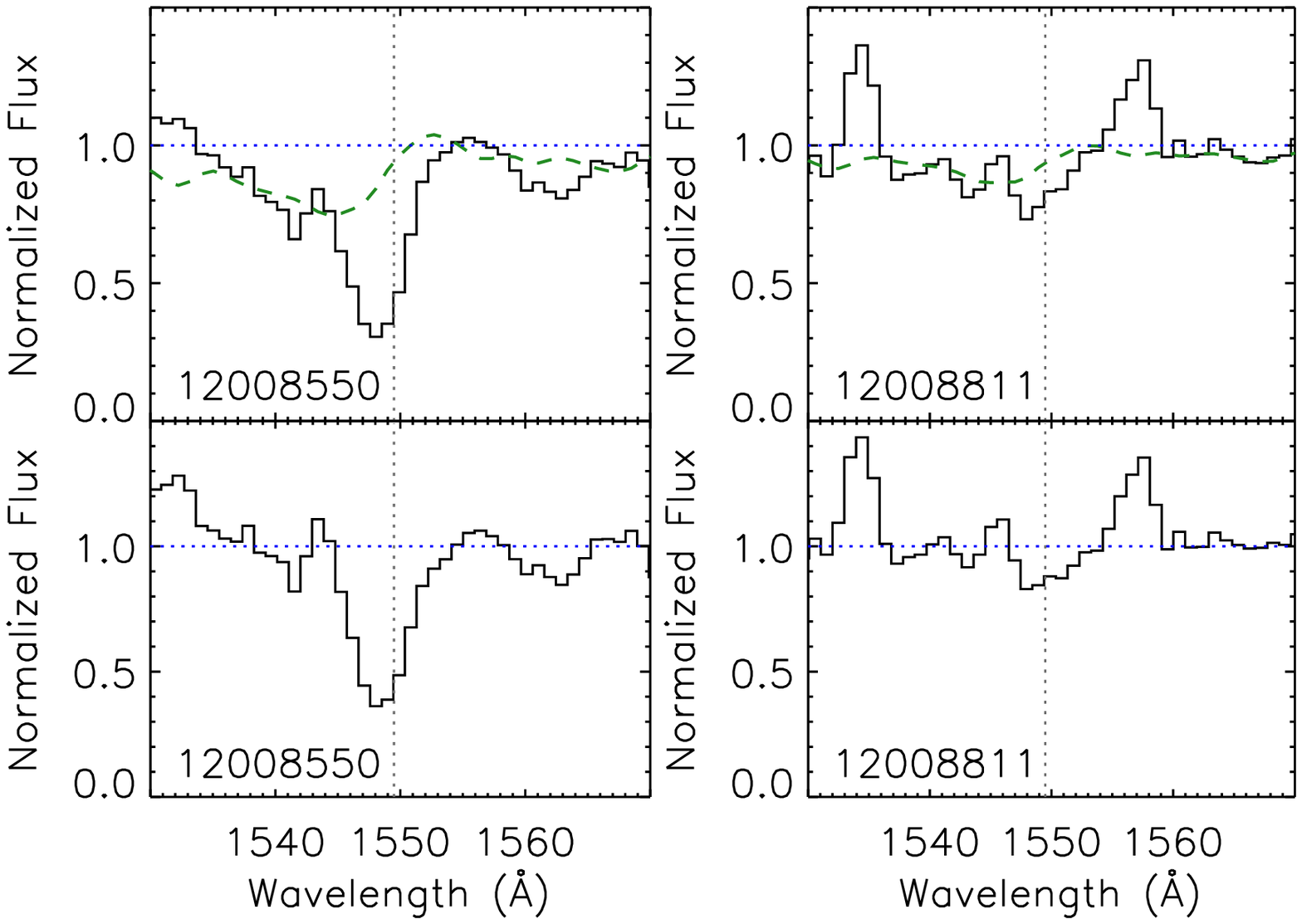}
\includegraphics[width=0.5\linewidth]{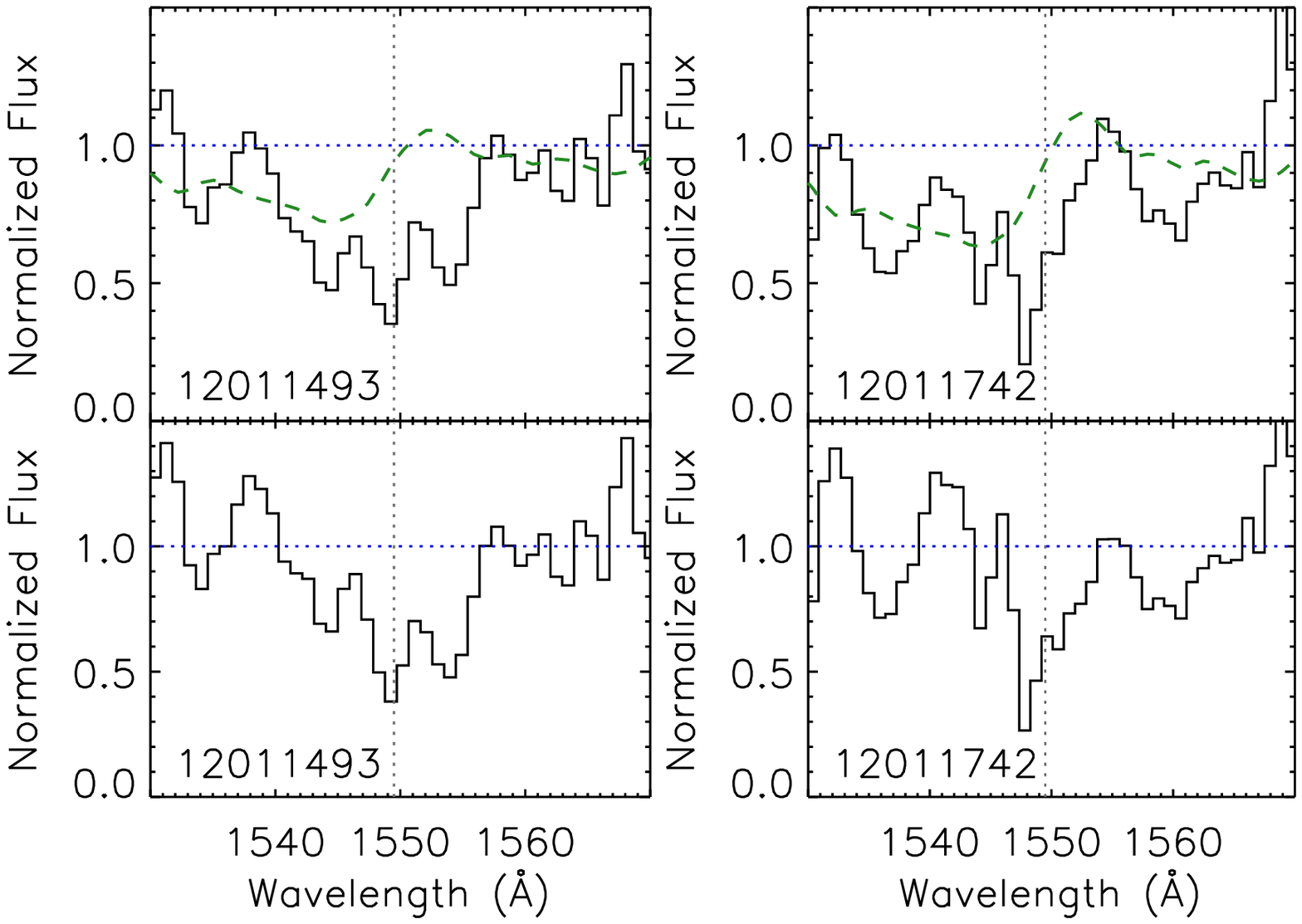}
\includegraphics[width=0.5\linewidth]{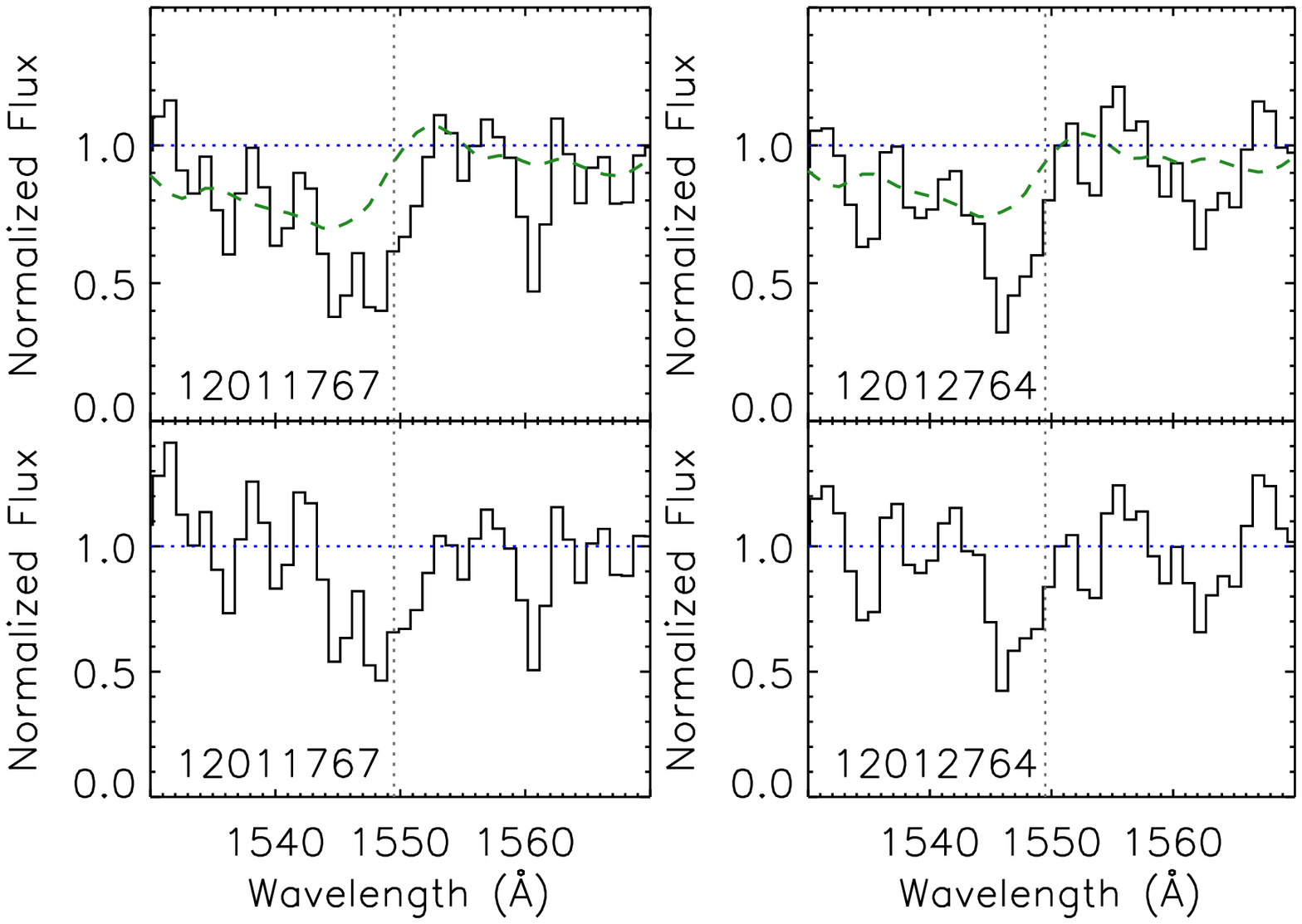}
\includegraphics[width=0.5\linewidth]{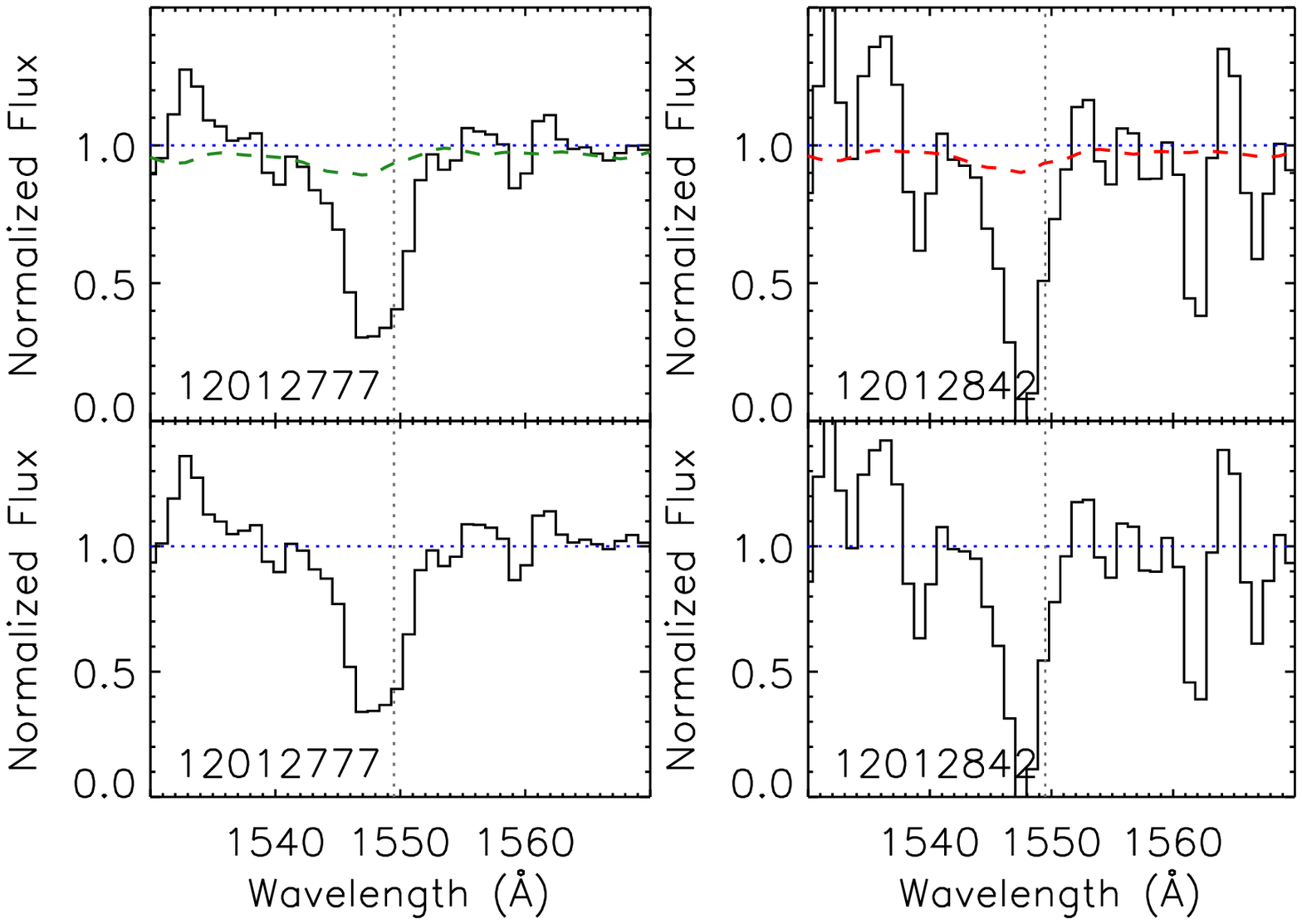}
\includegraphics[width=0.5\linewidth]{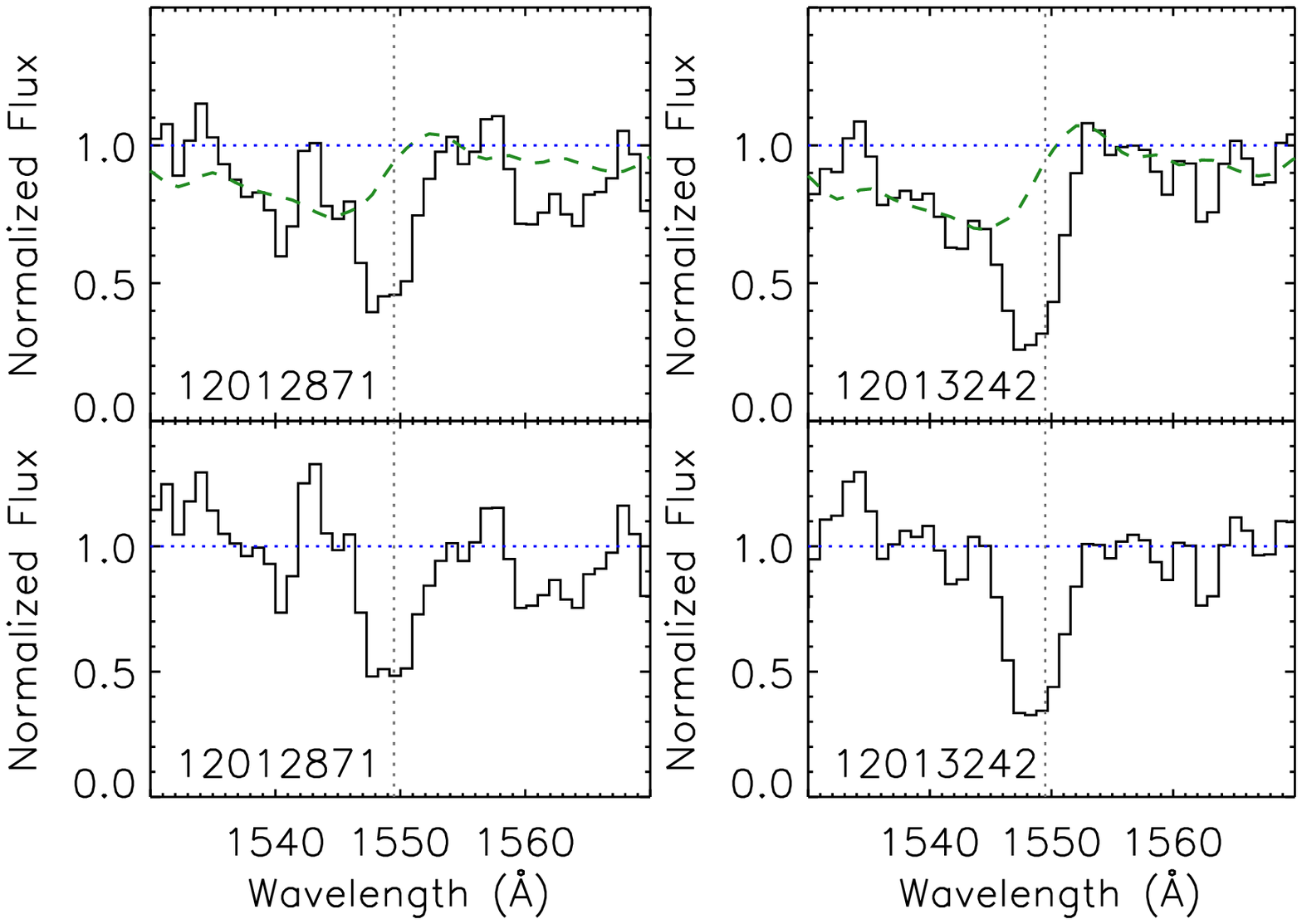}
\caption{Continuum-normalized spectra (top) and spectra normalized by the best-fit stellar model (bottom) for 32 objects with continuum $S/N>5$ and \textrm{C}~\textsc{iv} EW measurements $>$ 3$\sigma$, showing individual detection of \textrm{C}~\textsc{iv} absorption. The blue horizontal dotted line indicates the continuum level, the gray vertical dotted line suggests the rest-frame wavelength of blended \textrm{C}~\textsc{iv} doublet in the 1:1 case (1549.5 $\mbox{\AA}$). The dashed line shows the best-fit stellar model for each object. A green dashed line indicates that the object was bracketed with neighboring metallicities, which were linearly combined to determine the best-fit stellar model. A red dashed line indicates that the object failed to be bracketed with neighboring metallicities, so the best-fit stellar model from \citet{Leitherer2010} was either $0.05Z_{\sun}$ or $2.0Z_{\sun}$.}
\label{fig:civpro}
\end{figure*}

\begin{Contfigure*}
\includegraphics[width=0.5\linewidth]{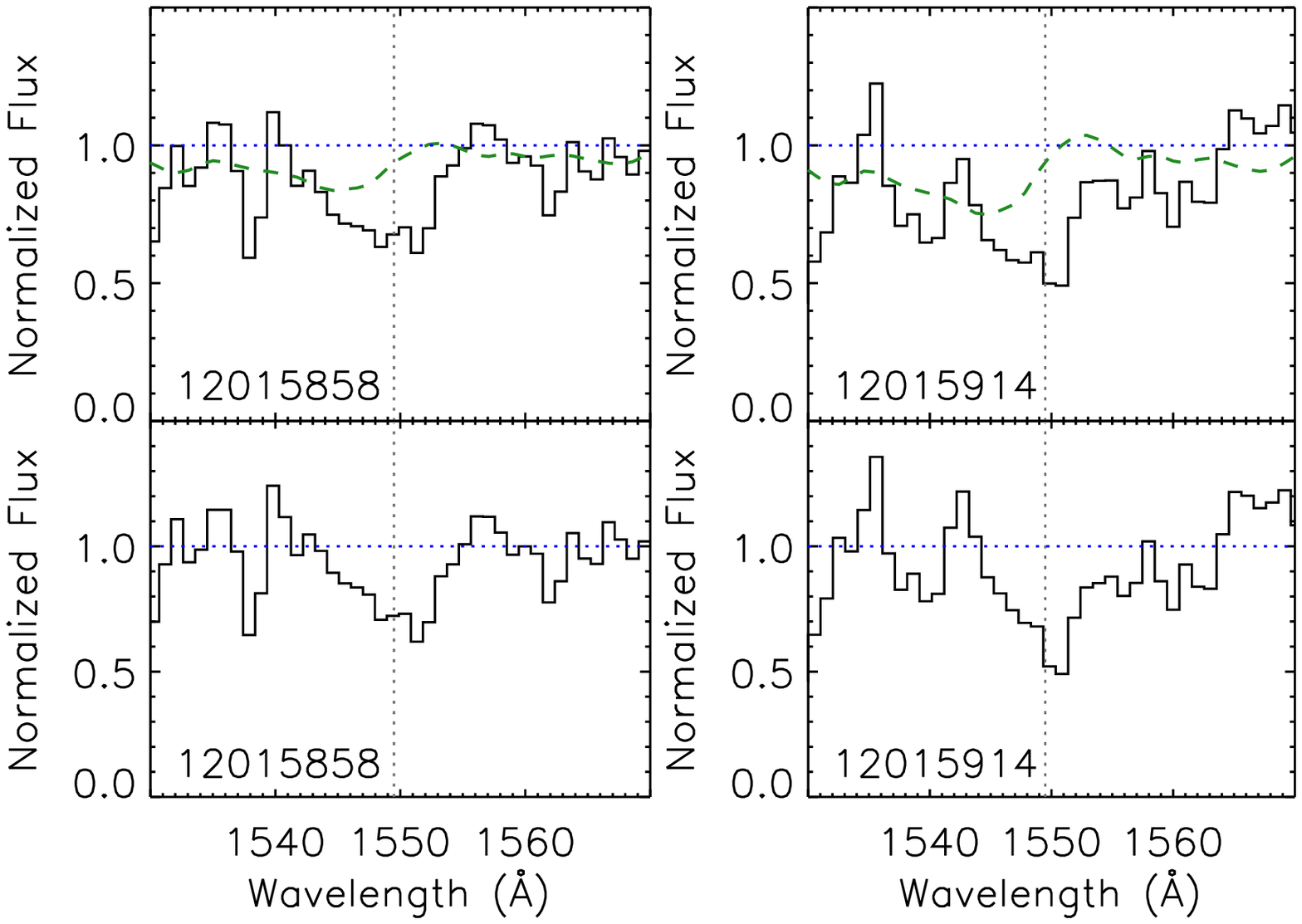}
\includegraphics[width=0.5\linewidth]{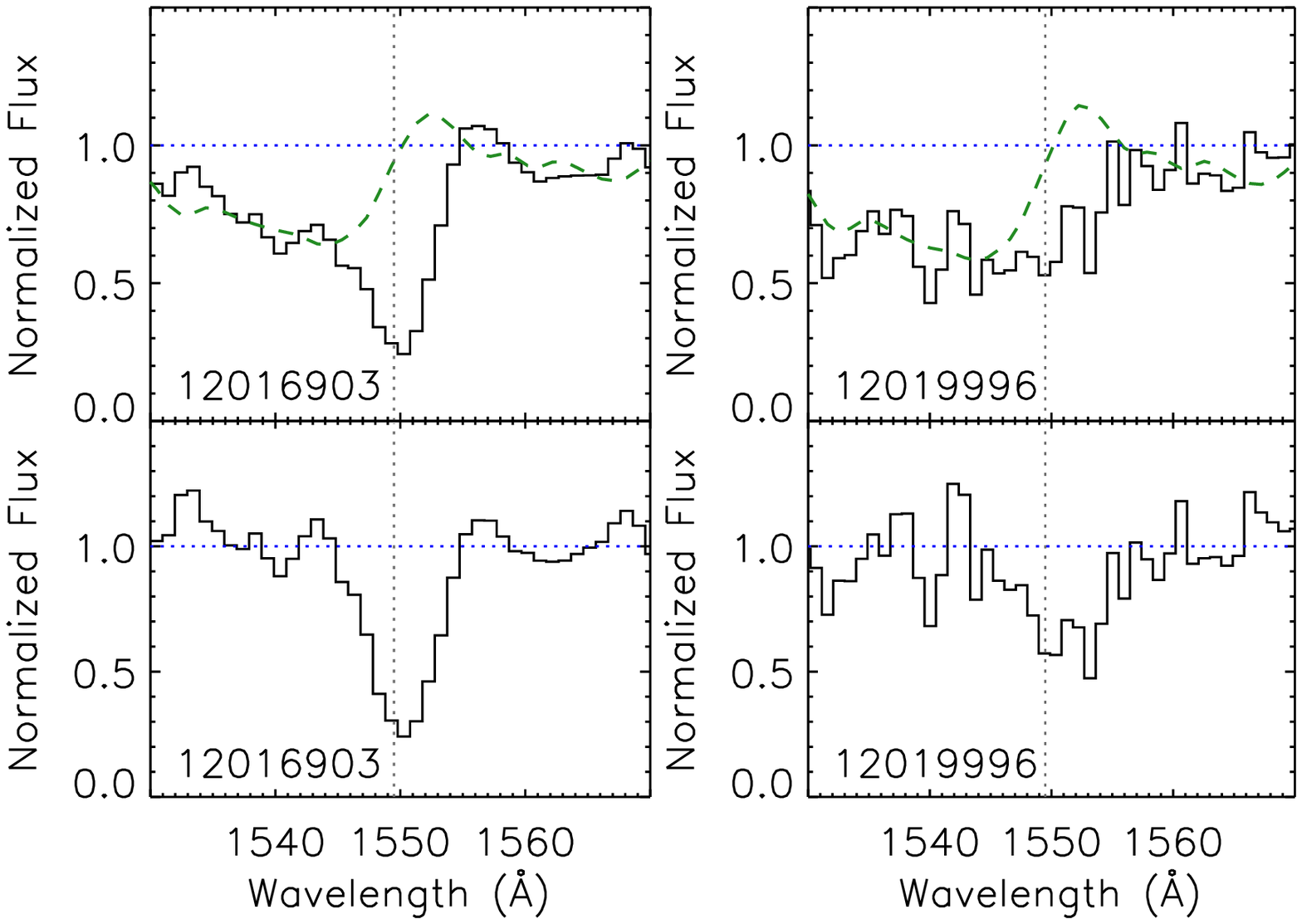}
\includegraphics[width=0.5\linewidth]{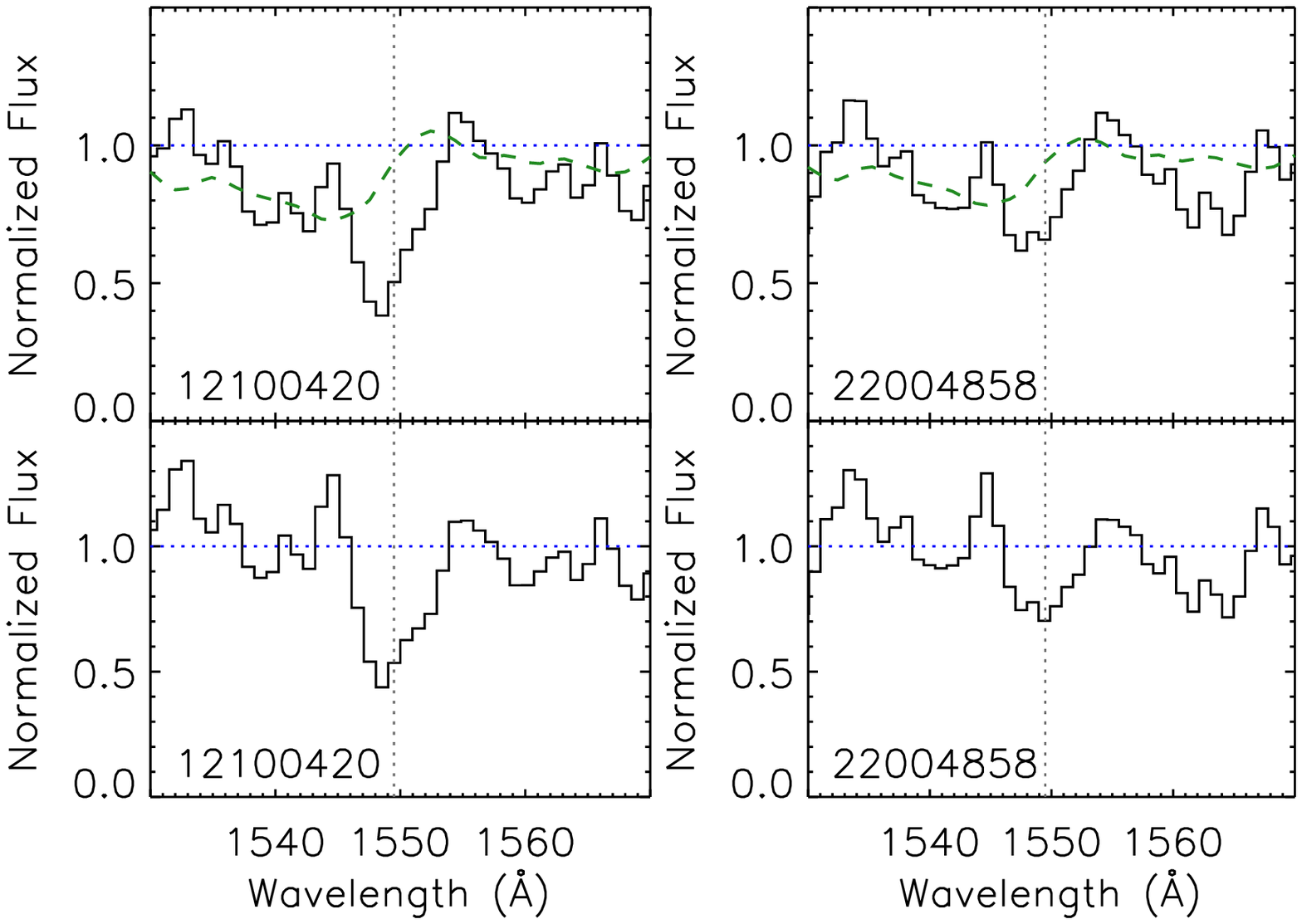}
\includegraphics[width=0.5\linewidth]{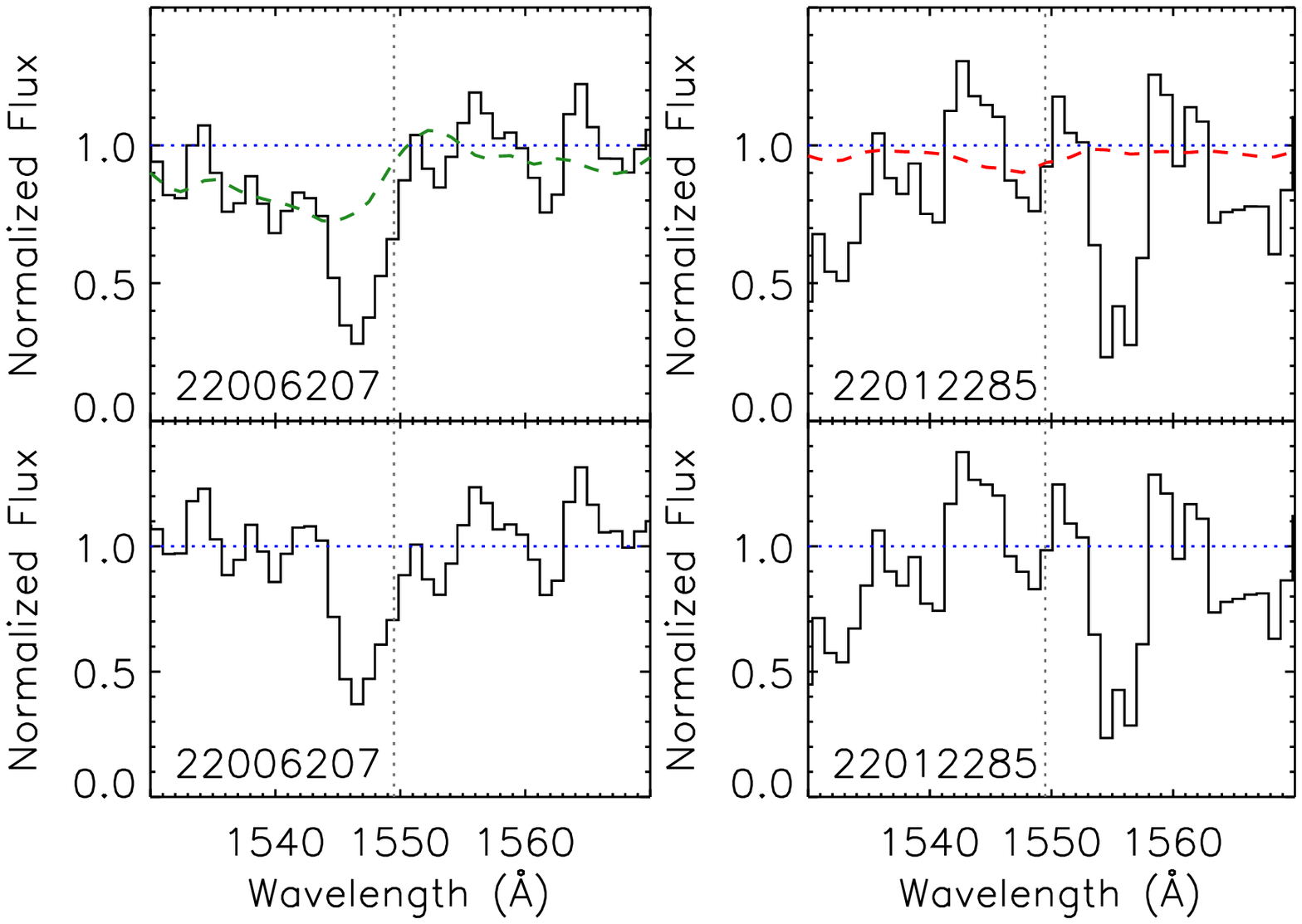}
\includegraphics[width=0.5\linewidth]{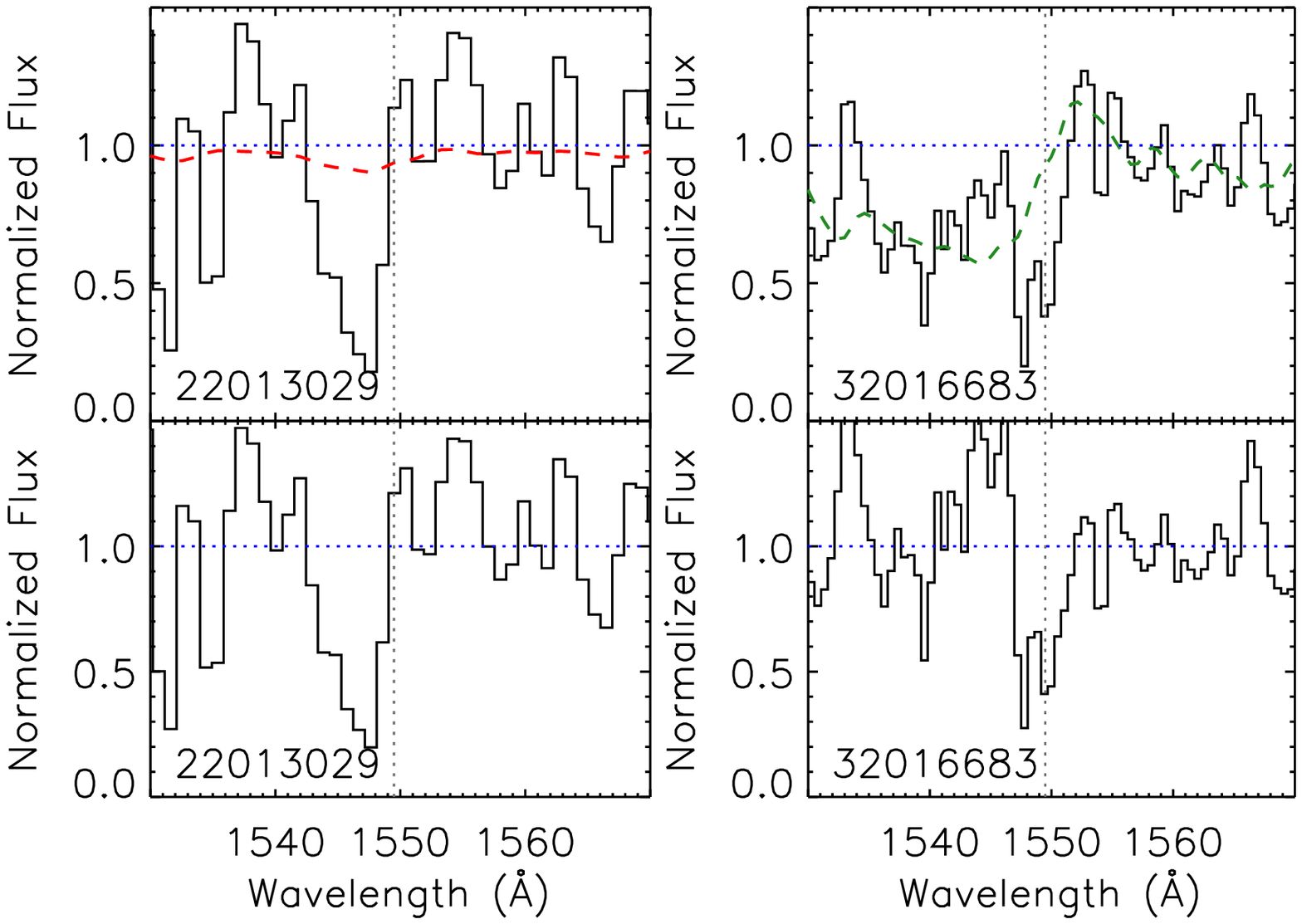}
\includegraphics[width=0.5\linewidth]{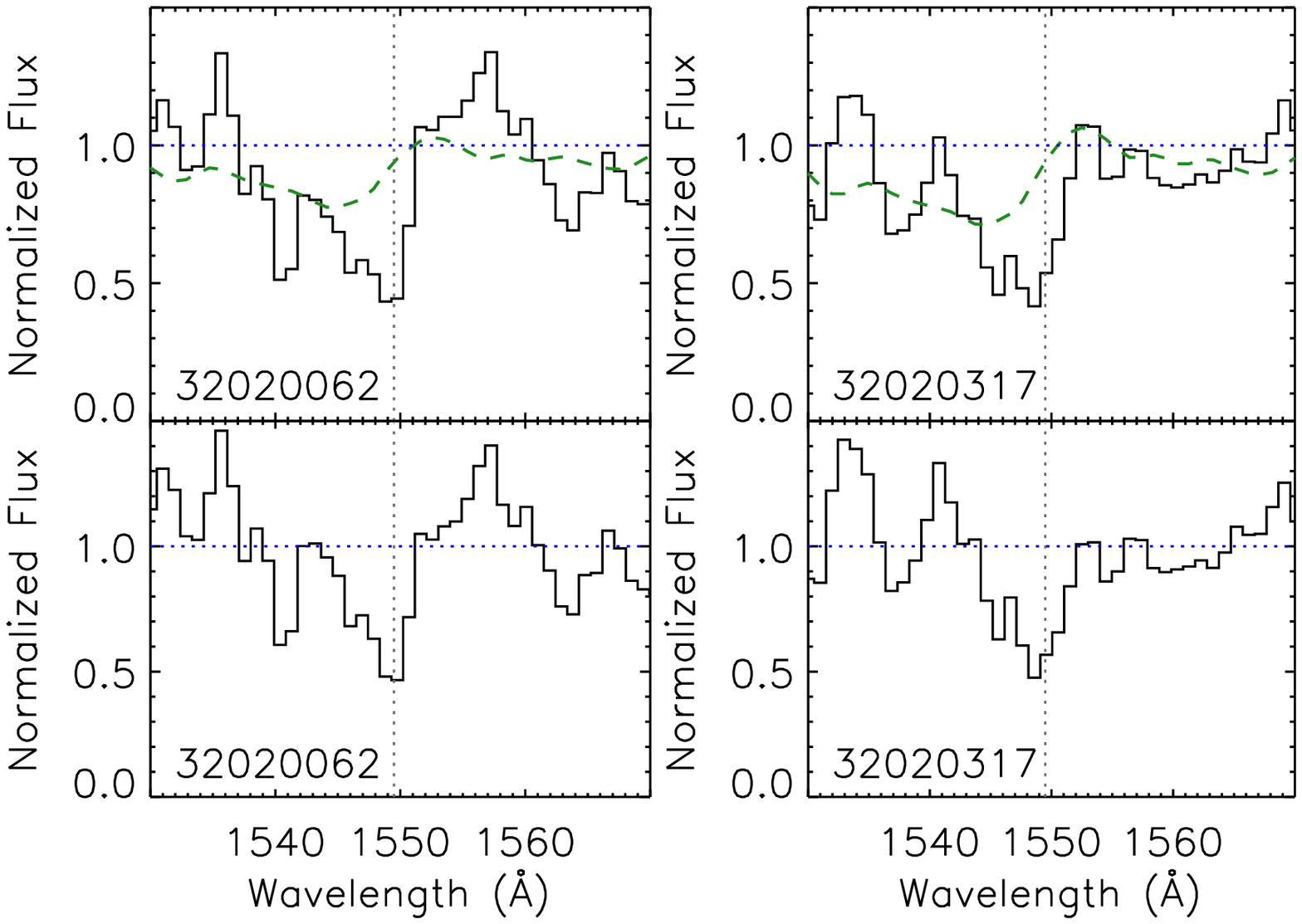}

\caption{Continuum-normalized spectra (top) and spectra normalized by the best-fit stellar model (bottom) for 32 objects with continuum $S/N>5$ and \textrm{C}~\textsc{iv} EW measurements $>$ 3$\sigma$, showing individual detection of \textrm{C}~\textsc{iv} absorption. The blue horizontal dotted line indicates the continuum level, the gray vertical dotted line suggests the rest-frame wavelength of blended \textrm{C}~\textsc{iv} doublet in the 1:1 case (1549.5 $\mbox{\AA}$). The dashed line shows the best-fit stellar model for each object. A green dashed line indicates that the object was bracketed with neighboring metallicities, which were linearly combined to determine the best-fit stellar model. A red dashed line indicates that the object failed to be bracketed with neighboring metallicities, so the best-fit stellar model from \citet{Leitherer2010} was either $0.05Z_{\sun}$ or $2.0Z_{\sun}$.}
\end{Contfigure*}

\begin{Contfigure*}
\includegraphics[width=0.5\linewidth]{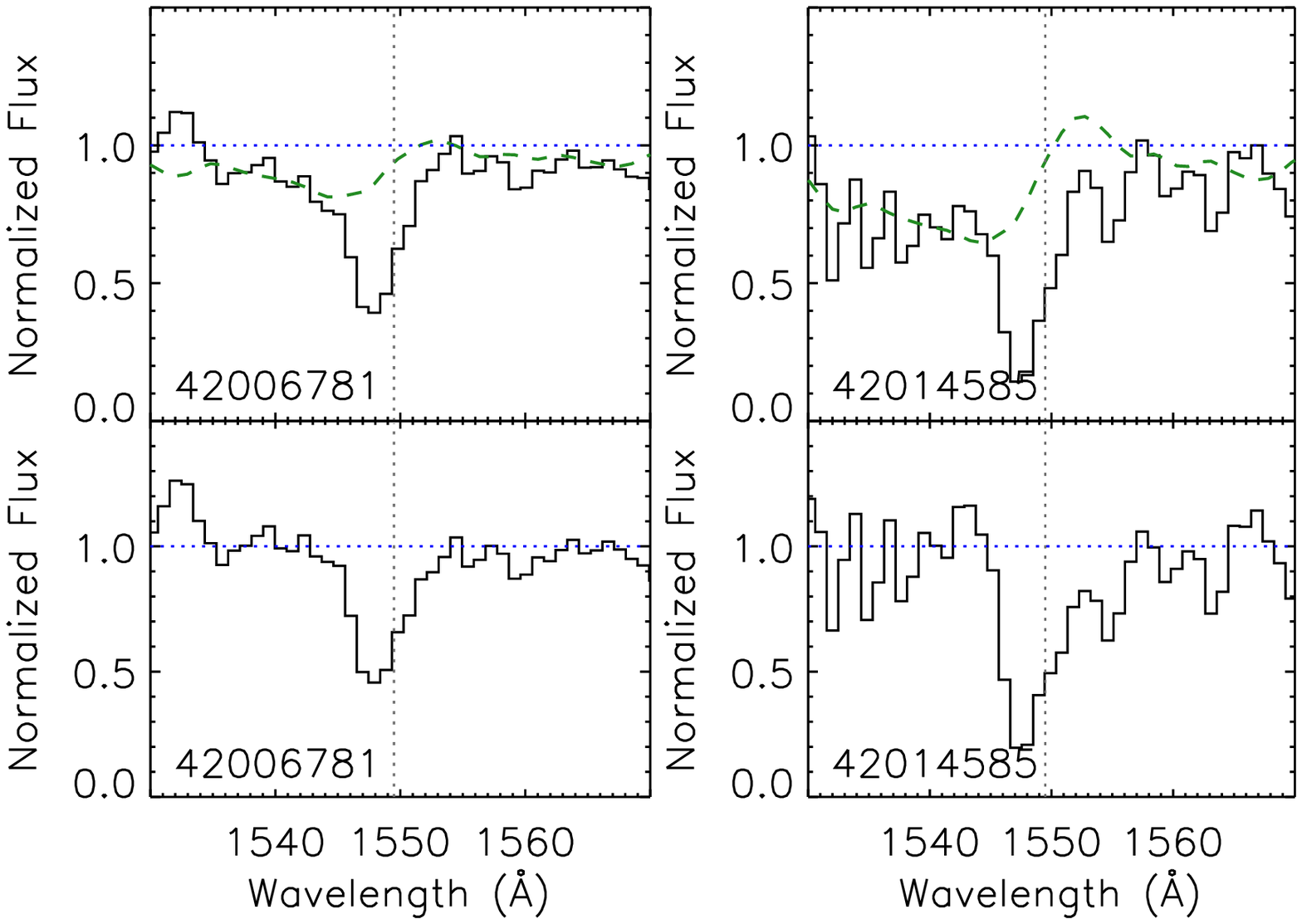}
\includegraphics[width=0.5\linewidth]{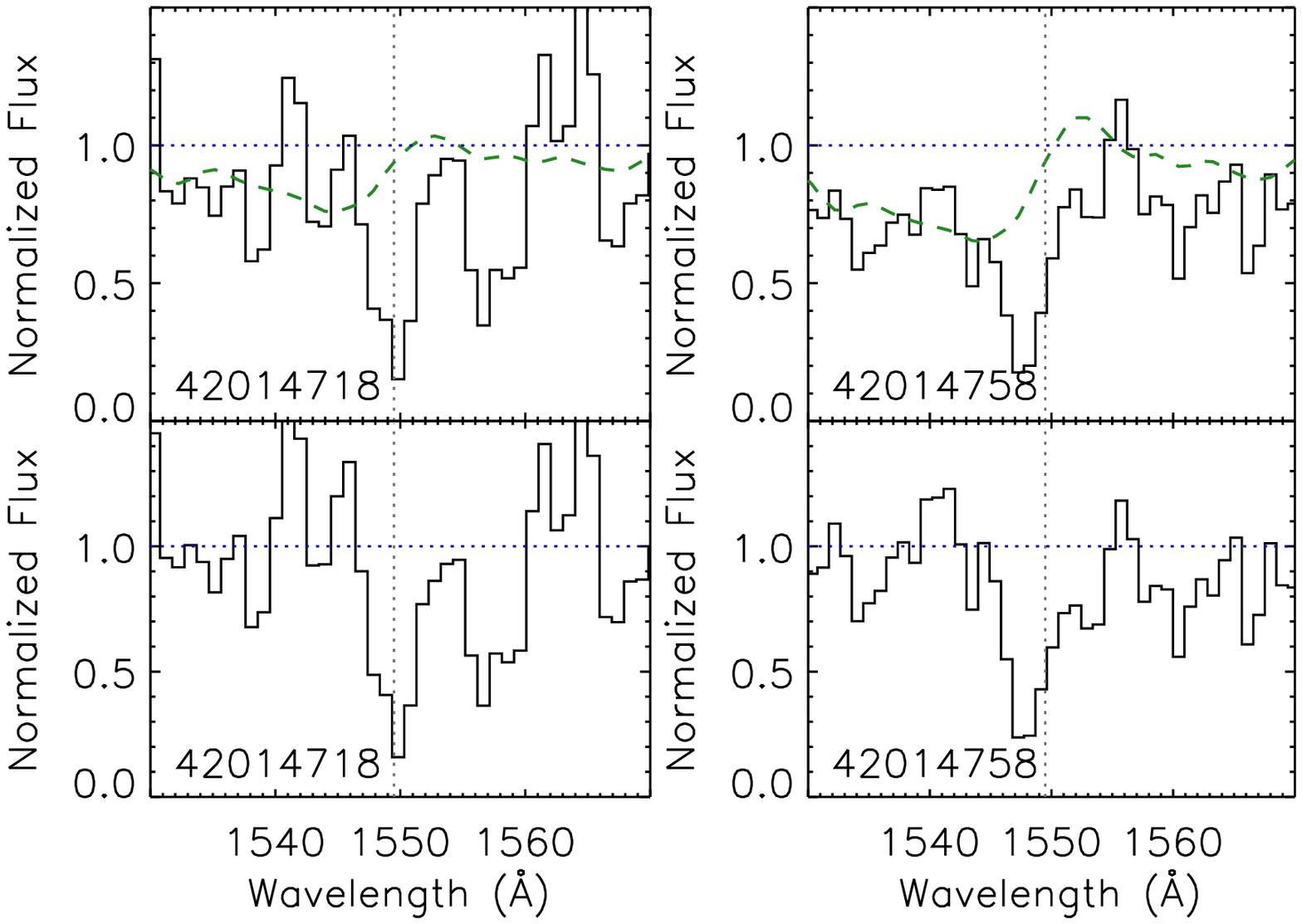}
\includegraphics[width=0.5\linewidth]{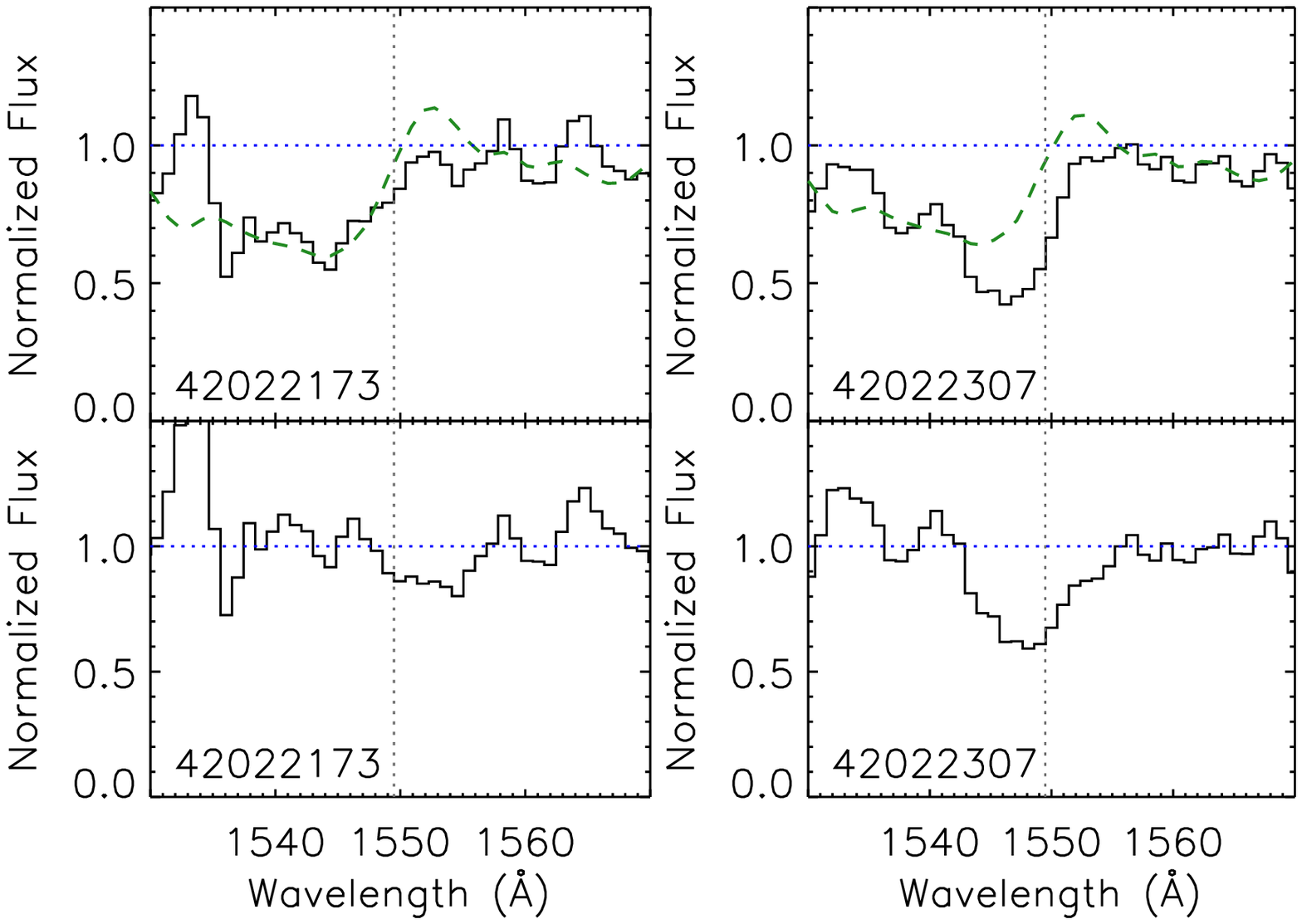}
\includegraphics[width=0.5\linewidth]{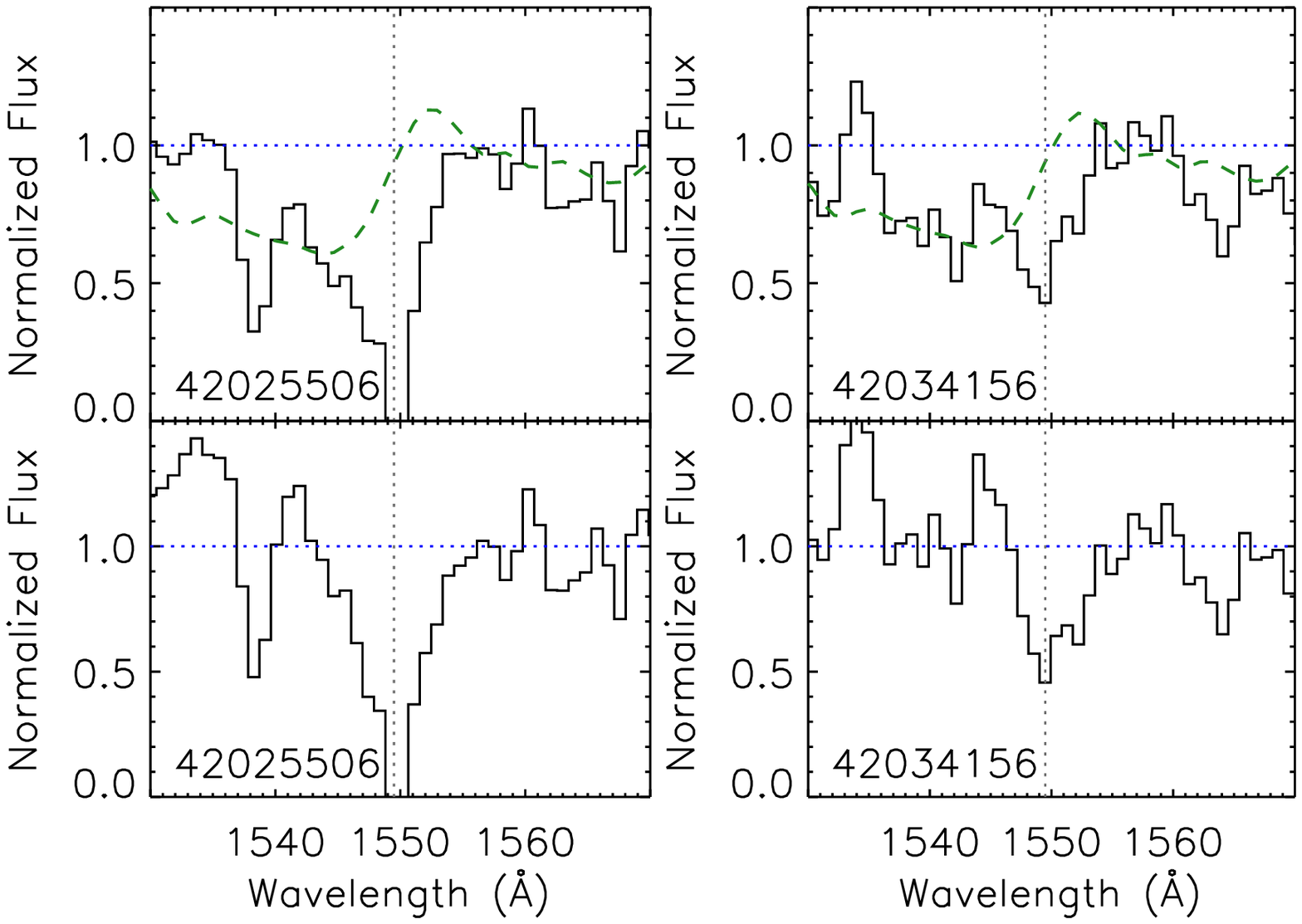}
\caption{Continuum-normalized spectra (top) and spectra normalized by the best-fit stellar model (bottom) for 32 objects with continuum $S/N>5$ and \textrm{C}~\textsc{iv} EW measurements $>$ 3$\sigma$, showing individual detection of \textrm{C}~\textsc{iv} absorption. The blue horizontal dotted line indicates the continuum level, the gray vertical dotted line suggests the rest-frame wavelength of blended \textrm{C}~\textsc{iv} doublet in the 1:1 case (1549.5 $\mbox{\AA}$). The dashed line shows the best-fit stellar model for each object. A green dashed line indicates that the object was bracketed with neighboring metallicities, which were linearly combined to determine the best-fit stellar model. A red dashed line indicates that the object failed to be bracketed with neighboring metallicities, so the best-fit stellar model from \citet{Leitherer2010} was either $0.05Z_{\sun}$ or $2.0Z_{\sun}$.}
\end{Contfigure*}

\begin{figure*}[t!]
\includegraphics[width=0.5\linewidth]{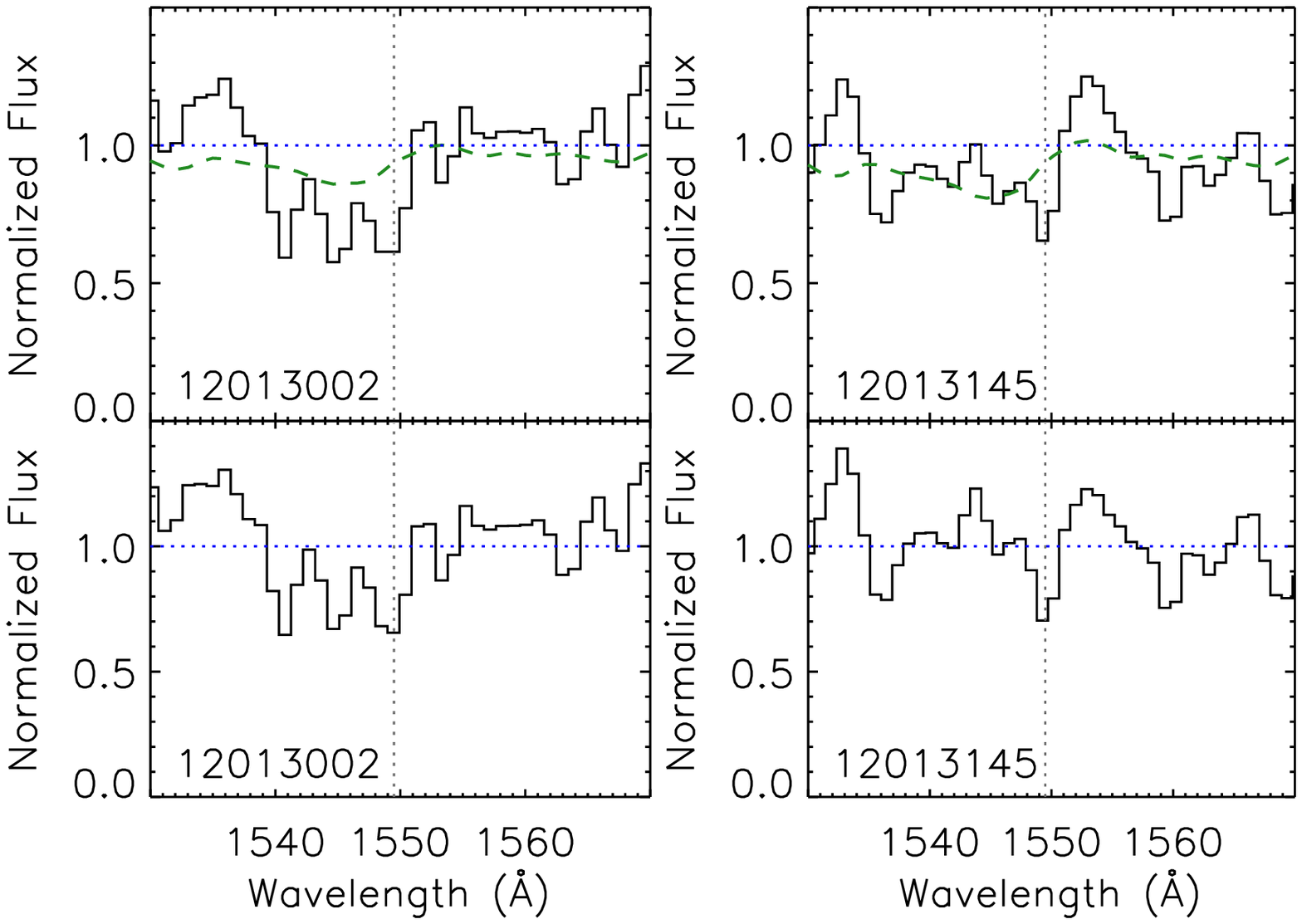}
\includegraphics[width=0.5\linewidth]{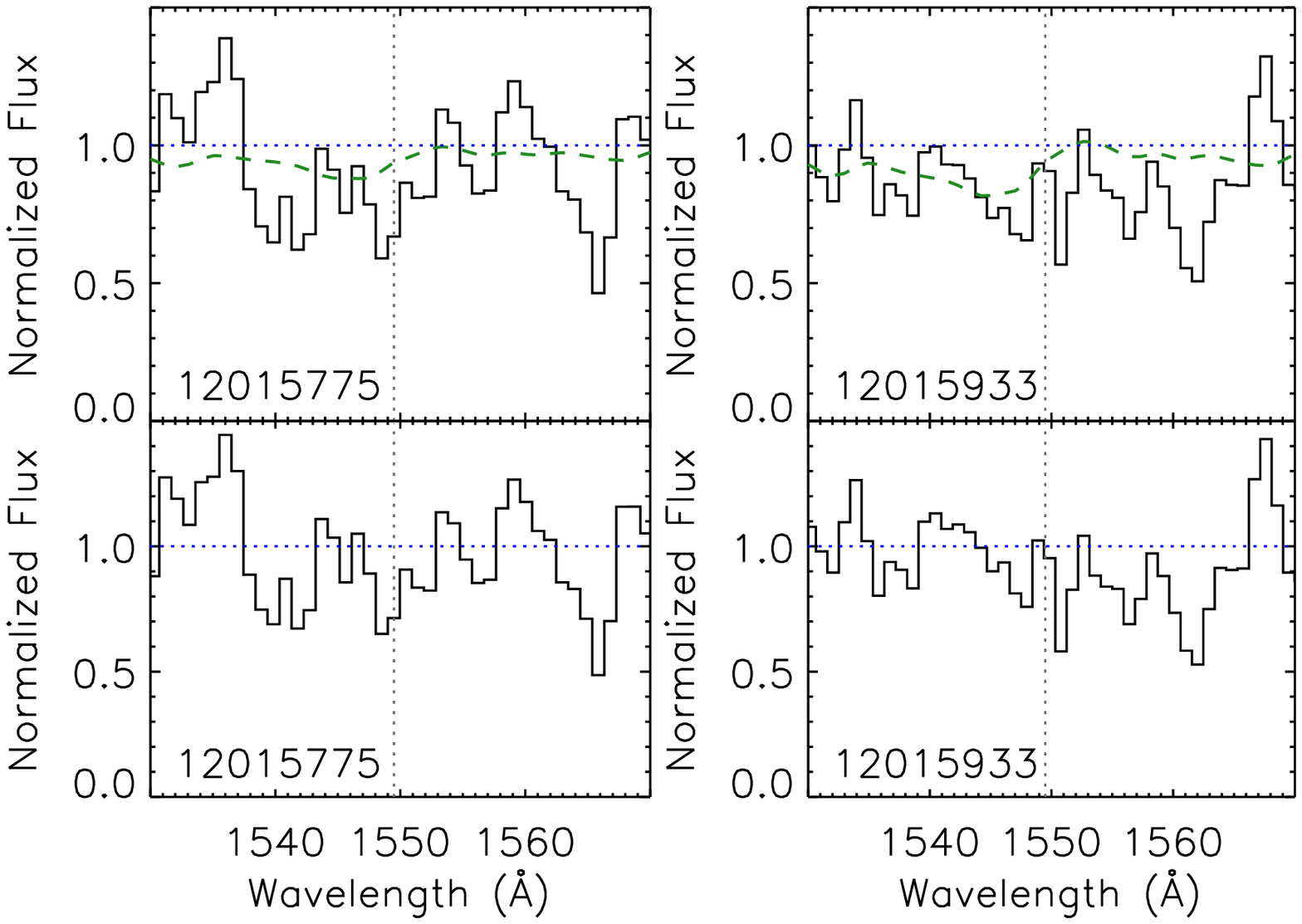}
\includegraphics[width=0.5\linewidth]{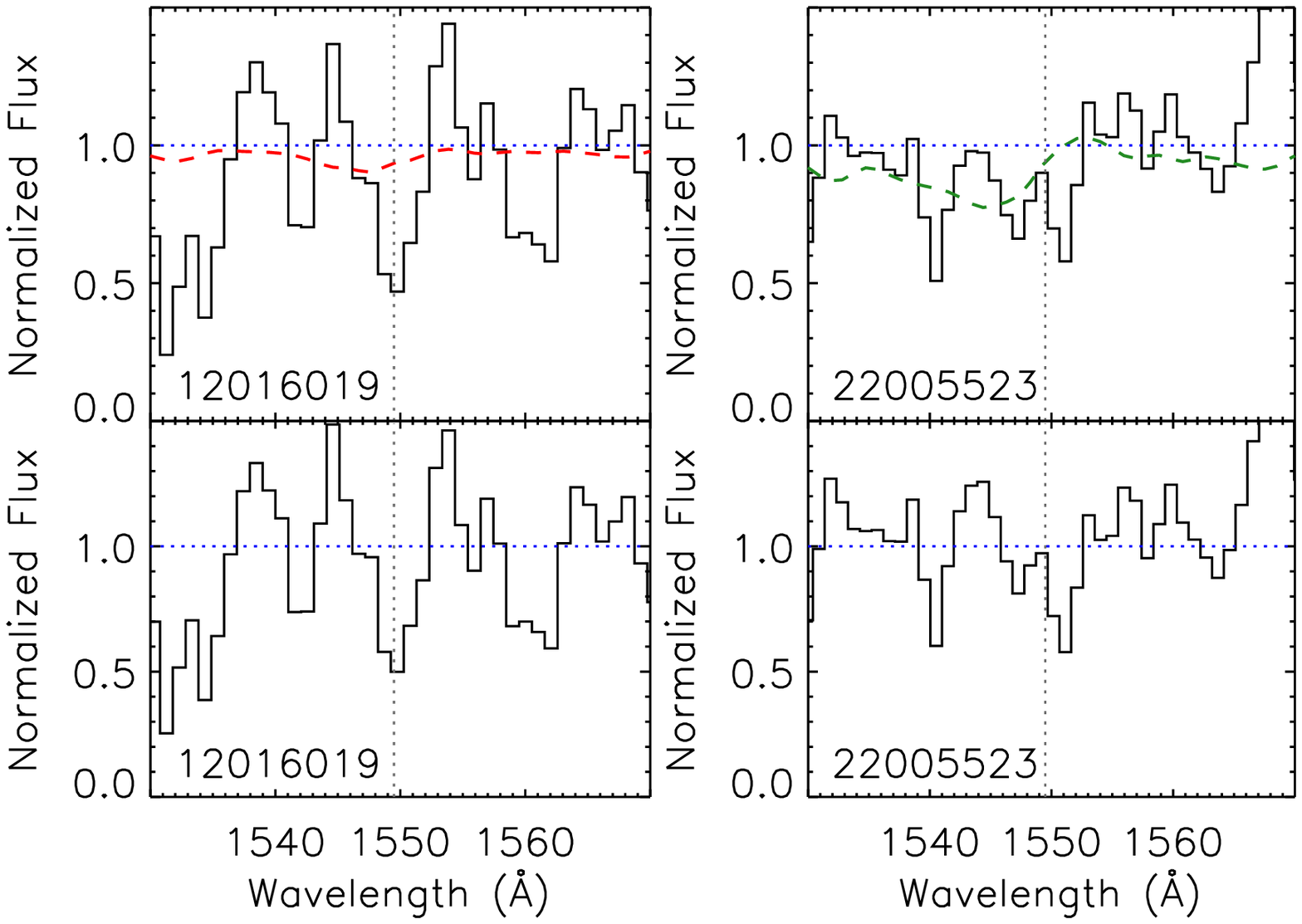}
\includegraphics[width=0.5\linewidth]{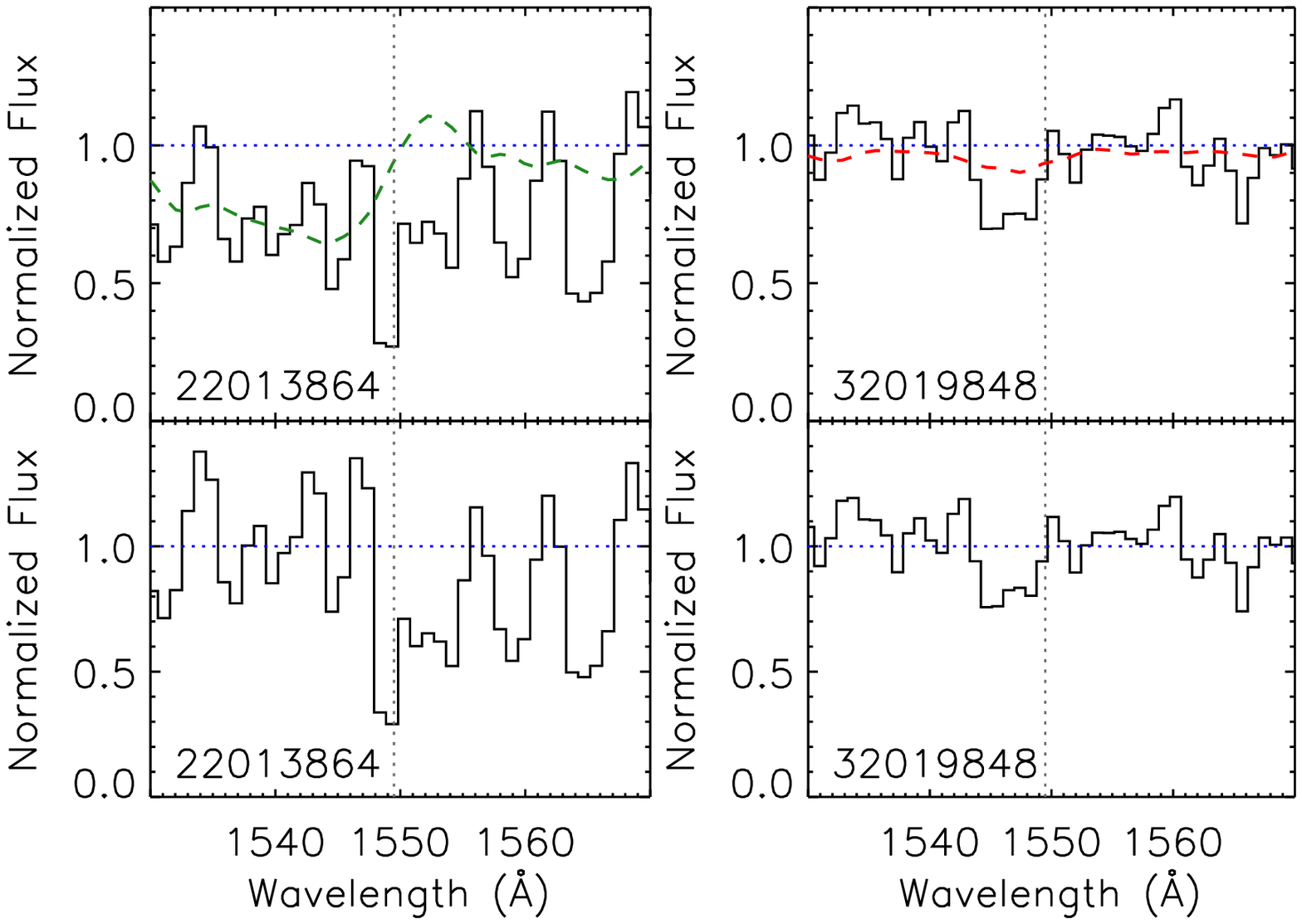}
\includegraphics[width=0.5\linewidth]{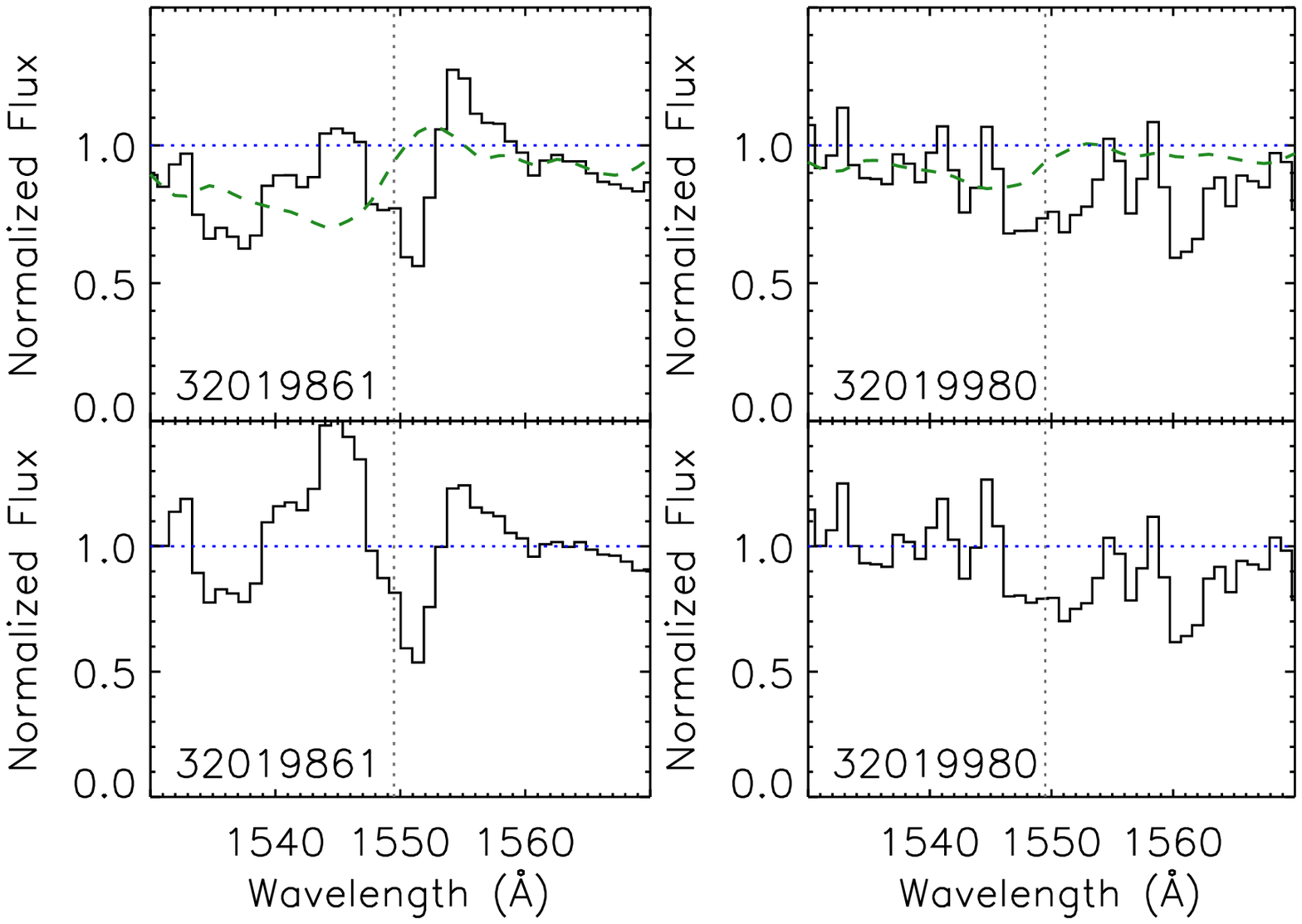}
\includegraphics[width=0.5\linewidth]{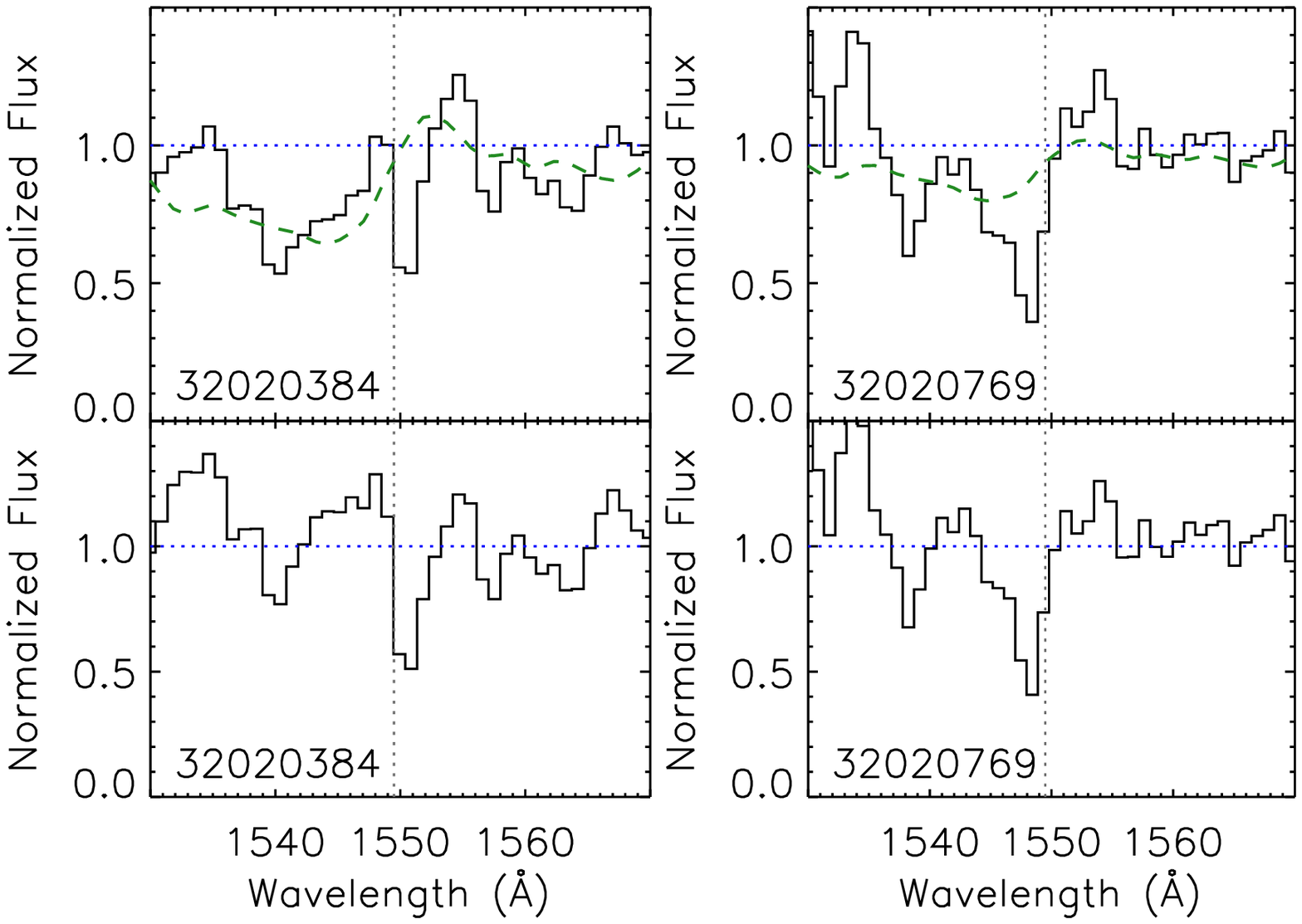}
\caption{Continuum-normalized spectra (top) and spectra normalized by the best-fit stellar model (bottom) for 14 objects with continuum $S/N>5$ and \textrm{C}~\textsc{iv} EW measurements $<$ 3$\sigma$, showing individual \textrm{C}~\textsc{iv} absorption profiles. Legends are the same as in Figure~\ref{fig:civpro}.}
\label{fig:civprond}
\end{figure*}

\begin{Contfigure*}
\centering
\includegraphics[width=0.5\linewidth]{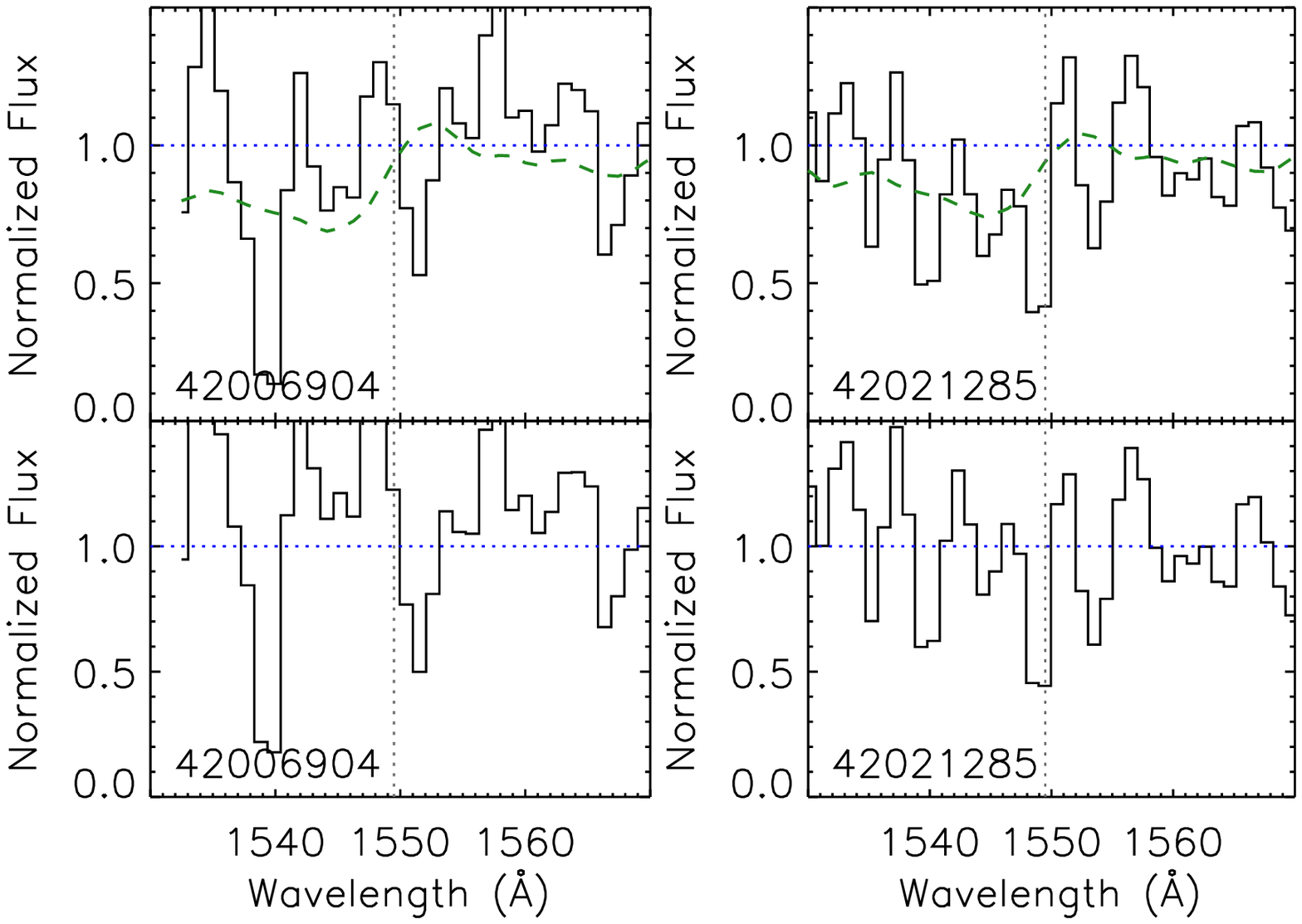}
\caption{Continuum-normalized spectra (top) and spectra normalized by the best-fit stellar model (bottom) for 14 objects with continuum $S/N>5$ and \textrm{C}~\textsc{iv} EW measurements $<$ 3$\sigma$, showing individual \textrm{C}~\textsc{iv} absorption profiles. Legends are the same as in Figure~\ref{fig:civpro}.}
\end{Contfigure*}

As shown in Equation \ref{eq:residual}, the residual, $res$ is defined as:
\begin{equation}
\label{eq:residual}
res=\frac{\sum\limits_{i}{\frac{y_{model,i}-y_{data,i}}{{y^{2}_{err,i}}}}}{\sum\frac{1}{{y^{2}_{err,i}}}}
\end{equation}
where $y_{model,i}$, $y_{data,i}$ and $y_{err,i}$ are, respectively, the continuum-normalized stellar model from \citet{Leitherer2010}, the science spectrum and the corresponding error spectrum at pixel $i$.

As presented in the top panel of Figure~\ref{fig:meastep}, we constructed a best-fit stellar model at each wavelength from the linear combination of the two models (red for 0.2$Z_{\sun}$ and blue for 0.4$Z_{\sun}$) with P-Cygni absorption depths (i.e., metallicities) that bracketed the observed profile.

This best-fit model, $Bestfit(\lambda$), is characterized as:
\begin{equation}
\begin{aligned}
\label{eq:wingfit}
Best fit(\lambda)&=\frac{model_{1}(\lambda)\vert\frac{1}{res_{1}}\vert+model_{2}(\lambda)\vert\frac{1}{res_{2}}\vert}{\vert\frac{1}{res_{1}}\vert+\vert\frac{1}{res_{2}}\vert} \\
&=\frac{model_{1}(\lambda)\vert{res_{2}}\vert+model_{2}(\lambda)\vert{res_{1}}\vert}{\vert{res_{1}}\vert+\vert{res_{2}}\vert} 
\end{aligned}
\end{equation}
where $res_{1}$ and $res_{2}$ are the residuals for $model_{1}$ (red), in the top panel of Figure~\ref{fig:meastep} and $model_{2}$ (blue), respectively. As shown in Equation \ref{eq:wingfit}, the linear combination weights were determined from the inverse residual between model and data. In the top panel of Figure~\ref{fig:meastep}, the residuals for the 0.2$Z_{\sun}$ and 0.4$Z_{\sun}$ models are 0.0624 and -0.0659, respectively. 

While bulk of the sample (93 objects with \textrm{C}~\textsc{iv} coverage) is best described by models with $0.05Z_{\sun} < Z < 2.0Z_{\sun}$,\footnote{Although the determination of best-fit stellar models is not robust for objects with continuum $S/N < 5$, the composite resulting from including these low $S/N$ spectra shows no systematic bias in interstellar \textrm{C}~\textsc{iv} absorption.} 22 objects failed to be bracketed with 
neighboring metallicities, indicating that the \textrm{C}~\textsc{iv} wing at $1535\mbox{\AA}$ to $1544\mbox{\AA}$ was either too shallow for the lowest metallicity or too deep for the highest. In such cases, we chose the best-fit to be either $0.05Z_{\sun}$ (18 objects) or $2.0Z_{\sun}$ (4 objects) for objects with, respectively, the shallowest and deepest stellar \textrm{C}~\textsc{iv} absorption. We then divided both the continuum-normalized spectra and the error spectra by the corresponding 
best-fit stellar models and, accordingly, isolated the interstellar \textrm{C}~\textsc{iv} absorption component.

Since the \textrm{C}~\textsc{iv}$\lambda\lambda$1548,1550 doublet was not resolved in these low resolution spectra, we deblended the interstellar \textrm{C}~\textsc{iv} profile into two Gaussian profiles for the 46 objects with continuum $S/N$ $>5$. We used the IDL program MPFIT \citep{Mark2009} with the initial starting values 
of continuum flux level, line centroid, EW and Gaussian FWHM for each component estimated from the program $splot$ in IRAF. We fixed the doublet wavelengths at the rest-wavelength ratio and forced the FWHM of each doublet member to be identical as well. The bottom panel of Figure~\ref{fig:meastep} shows the fitted individual \textrm{C}~\textsc{iv} doublet members (blue and magenta dotted lines) and the overall fit to the absorption profile (red dotted line) in the 1:1 case.

We iterated the fitting over a narrower wavelength range: centroid$-4\sigma < \lambda < $centroid$+4\sigma$, where the centroid and $\sigma$ were, respectively, the returned central wavelength and standard deviation of the best-fit Gaussian profile from the initial MPFIT fit to the \textrm{C}~\textsc{iv} doublet over $1540\mbox{\AA}$ to $1560\mbox{\AA}$. 
We then determined the significance of the \textrm{C}~\textsc{iv} EW in each object and identified 34 out of 46 objects with \textrm{C}~\textsc{iv} detection $>$ 3$\sigma$ in the sample for the study of individual velocity shifts. 

There were 2 objects, 32019861 and 32020769, showing excessive absorption on the red side of \textrm{C}~\textsc{iv} (red asymmetry). In either the 1:1 or 2:1 case, profiles with \textrm{C}~\textsc{iv}$\lambda$1550 stronger than \textrm{C}~\textsc{iv}$\lambda$1548 are unphysical. Possible sources of systematic error leading to this situation are associated with continuum fitting, best-fit stellar model determination and sky noise. We flagged and excluded these two objects from further study (gray squares in the top-right and the bottom-left panels in Figure~\ref{fig:galprop}). The final sample of robust \textrm{C}~\textsc{iv} detections includes 32 objects, of which the continuum-normalized spectra, and the spectra normalized by the best-fit stellar model are shown in Figure~\ref{fig:civpro}.\footnote{Dividing out the stellar absorption has the largest effect on the blue side of the \textrm{C}~\textsc{iv} profile, tending to make the inferred centroid of interstellar \textrm{C}~\textsc{iv} more redshifted compared to that of the overall \textrm{C}~\textsc{iv} absorption profile. Normalizing by the stellar model changes the inferred \textrm{C}~\textsc{iv} centroid velocities by a few tens up to $\sim$150 km $\mbox{s}^{-1}$, depending on the relative strengths of stellar and interstellar \textrm{C}~\textsc{iv} absorption. Division of the best-fit stellar model minimizes the contamination from stellar absorption and thus provides a more robust description of interstellar \textrm{C}~\textsc{iv} absorption.} The same sets of spectra are shown in Figure~\ref{fig:civprond} for the 14 objects with continuum $S/N > 5$ and \textrm{C}~\textsc{iv} EW $<$ 3$\sigma$.

\subsection{Measurement of Low-Ionization Lines}
\label{sec:low-ion}

Multiple far-UV low-ionization lines fall in the vicinity of \textrm{C}~\textsc{iv}, and are covered in our LRIS spectra. These include \textrm{Si}~\textsc{ii}$\lambda1526$, \textrm{Fe}~\textsc{ii}$\lambda1608$ and \textrm{Al}~\textsc{ii}$\lambda1670$. We measure these features and compare the kinematics of different phases of interstellar gas.

We defined our far-UV low-ionization line sample in an analogous manner to that of \textrm{C}~\textsc{iv}. We selected objects from the sample of 93 galaxies with \textrm{C}~\textsc{iv} coverage and required continuum $S/N$ $>$5 in the vicinity of each low-ionization feature. Accordingly, there were 46 objects in the \textrm{Si}~\textsc{ii}$\lambda1526$ sample and 69 in the \textrm{Fe}~\textsc{ii}$\lambda1608$ and \textrm{Al}~\textsc{ii}$\lambda1670$ samples. The continuum $S/N$ was calculated from 1570$\mbox{\AA}$ to 1590$\mbox{\AA}$ for \textrm{Si}~\textsc{ii}, and from 1690$\mbox{\AA}$ to 1750$\mbox{\AA}$ for \textrm{Fe}~\textsc{ii} and \textrm{Al}~\textsc{ii}. Since all of these features are singlets, we simply fit one Gaussian profile to each line. These profile fits resulted in 22 objects in the \textrm{Si}~\textsc{ii} sample, 14 in the \textrm{Fe}~\textsc{ii} sample and 44 in the \textrm{Al}~\textsc{ii} sample where the absorption lines were significantly detected (EW $>3\sigma$).

To apply a consistent methodology of simple Gaussian profile fitting for analyzing both near-UV and far-UV features, we also re-measured the near-UV features presented in \citet{Martin2012} for the 93 objects with \textrm{C}~\textsc{iv} coverage. These features include \textrm{Mg}~\textsc{ii}$\lambda\lambda$2796, 2803 and \textrm{Fe}~\textsc{ii}$\lambda$2344, $\lambda$2374, $\lambda$2587. To trace the kinematics of low-ionization absorption, \citet{Martin2012} fit the troughs of \textrm{Fe}~\textsc{ii}$\lambda$2250, 2261, 2344, 2374, 2587 with profiles of the form $I(\lambda)=I_{0}e^{-\tau(\lambda)}$, where $\tau(\lambda)$ is a Gaussian parameterized by the central wavelength, $\lambda_{0}$ and the Doppler parameter, b. The central wavelengths of these lines were tied together to define a single-component Doppler shift $\mbox{V}_{1}$, and the optical depths were also tied by the ratio of their oscillator strengths. \textrm{Fe}~\textsc{ii}$\lambda$ 2382 and $\lambda$2600 may suffer from significant emission filling and were excluded from the fitting (see the text about \textrm{Fe}~\textsc{ii}$\lambda$1608 in Section \ref{sec:civnuv}). 

In our simple Gaussian fits of the near-UV \textrm{Fe}~\textsc{ii} features, we further excluded the weak transitions \textrm{Fe}~\textsc{ii}$\lambda$2250, $\lambda$2261 (with oscillator strengths 14.2 and 13.3 times lower than that of \textrm{Fe}~\textsc{ii}$\lambda$2374, respectively) as they provided little constraint on the velocity shift. We fixed the wavelength centroid ratios at their rest-frame values and assumed identical profile widths for all fitted near-UV \textrm{Fe}~\textsc{ii} lines.\footnote{Tying the centroids introduced no systematic offset to the velocity measurements of the near-UV \textrm{Fe}~\textsc{ii} features. Velocity shifts inferred from tied centroids were consistent with those based on fits to individual near-UV \textrm{Fe}~\textsc{ii} profiles. The resulting \textrm{Fe}~\textsc{ii} velocity shifts were consistent with those in \citet{Martin2012}}. On the other hand, we set the centroids and widths of \textrm{Mg}~\textsc{ii} to float freely and fit the doublet wherever the $\chi^{2}$ reaches a minimum, to accommodate the effects of \textrm{Mg}~\textsc{ii} emission filling. Specifically, the red parts of the \textrm{Mg}~\textsc{ii}$\lambda$2796 and \textrm{Mg}~\textsc{ii}$\lambda$2803 absorption profiles may be washed out by different amounts due to \textrm{Mg}~\textsc{ii} resonant emission. Therefore, the centroids of the doublet members do not necessarily show a fixed wavelength ratio and the widths may also not be the same \citep{Martin2012}. Table ~\ref{tab:vplot} lists the velocity shift of the measured lines (\textrm{C}~\textsc{iv}, \textrm{Al}~\textsc{ii}$\lambda$1670, \textrm{Si}~\textsc{ii}$\lambda$1526 and near-UV \textrm{Fe}~\textsc{ii}) of individual objects derived from Gaussian fitting, as well as the near-UV \textrm{Fe}~\textsc{ii} velocity shifts from \citet{Martin2012}.

\section{Kinematics}
\label{sec:kine}

To investigate whether the high-ionization kinematics differ from those of low-ionization lines, we compare the velocity shifts of \textrm{C}~\textsc{iv} and near-UV \textrm{Fe}~\textsc{ii} and \textrm{Mg}~\textsc{ii} lines in Section \ref{sec:civnuv}. We also compare the low-ionization kinematics traced by both far-UV and near-UV lines, and discuss possible interpretations for the observed discrepancies in Section \ref{sec:civfuv}.

\subsection{\textrm{C}~\textsc{iv} vs. Near-UV lines}
\label{sec:civnuv}

\begin{figure}
\includegraphics[width=1.0\linewidth]{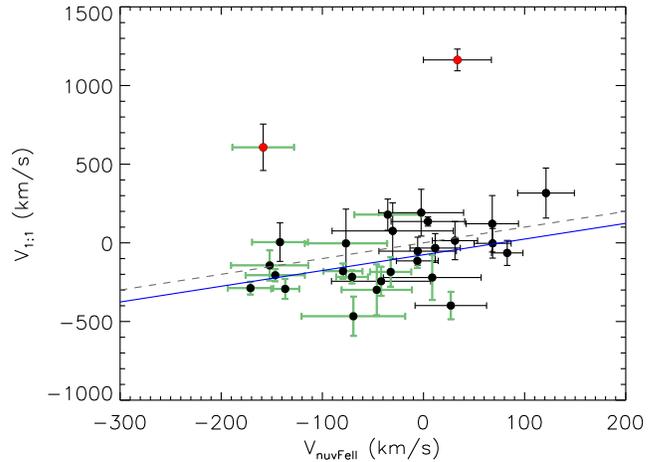}
\caption{Comparison between \textrm{C}~\textsc{iv} and near-UV \textrm{Fe}~\textsc{ii} centroid velocity shifts. There are 28 objects with continuum $S/N>5$ and EW detections of both \textrm{C}~\textsc{iv} and near-UV \textrm{Fe}~\textsc{ii} $>3\sigma$. The 1:1 doublet ratio is assumed in modeling the \textrm{C}~\textsc{iv} feature. The median velocity shift for \textrm{C}~\textsc{iv} and near-UV \textrm{Fe}~\textsc{ii} (excluding 2 outliers, 22012285 and 42022173, marked in red circles) is -64 km $\mbox{s}^{-1}$ and -30 km $\mbox{s}^{-1}$, respectively. The green error bars indicate 1-sigma outflows for corresponding transitions. The gray dashed line marks where $\mbox{V}_{1:1}$ and $\mbox{V}_{\textrm{nuvFe}~\textsc{ii}}$ are equal, and the blue solid line presents the best linear fit with the slope fixed at 1. The offset between the gray and blue lines is -76 $\pm$ 26 km $\mbox{s}^{-1}$ ($\mbox{V}_{1:1}-\mbox{V}_{\textrm{nuvFe}~\textsc{ii}}$). The uncertainty on the best-fit intercept were estimated from Monte Carlo methods.}
\label{fig:vplot}
\end{figure}

In order to compare the kinematics of \textrm{C}~\textsc{iv} with those of the low-ionization lines, we calculated the velocity shifts for \textrm{C}~\textsc{iv} in the 1:1 case and the near-UV \textrm{Fe}~\textsc{ii} lines. We define these velocity shifts, respectively, as $\mbox{V}_{1:1}$ and $\mbox{V}_{\textrm{nuvFe}~\textsc{ii}}$. In Figure~\ref{fig:vplot}, the black circles (28 objects) represent objects with robust measurements for both \textrm{C}~\textsc{iv} and near-UV \textrm{Fe}~\textsc{ii} features (continuum $S/N>5$ and EW $\geqslant 3\sigma$). Two objects stand out as having significantly redshifted \textrm{C}~\textsc{iv} centroid velocities. One object, 22012285, is extremely redshifted 
with a velocity of +1162 km $\mbox{s}^{-1}$. We interpret the \textrm{C}~\textsc{iv} redshift of this object with caution, as the absorption (while formally significant), may be caused by systematic oversubtraction of the sky spectrum. 
Another object, 42022173 has a \textrm{C}~\textsc{iv} velocity of +607 km $\mbox{s}^{-1}$ which could be due to systematics in the normalization of the stellar wind component. The best-fit stellar model for this object (bracketed by models with 0.4$Z_{\sun}$ and 1.0$Z_{\sun}$), while based on the blueshifted stellar absorption profile, overproduces \textrm{C}~\textsc{iv} emission at $\sim1552\mbox{\AA}$. Therefore, after division by the best-fit model, some apparent interstellar absorption remains on the red side of the \textrm{C}~\textsc{iv} profile. It is not clear whether this absorption is real or an artifact of the mismatch between model and data. Given the potentially large uncertainties associated with these two objects, we removed them from the sample when deriving the statistical properties.

\begin{figure}
\includegraphics[width=1.0\linewidth]{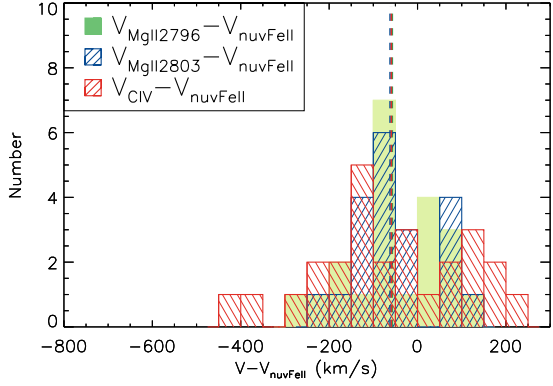}
\caption{Histogram of \textrm{C}~\textsc{iv}, \textrm{Mg}~\textsc{ii}, and \textrm{Fe}~\textsc{ii} velocity differences. Velocity differences are only plotted for objects with continuum $S/N>5$ and EW$>$ 3$\sigma$ detection for \textrm{C}~\textsc{iv} and near-UV \textrm{Fe}~\textsc{ii} lines (red histogram, 26 objects, excluding 22012285 and 42022173), and \textrm{Mg}~\textsc{ii} and near-UV \textrm{Fe}~\textsc{ii} lines (green and blue histograms, 21 and 20 objects, respectively). The vertical colored dashed lines show the median velocity difference of corresponding pairs, which are -65 km $\mbox{s}^{-1}$, -57 km $\mbox{s}^{-1}$ and -63 km $\mbox{s}^{-1}$ for \textrm{C}~\textsc{iv}, \textrm{Mg}~\textsc{ii}$\lambda$2796 and \textrm{Mg}~\textsc{ii}$\lambda$2803 with respect to near-UV \textrm{Fe}~\textsc{ii}, respectively.}
\label{fig:hist}
\end{figure}

\begin{figure}
\includegraphics[width=1.0\linewidth]{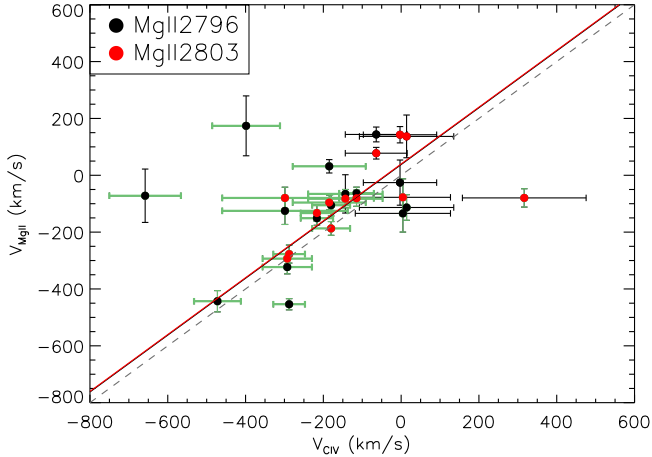}
\caption{Comparison of \textrm{C}~\textsc{iv} and \textrm{Mg}~\textsc{ii} velocity shifts. Red circles represent \textrm{Mg}~\textsc{ii}$\lambda$2796 (15 objects) and black circles represent \textrm{Mg}~\textsc{ii}$\lambda$2803 (13 objects). All objects shown here have continuum $S/N$ $>5$ and line EWs $>$ 3$\sigma$ for both \textrm{C}~\textsc{iv} and the corresponding \textrm{Mg}~\textsc{ii} doublet member. The green error bars indicate 1-sigma outflows for corresponding transitions. The red and black solid lines are the best-fit linear models for \textrm{Mg}~\textsc{ii}$\lambda$2796 and \textrm{Mg}~\textsc{ii}$\lambda$2803, respectively. The intercept for the doublet members are +38 $\pm$ 43 km $\mbox{s}^{-1}$ and +40 $\pm$ 18 km $\mbox{s}^{-1}$. The gray dashed line marks the 1:1 relation.}
\label{fig:mgscat}
\end{figure}

\begin{figure}
\includegraphics[width=1.0\linewidth]{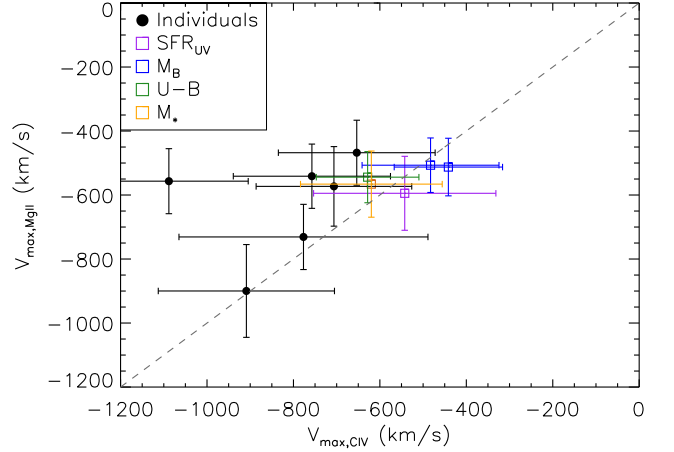}
\caption{Comparison between \textrm{C}~\textsc{iv} and \textrm{Mg}~\textsc{ii} $\mbox{V}_{max}$ for resolved blue wings. The black dots represent individual objects with the highest continuum $S/N$ ($>10.0$), and the colored open squares are from the composite spectra (purple, blue, green, yellow for SFR, $\mbox{M}_{B}$, $\ub$ and $\mbox{M}_{*}$, respectively) described in Section ~\ref{sec:galprop}. The gray dashed line indicates where the two velocities are equal, and all objects fall close to the 1:1 line but the outlier 42022307.}
\label{fig:vmax}
\end{figure}

After eliminating these 2 outliers, we find that 50$\%$ (27$\%$) of the sample of 26 objects indicate 1$\sigma$ (3$\sigma$) outflows. Moreover, this sample shows that \textrm{C}~\textsc{iv} (median velocity shift -64 km $\mbox{s}^{-1}$) is more blueshifted than the near-UV \textrm{Fe}~\textsc{ii} lines (median velocity shift -30 km $\mbox{s}^{-1}$), with a median velocity difference, $\mbox{V}_{1:1}-\mbox{V}_{\textrm{nuvFe}~\textsc{ii}}$, of -65 km $\mbox{s}^{-1}$. Furthermore, the best-fit linear model (measurement errors considered) gives an intercept of -76 $\pm$ 26 km $\mbox{s}^{-1}$ with a fixed slope at 1, which confirms the larger blueshift of \textrm{C}~\textsc{iv} at the $\sim$ 3-sigma level. The error bar on the best-fit intercept was estimated using Monte Carlo methods. Specifically, we bootstrap resampled the velocity shifts, $\mbox{V}_{1:1}$ and $\mbox{V}_{\textrm{nuvFe}~\textsc{ii}}$, of these 26 objects and performed a inverse-squared weighted linear regression to the bootstrap sample. The process was repeated 100 times and the standard deviation of the intercept distribution was taken as the corresponding uncertainty. Finally, at least as large a fraction of \textrm{C}~\textsc{iv} absorption profiles shows a blueshift with respect to the systemic velocity (69 $\%$ for $\mbox{V}_{1:1}<$ 0 km $\mbox{s}^{-1}$), compared with 65 $\%$ of the near-UV \textrm{Fe}~\textsc{ii} profiles \citep{Martin2012}.

Although the larger blueshift of \textrm{C}~\textsc{iv} may suggest that the highly-ionized gas is in fact traveling at a slightly higher speed, a close investigation of the kinematics of \textrm{Mg}~\textsc{ii} absorption reveals another possibility. Figure~\ref{fig:hist} shows histograms of the velocity differences between \textrm{C}~\textsc{iv} and near-UV \textrm{Fe}~\textsc{ii} ($\mbox{V}_{1:1}-\mbox{V}_{\textrm{nuvFe}~\textsc{ii}}$, red, 26 objects, excluding 22012285 and 42022173) and individual \textrm{Mg}~\textsc{ii} doublet members and near-UV \textrm{Fe}~\textsc{ii} ($\mbox{V}_{\textrm{Mg}~\textsc{ii}\lambda2796}-\mbox{V}_{\textrm{nuvFe}~\textsc{ii}}$, green, 21 objects and $\mbox{V}_{\textrm{Mg}~\textsc{ii}\lambda2803}-\mbox{V}_{\textrm{nuvFe}~\textsc{ii}}$, blue, 20 objects), where all the lines have a continuum $S/N$ $>$ 5 and $\geqslant$ 3$\sigma$ absorption detection. We note that the sample here comparing \textrm{C}~\textsc{iv} and near-UV \textrm{Fe}~\textsc{ii} is the same as that used in Figure~\ref{fig:vplot}. The dashed lines of corresponding colors indicate the median velocity shift between \textrm{C}~\textsc{iv} and \textrm{Fe}~\textsc{ii}, between \textrm{Mg}~\textsc{ii}$\lambda$2796 and \textrm{Fe}~\textsc{ii} and between \textrm{Mg}~\textsc{ii}$\lambda$2803 and \textrm{Fe}~\textsc{ii}, respectively. The median values of these three distributions are very close to each other, suggesting that the velocity shift of \textrm{C}~\textsc{iv} is nearly identical to that of \textrm{Mg}~\textsc{ii}. 

Moreover, when directly comparing the velocity shifts derived from \textrm{C}~\textsc{iv} and \textrm{Mg}~\textsc{ii} (Figure~\ref{fig:mgscat}), we find that the data points scatter around the 1:1 line despite a few outliers. We note that while the best-fit intercepts give $\sim$ +40 km $\mbox{s}^{-1}$ for both \textrm{Mg}~\textsc{ii} doublet members (+38 $\pm$ 43 km $\mbox{s}^{-1}$ and +40 $\pm$ 18 km $\mbox{s}^{-1}$ for \textrm{Mg}~\textsc{ii}$\lambda$2796 and \textrm{Mg}~\textsc{ii}$\lambda$2803, respectively), this result does not necessarily suggest that \textrm{C}~\textsc{iv} is more blueshifted than \textrm{Mg}~\textsc{ii}, given the small size of the sample. In fact, we find that the blueshift (derived from centroid velocities) of \textrm{C}~\textsc{iv} is on average similar to that of \textrm{Mg}~\textsc{ii} when combining both Figure~\ref{fig:hist} and \ref{fig:mgscat}. As the \textrm{Mg}~\textsc{ii} profile is significantly affected by interstellar emission filling, a natural conclusion is that \textrm{C}~\textsc{iv}, another resonant line showing a similar velocity shift, may be similarly affected.

For a better understanding of the blueshift of \textrm{C}~\textsc{iv}, we further studied the blue wing of the absorption trough. We focused on the extent to which measurable absorption extends blueward of the line centroid by calculating the maximum velocity, $\mbox{V}_{max}$. $\mbox{V}_{max}$ is defined with respect to the blue side of the absorption profile where the measured flux meets with the spectral continuum at the $1\sigma$ level \citep{Martin2012}, and represents the largest velocity at which there is measurable absorption, regardless of the shape of the absorption trough. 

To obtain $\mbox{V}_{max}$ and the associated error bars, we perturbed the spectrum at each wavelength with a Gaussian random variable determined by the corresponding error spectrum, measured $\mbox{V}_{max}$ in the perturbed spectrum, and then repeated the process 500 times for each object. We then calculated the average $\mbox{V}_{max}$ and assigned the error bars as the larger of the standard deviation of the $\mbox{V}_{max}$ distribution or half the pixel width \citep{Martin2012}. We determined $\mbox{V}_{max}$ for \textrm{C}~\textsc{iv} and \textrm{Mg}~\textsc{ii} in both the composite spectra (in different bins of galaxy properties, see Section \ref{sec:galprop}) as well as 10 individual objects with the highest continuum $S/N$, as plotted in Figure~\ref{fig:vmax}. Note that we only included $\mbox{V}_{max}$ for the resolved line wings which required $\mbox{V}_{max}$ to be more significantly blueshifted than -282 km $\mbox{s}^{-1}$ for the 600-line spectra, or -435 km $\mbox{s}^{-1}$ for the 400-line spectra. \footnote{While $\mbox{V}_{max}$ is less susceptible to emission filling, many objects with continuum $S/N$ $>$ 5 and robust line measurements do not have resolved $\mbox{V}_{max}$ due to the low resolution of the spectra and the intrinsic widths of the absorption lines. Therefore, using $\mbox{V}_{max}$ as the main velocity indicator in our sample would significantly reduce the sample size and prevent us from drawing any statistical conclusions.} Nearly all the points lie on the 1:1 line, indicating that the $\mbox{V}_{max}$ for \textrm{C}~\textsc{iv} and for \textrm{Mg}~\textsc{ii} are statistically the same. The only outlier is 42022307, which has a $\mbox{V}_{max,\textrm{C}~\textsc{iv}} < $ -1000 km $\mbox{s}^{-1}$. We flagged this object since the \textrm{C}~\textsc{iv} absorption is unusually broad, and the \textrm{Mg}~\textsc{ii} measurement is also not robust due to a sky line residing just bluewards of \textrm{Mg}~\textsc{ii}$\lambda$2796.

In summary, the small but significant blueshift of \textrm{C}~\textsc{iv} relative to near-UV \textrm{Fe}~\textsc{ii} suggests either the highly-ionized gas is faster-moving, or else the kinematics of low and high ions are in general consistent while the red side of the interstellar absorption trough is filled up from the \textrm{C}~\textsc{iv} resonant emission. Although we favor the latter case as \textrm{C}~\textsc{iv} shows a similar blueshift to that of \textrm{Mg}~\textsc{ii}, further studies at both higher resolution and $S/N$ are needed to break the degeneracy.

\subsection{Far-UV vs. Near-UV lines}
\label{sec:civfuv}

\begin{figure}
\includegraphics[width=1.0\linewidth]{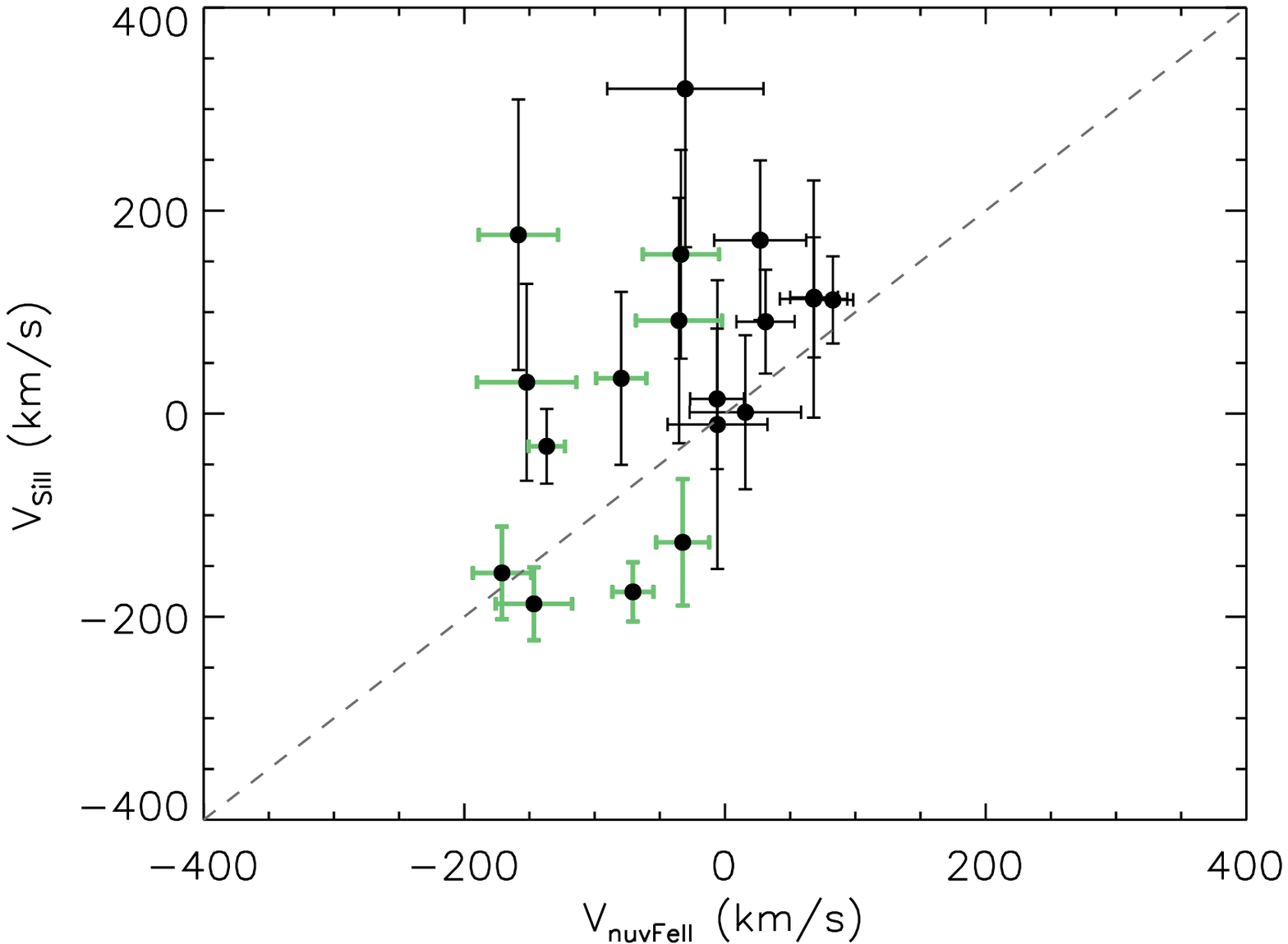}
\label{fig:si}
\includegraphics[width=1.0\linewidth]{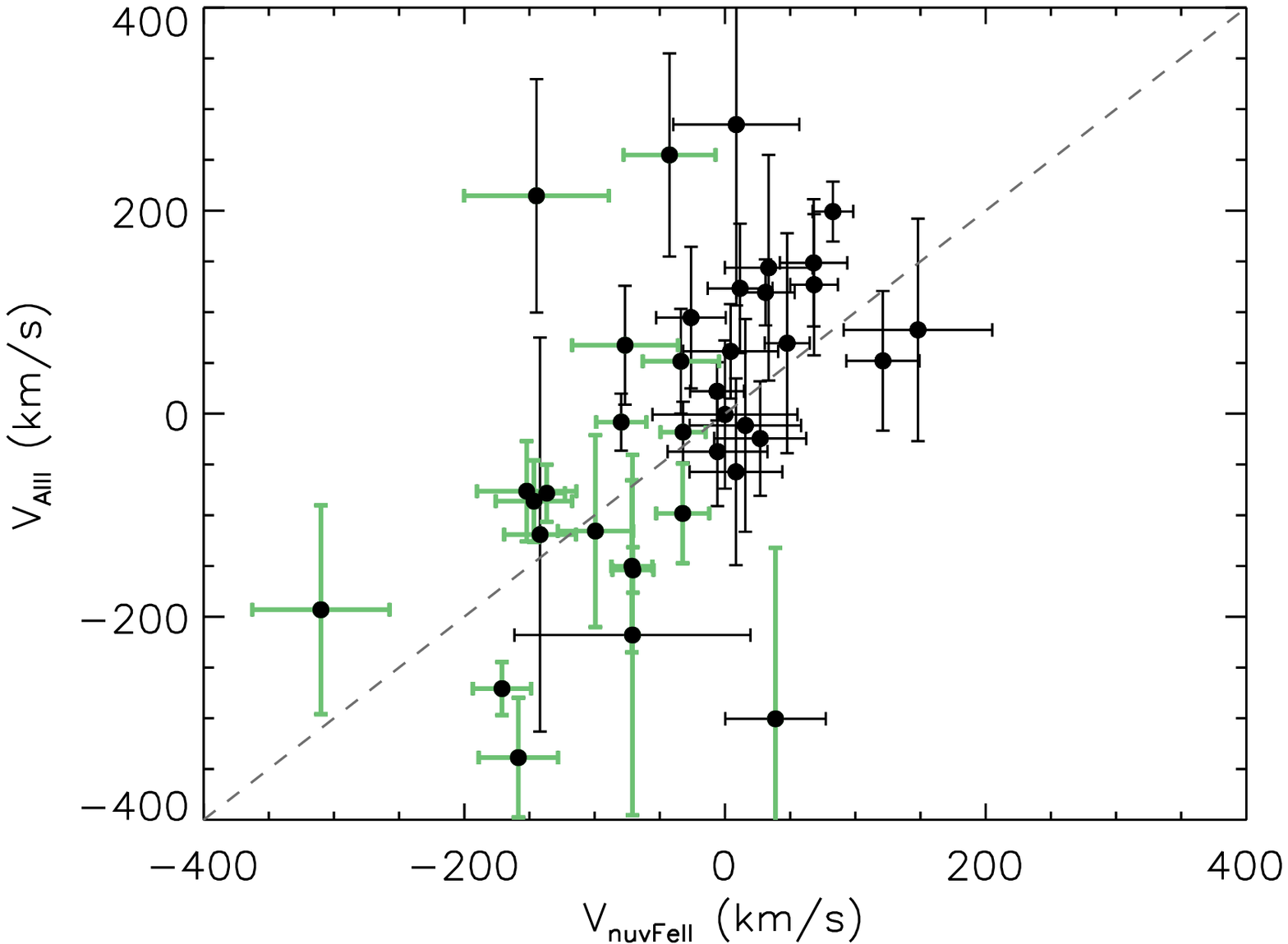}
\includegraphics[width=1.0\linewidth]{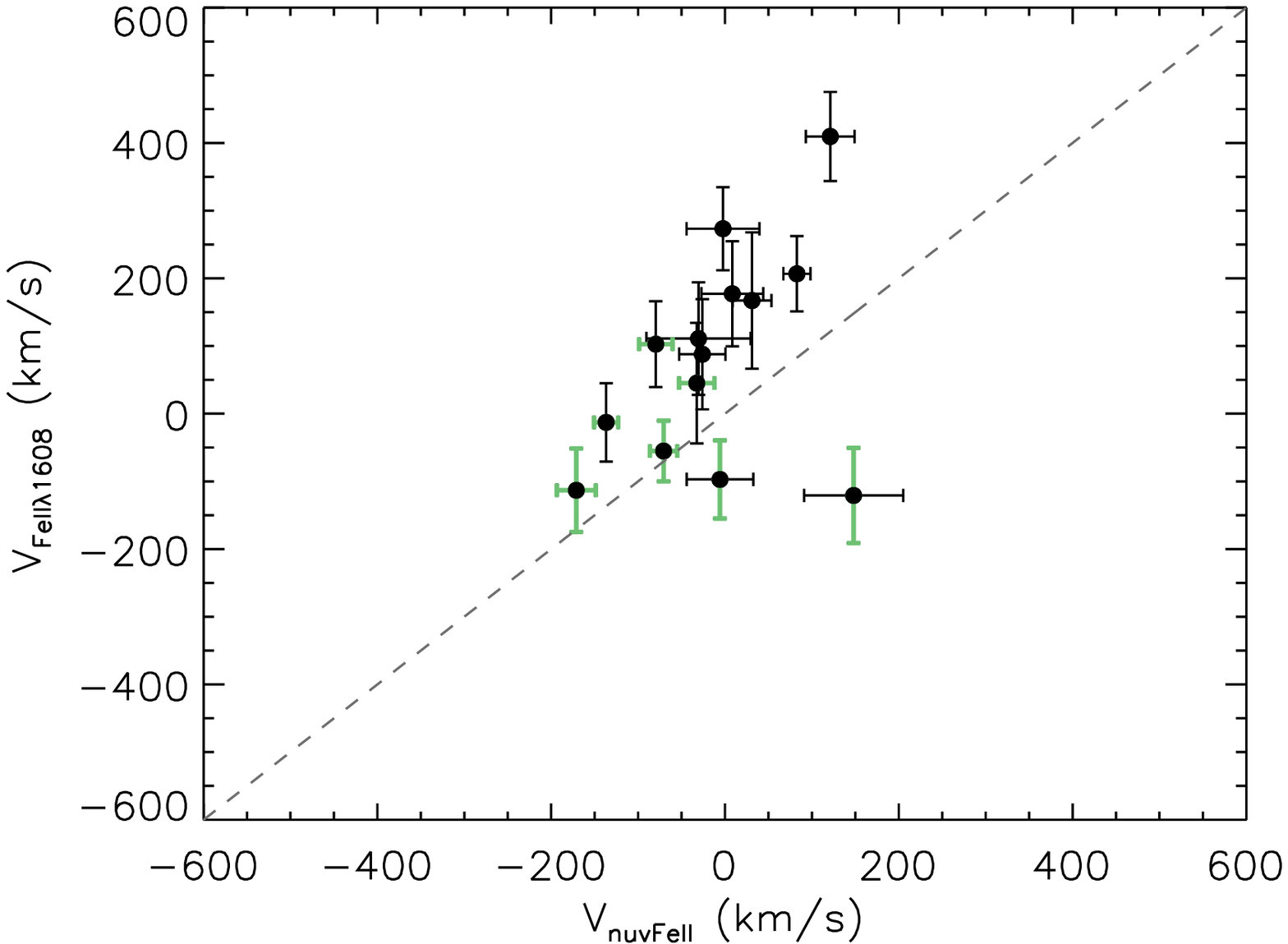}
\label{fig:fe}
\caption{\textrm{Si}~\textsc{ii}$\lambda$1526 (top), \textrm{Al}~\textsc{ii}$\lambda$1670 (middle), and \textrm{Fe}~\textsc{ii}$\lambda$1608 (bottom) velocity shifts compared with those from near-UV \textrm{Fe}~\textsc{ii} lines. Legends are the same as in Figure~\ref{fig:vplot}. The median velocity for \textrm{Si}~\textsc{ii}$\lambda$1526, \textrm{Al}~\textsc{ii}$\lambda$1670 and \textrm{Fe}~\textsc{ii}$\lambda$1608 is +35 km $\mbox{s}^{-1}$, -5 km $\mbox{s}^{-1}$ and +95 km $\mbox{s}^{-1}$, respectively. The intercept for the best linear fits with slope fixed at 1 for those lines are +20 $\pm$ 28 km $\mbox{s}^{-1}$, +27 $\pm$ 22 km $\mbox{s}^{-1}$ and +95 $\pm$ 33 km $\mbox{s}^{-1}$, respectively.}
\label{fig:vplotside}
\end{figure}

\begin{figure}
\includegraphics[width=1.0\linewidth]{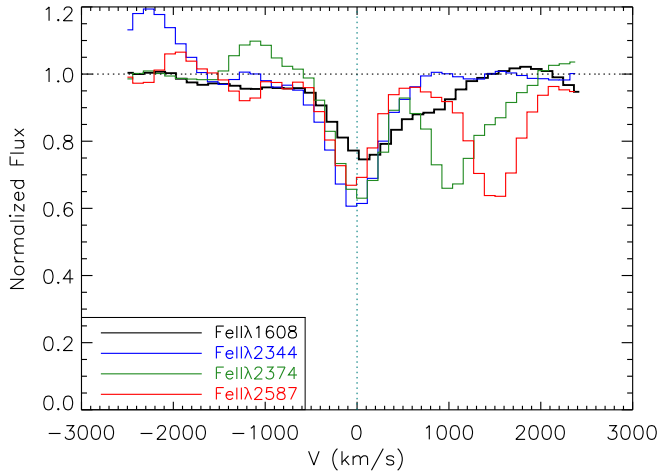}
\caption{The stacked velocity profile of \textrm{Fe}~\textsc{ii}$\lambda$1608 and \textrm{Fe}~\textsc{ii}$\lambda$2344, 2374, 2587 for objects observed in the 400-line masks with continuum $S/N$ $>$ 5 (65 objects). The black line shows the absorption profile of \textrm{Fe}~\textsc{ii}$\lambda$1608, and those of \textrm{Fe}~\textsc{ii}$\lambda$2344, \textrm{Fe}~\textsc{ii}$\lambda$2374 and \textrm{Fe}~\textsc{ii}$\lambda$2587 are shown in blue, green and red, respectively. The dotted horizontal line marks the expected continuum level, and the dotted vertical line indicates the systemic velocity.}
\label{fig:fe1608}
\end{figure}

In order to establish a velocity frame traced by far-UV absorption lines, we also plotted \textrm{Si}~\textsc{ii}$\lambda$1526, \textrm{Fe}~\textsc{ii}$\lambda$1608 and \textrm{Al}~\textsc{ii}$\lambda$1670 against the near-UV \textrm{Fe}~\textsc{ii} lines. From Figure~\ref{fig:vplotside}, we find that \textrm{Al}~\textsc{ii}$\lambda$1670 scatters around the 1:1 line (median velocity shift -5 km $\mbox{s}^{-1}$) while \textrm{Si}~\textsc{ii}$\lambda$1526 displays a systematic redshift relative to the near-UV \textrm{Fe}~\textsc{ii} (median velocity shift +35 km $\mbox{s}^{-1}$). This offset is caused by the larger error bars associated with the objects at the top left corner of the top panel of Figure~\ref {fig:vplotside}. In fact, the intercepts for the best-fit models with slope fixed at 1.0 for \textrm{Si}~\textsc{ii}$\lambda$1526 and \textrm{Al}~\textsc{ii}$\lambda$1670 are, respectively, +20 km $\pm$ 28 km $\mbox{s}^{-1}$ and +27 $\pm$ 22 km $\mbox{s}^{-1}$, which suggests that both \textrm{Si}~\textsc{ii} and \textrm{Al}~\textsc{ii} have velocities that are roughly consistent on average with those of the near-UV \textrm{Fe}~\textsc{ii} lines. Therefore, we claim that the velocity shifts traced by far-UV \textrm{Si}~\textsc{ii} and \textrm{Al}~\textsc{ii} in general agree with that traced by near-UV \textrm{Fe}~\textsc{ii} lines despite larger uncertainties associated with the velocity measurements of \textrm{Si}~\textsc{ii}$\lambda$1526.

It is worth noting that \textrm{Fe}~\textsc{ii}$\lambda$1608, on an individual basis, appears to be significantly more redshifted relative to \textrm{C}~\textsc{iv} (median $\mbox{V}_{\textrm{Fe}~\textsc{ii}\lambda1608}-\mbox{V}_{1:1}$ = +161 km $\mbox{s}^{-1}$), and near-UV \textrm{Fe}~\textsc{ii} lines (median $\mbox{V}_{\textrm{Fe}~\textsc{ii}\lambda1608}-\mbox{V}_{\textrm{nuvFe}~\textsc{ii}}$ = +124 km $\mbox{s}^{-1}$ median). To investigate the reason for the relative redshift of \textrm{Fe}~\textsc{ii}$\lambda$1608, we constructed a composite spectrum and compared the absorption profiles of both \textrm{Fe}~\textsc{ii}$\lambda$1608 and the near-UV \textrm{Fe}~\textsc{ii} lines. We selected a subset of 65 galaxies, which were observed with 400-line masks (to ensure the same resolution), with continuum $S/N >5$ in the vicinity of the near-UV \textrm{Fe}~\textsc{ii} features from 2400 $\mbox{\AA}$ to 2550 $\mbox{\AA}$. We then stacked individual continuum-normalized spectra, and extracted the median value at each wavelength to create a composite spectrum. In Figure~\ref{fig:fe1608}, \textrm{Fe}~\textsc{ii}$\lambda$1608, 2344, 2374 and 2587 are plotted with respect to their rest wavelengths using black, blue, green and red lines, respectively. While there is an agreement among the velocity profiles of \textrm{Fe}~\textsc{ii}$\lambda$2344, 2374, 2587, \textrm{Fe}~\textsc{ii}$\lambda$1608 shows an extended wing on the red side at $\sim$+800km $\mbox{s}^{-1}$, making the fitted centroid redshifted compared to the other \textrm{Fe}~\textsc{ii} lines. 

One plausible interpretation of this profile is that \textrm{Fe}~\textsc{ii}$\lambda$1608 suffers less emission filling than the other near-UV \textrm{Fe}~\textsc{ii} lines. The continuum photons emitted by galaxies can be absorbed by the interstellar medium (ISM) and re-emitted in random directions. The electrons, after getting excited to a higher energy level from the ground state, can either decay back to the ground state (resonant emission) or to an excited ground state (fluorescent emission). The resonant emission cannot cancel out the resonant absorption exactly because of different Doppler velocities of the emitting and absorbing gas, thus we have to consider the extent to which the resonant emission, usually expected in galaxy spectra, fills the absorption trough and results in a more blueshifted absorption profile than it would otherwise \citep{Prochaska2011,Martin2012}. Based on the Einstein A coefficients of the resonant and fluorescent emission, following the absorption in the \textrm{Fe}~\textsc{ii}$\lambda$1608 resonant transition, only 26$\%$ of the photons are redirected into into fluorescent emission lines at Fe$\mbox{II}^{*}\lambda$1618.47 (22$\%$) and Fe$\mbox{II}^{*}\lambda$1625.91 (4$\%$). In comparison, \textrm{Fe}~\textsc{ii}$\lambda$2382 absorption results in no fluorescent emissions (i.e., all photons return to the ground state exclusively through resonant emission), and the fractions of fluorescent emission of the \textrm{Fe}~\textsc{ii}$\lambda$2344, 2374, 2587 and 2600
resonant transitions are 35$\%$, 86$\%$, 67$\%$, 13$\%$, respectively. \footnote{While theoretical wind models \citep[e.g., ][]{Prochaska2011} predict a difference in kinematics among the near-UV \textrm{Fe}~\textsc{ii} lines (i.e., \textrm{Fe}~\textsc{ii}$\lambda$2382 and \textrm{Fe}~\textsc{ii}$\lambda$2600 being more blueshifted than \textrm{Fe}~\textsc{ii}$\lambda$2374 and \textrm{Fe}~\textsc{ii}$\lambda$2587), we do not observe this trend in our sample. Instead, \textrm{Fe}~\textsc{ii}$\lambda$2382 and \textrm{Fe}~\textsc{ii}$\lambda$2600 display similar centroid velocities on average to those of \textrm{Fe}~\textsc{ii}$\lambda$2344, \textrm{Fe}~\textsc{ii}$\lambda$2374 and \textrm{Fe}~\textsc{ii}$\lambda$2587, suggesting that \textrm{Fe}~\textsc{ii} features are less subject to emission filling than \textrm{Mg}~\textsc{ii} \citep[see also][for a full discussion of emission filling in \textrm{Fe}~\textsc{ii} and \textrm{Mg}~\textsc{ii} features]{Erb2012}.} Thus emission filling is apparently more significant for the \textrm{Fe}~\textsc{ii}$\lambda$1608 profile than for the near-UV \textrm{Fe}~\textsc{ii} lines used in \citet{Martin2012}, and is only slightly less of an issue than for the lines excluded (i.e., \textrm{Fe}~\textsc{ii}$\lambda$2382 and \textrm{Fe}~\textsc{ii}$\lambda$2600). Since emission filling should affect \textrm{Fe}~\textsc{ii}$\lambda$1608 more than the other lines plotted, and, if anything, should result in a larger blueshift due to more emission filling on the red side of the absorption profile (in contract to what is observed), this process clearly does not provide an explanation for the discrepant kinematics between \textrm{Fe}~\textsc{ii}$\lambda$1608 and the near-UV \textrm{Fe}~\textsc{ii} features.

Another possible explanation is additional absorption by a nearby resonant transition, i.e., \textrm{Fe}~\textsc{ii}$\lambda$1611, at the red edge of \textrm{Fe}~\textsc{ii}$\lambda$1608. \textrm{Fe}~\textsc{ii}$\lambda$1611 has an oscillator strength $\sim$ 2.5$\%$ that of \textrm{Fe}~\textsc{ii}$\lambda$1608 and gives additional absorption at $\sim$+500km $\mbox{s}^{-1}$ when blended with the \textrm{Fe}~\textsc{ii}$\lambda$1608 line. Meanwhile, inspection of the \citet{Leitherer2010} far-UV spectra indicates that stellar photospheric \textrm{Fe}~\textsc{iv} absorption at $\sim$ 1610$\mbox{\AA}$ may also contribute to the redshifted profile of \textrm{Fe}~\textsc{ii}$\lambda$1608 \citep[][private communication]{Dean1985,Leitherer2010}. Due to the potential contamination from \textrm{Fe}~\textsc{ii}$\lambda$1611 (more than 2.5$\%$ if the \textrm{Fe}~\textsc{ii} lines are not optically thin), \textrm{Fe}~\textsc{iv} and possibly additional unidentified absorption lines, we do not use \textrm{Fe}~\textsc{ii}$\lambda$1608 for determining low-ionization absorption kinematics.

\section{Galaxy Properties}
\label{sec:galprop}

\begin{figure*}
\includegraphics[width=0.5\linewidth]{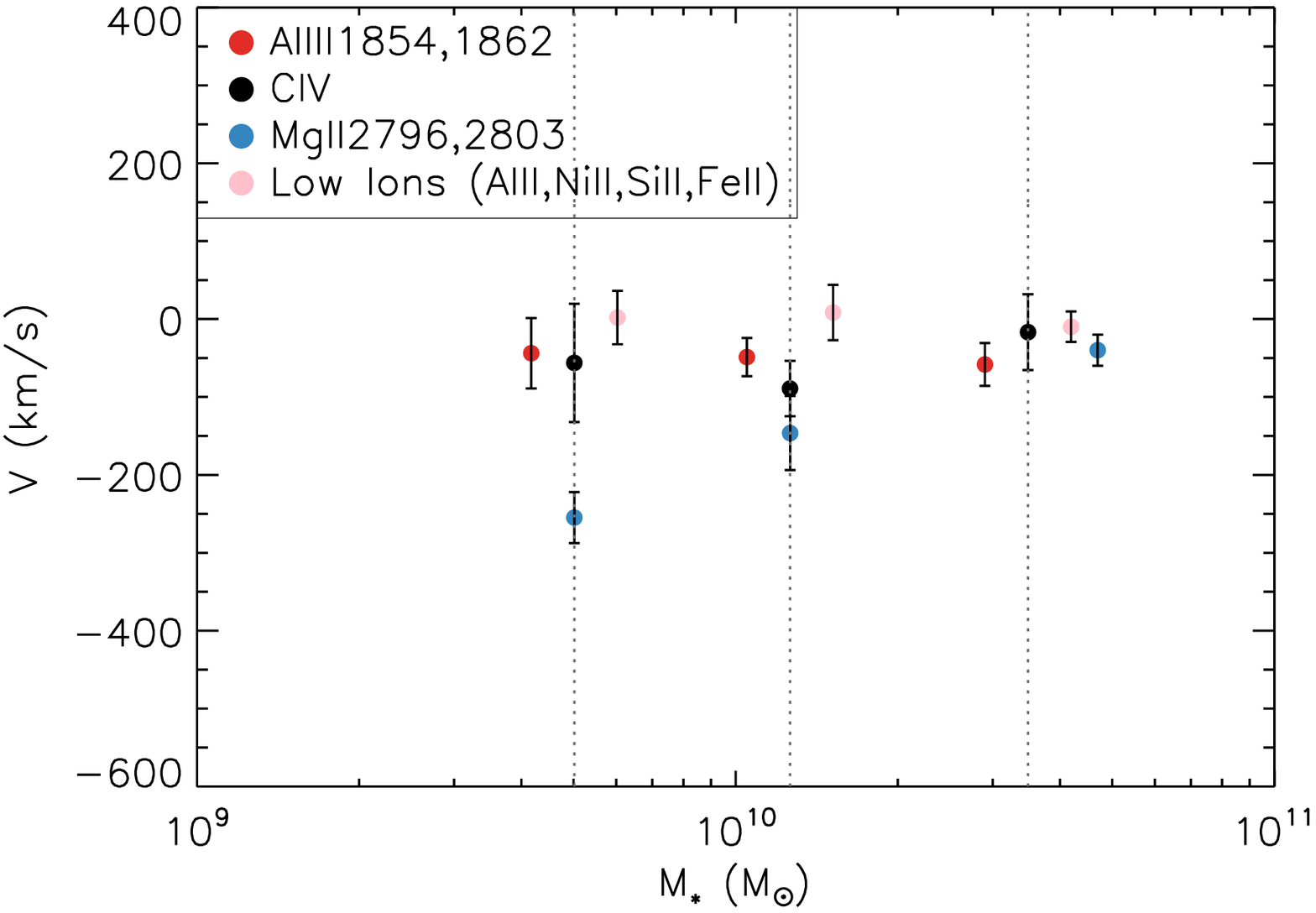}
\includegraphics[width=0.5\linewidth]{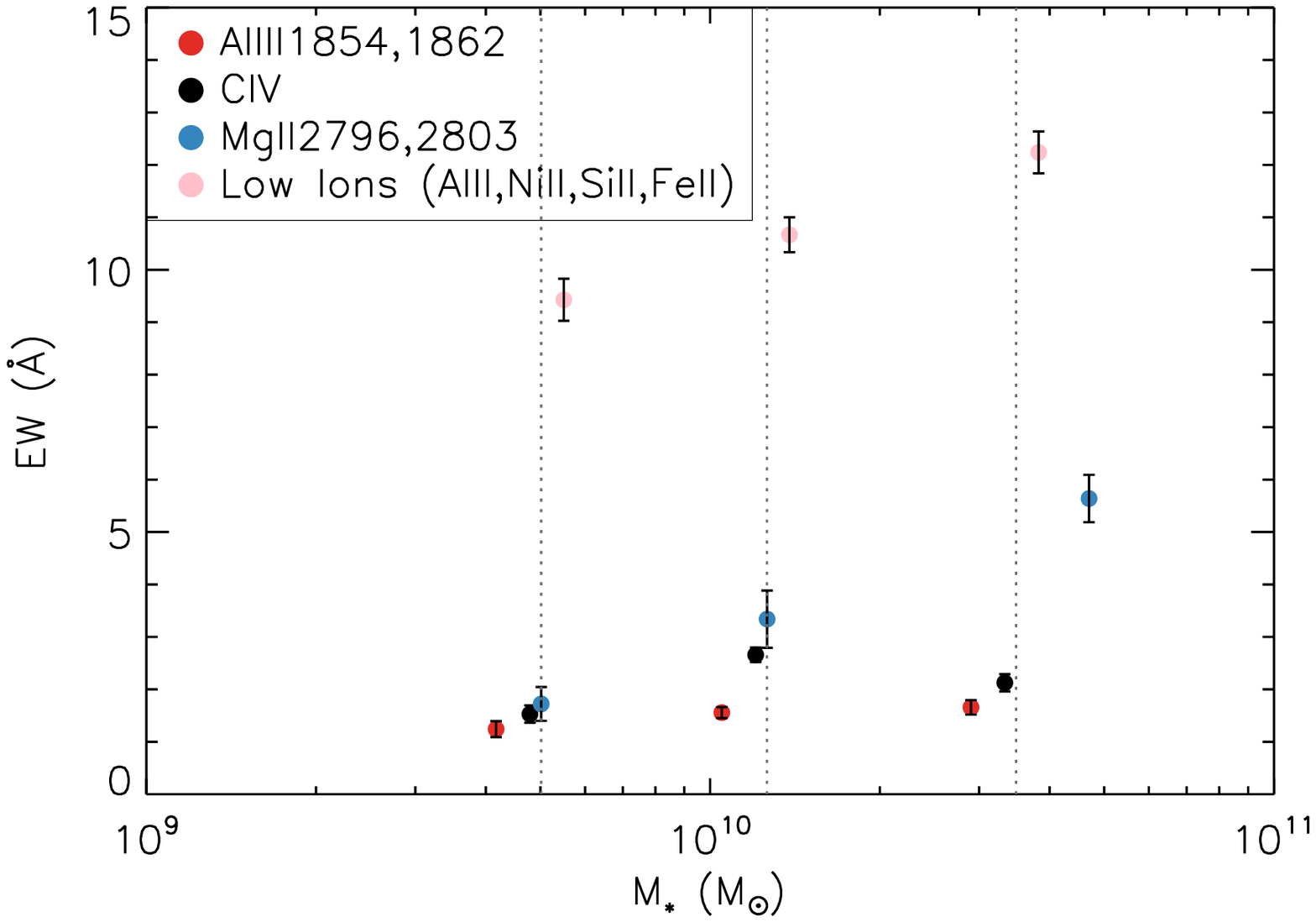}
\caption{Interstellar absorption velocity shift and EW vs. stellar mass in the composite spectra. The median stellar mass of each bin is indicated with the vertical gray dashed lines, and the data points are offset horizontally for display purposes. The black points represent \textrm{C}~\textsc{iv} measurements, while the red points indicate the \textrm{Al}~\textsc{iii} doublet. The light pink points indicate a set of low-ionization lines (\textrm{Si}~\textsc{ii}$\lambda$1526, \textrm{Al}~\textsc{ii}$\lambda$1670, \textrm{Si}~\textsc{ii}$\lambda\lambda$1741,1751, \textrm{Si}~\textsc{ii}$\lambda$1808 and near-UV \textrm{Fe}~\textsc{ii}), representing the average velocity shift of the total EW. Finally, the \textrm{Mg}~\textsc{ii} doublet, as isolated from the other low-ionization lines, is shown by blue points.}
\label{fig:ew-sm}
\end{figure*}

While the study of individual velocity shifts can only be performed on objects with high $S/N$ and significant line detections, composite spectra enable us to utilize the full sample and study the overall correlations between spectral and galaxy properties. Moreover, using composite spectra potentially enables the measurement of weak absorption lines by increasing continuum $S/N$ (continuum $S/N$ increased by a fact of $\sqrt{N}$, where $N$ is the number of objects that go into each composite spectrum). 

Within the coverage of the DEEP2/LRIS spectra, there reside multiple weak far-UV transition lines, including \textrm{Ni}~\textsc{ii}$\lambda\lambda$1741, 1751, \textrm{Si}~\textsc{ii}$\lambda$1808 and \textrm{Al}~\textsc{iii}$\lambda\lambda$1854, 1862, which we were not able to measure robustly in individual spectra. Composite spectra enabled the measurement of these weak lines in addition to the strong UV absorption lines we described before (\textrm{Si}~\textsc{ii}$\lambda$1526, \textrm{C}~\textsc{iv}$\lambda\lambda$1548, 1550, \textrm{Al}~\textsc{ii}$\lambda$1670, \textrm{Fe}~\textsc{ii}$\lambda$2344, $\lambda$2374, $\lambda$2587 and \textrm{Mg}~\textsc{ii}$\lambda\lambda$2796, 2803). In light of the fact that outflows have a multi-phase structure, we included \textrm{Al}~\textsc{iii} measurements here to explore the kinematics of an intermediate-ionization phase in the composite spectra.

To investigate the correlation between the kinematics and strength of both high- and low-ionization lines and galaxy properties, we divided all 93 objects with \textrm{C}~\textsc{iv} coverage into three bins in stellar mass, $\ub$ color and $B$-band luminosity, and two bins in SFR, SFR surface density and sSFR, given that only 25 objects in the AEGIS field have $\mbox{SFR}_{UV}$ measurements. Each bin contained nearly the same number of galaxies. To create the composite spectra, we used individual continuum-normalized spectra for measurements of the low-ionization lines and spectra normalized by the \citet{Leitherer2010} stellar models for measurements of \textrm{C}~\textsc{iv} and \textrm{Al}~\textsc{iii}. \textrm{Al}~\textsc{iii}, like \textrm{C}~\textsc{iv}, has a complex profile including both the absorption from stellar winds and the ISM. However, since the stellar wind component contributing to the \textrm{Al}~\textsc{iii} profile is several times weaker than that of \textrm{C}~\textsc{iv} \citep{Leitherer2010}, we simply used the blue wing of the \textrm{C}~\textsc{iv} feature to determine the stellar absorption affecting both \textrm{C}~\textsc{iv} and \textrm{Al}~\textsc{iii}. For each object, we divided the continuum-normalized spectrum by the best-fit stellar model determined from \textrm{C}~\textsc{iv} (described in Section \ref{sec:civ}). We then smoothed all 600-line spectra to the resolution of the 400-line spectra (for both continuum-normalized data and those normalized by the best-fit stellar models), and used the IRAF routine $scombine$ to extract the median value of the normalized spectra at each wavelength to create a normalized composite spectrum for each bin. To create the corresponding error spectrum for each composite spectrum, we bootstrap resampled each bin and perturbed each spectrum in the bootstrap sample according to its own error spectrum. The perturbed spectra in the bootstrap sample were then combined to create a new composite spectrum. The process was repeated 100 times and the standard deviation of these 100 fake composites at each wavelength was taken to create a composite error spectrum for each bin. With this approach, we accounted for both sample variance and measurement uncertainty. 

Both centroid velocity and $\mbox{V}_{max}$ can be used to probe the kinematics of interstellar gas. Although $\mbox{V}_{max}$ is not subject to potential emission filling, it is not resolved in all the bins of the stacked spectra, especially for \textrm{C}~\textsc{iv}. On the other hand, the centroid velocity is well measured for features in all stacked spectra. Accordingly, we used centroid velocity inferred from the absorption features for the kinematic analysis in this section.

We fit the \textrm{C}~\textsc{iv},  \textrm{Si}~\textsc{ii}$\lambda$1526, $\lambda$1808, \textrm{Al}~\textsc{ii}, \textrm{Ni}~\textsc{ii}, \textrm{Al}~\textsc{iii}, near-UV \textrm{Fe}~\textsc{ii} and \textrm{Mg}~\textsc{ii} features in the stacked spectra with Gaussian profiles using MPFIT to measure their centroids and EWs. We required all centroids of the doublet members and the near-UV \textrm{Fe}~\textsc{ii} lines to be fixed with the exception of \textrm{Mg}~\textsc{ii}, for the reason stated in Section ~\ref{sec:low-ion}. We also forced the FWHM of each \textrm{C}~\textsc{iv} and near-UV \textrm{Fe}~\textsc{ii} member to be the same. We allowed the width of doublet members to float freely for \textrm{Ni}~\textsc{ii} and \textrm{Al}~\textsc{iii}, since the profile of these weak features is significantly affected by noise, and thus may deviate from 1:1 width ratio.\footnote{The results do not change significantly if the FWHMs of the doublet members are fixed at 1:1 for the weak transitions.} 

Since we found no significant correlations between the spectral properties (blueshift and strength) of any lines listed above and $\ub$ color or $B$-band luminosity in the composite spectra, we only discuss stellar mass, SFR and sSFR in this section where systematic trends have been observed.
 
\subsection{Stellar Mass}
\label{sec:sm}

The plot of measured velocity shift and EW versus stellar mass is shown in Figure~\ref{fig:ew-sm}. The median stellar mass of each bin is plotted as a gray dashed line, and the data points are offset horizontally for display purposes. We show the summed EW and average velocity shift for all low-ionization lines (\textrm{Si}~\textsc{ii}$\lambda$1526, \textrm{Al}~\textsc{ii}$\lambda$1670, \textrm{Ni}~\textsc{ii}$\lambda\lambda$1741,1751, \textrm{Si}~\textsc{ii}$\lambda$1808 and near-UV \textrm{Fe}~\textsc{ii}) with the exception of \textrm{Mg}~\textsc{ii}. \textrm{Mg}~\textsc{ii} is plotted separately, as the trends it follows may be more significant due to emission filling.

In Figure~\ref{fig:ew-sm}, all lines except \textrm{Mg}~\textsc{ii} reside near the systemic velocity, and none but \textrm{Mg}~\textsc{ii} exhibits a trend between the velocity shift and stellar mass. It indicates that the blueshift of \textrm{C}~\textsc{iv}, \textrm{Al}~\textsc{iii} and the low-ionization lines, \textrm{Mg}~\textsc{ii} excluded, is not sensitive to stellar mass. \textrm{Mg}~\textsc{ii} displays a smaller blueshift and a larger EW in galaxies with higher stellar masses. Given that $\mbox{V}_{max,\textrm{Mg}~\textsc{ii}}$ for all three stellar mass bins are similar (-519 km $\mbox{s}^{-1}$, -566 km $\mbox{s}^{-1}$ and -538 km $\mbox{s}^{-1}$ for the high-, medium- and low-$\mbox{M}_{*}$ bins, respectively), this trend suggests that outflow velocity does not depend strongly on stellar mass \citep[See also ][]{Martin2012}. Rather, it results from the fact that resonant \textrm{Mg}~\textsc{ii} emission fills more of the absorption trough in less massive galaxies, making the observed profile more blueshifted while having a smaller EW. The strength of the low-ionization lines shows a positive correlation with stellar mass. Since massive galaxies tend to have more interstellar gas, larger covering fractions for the cold gas content in ISM lead to a deeper absorption trough for the low-ionization lines \citep{Erb2006}. \textrm{C}~\textsc{iv} and \textrm{Al}~\textsc{iii}, on the other hand, show no significant correlations in EW, suggesting that the strength of intermediate- and high-ionization absorption lines is independent of stellar mass.

\subsection{SFR}
\label{sec:sfr}

\begin{figure*}
\includegraphics[width=0.5\linewidth]{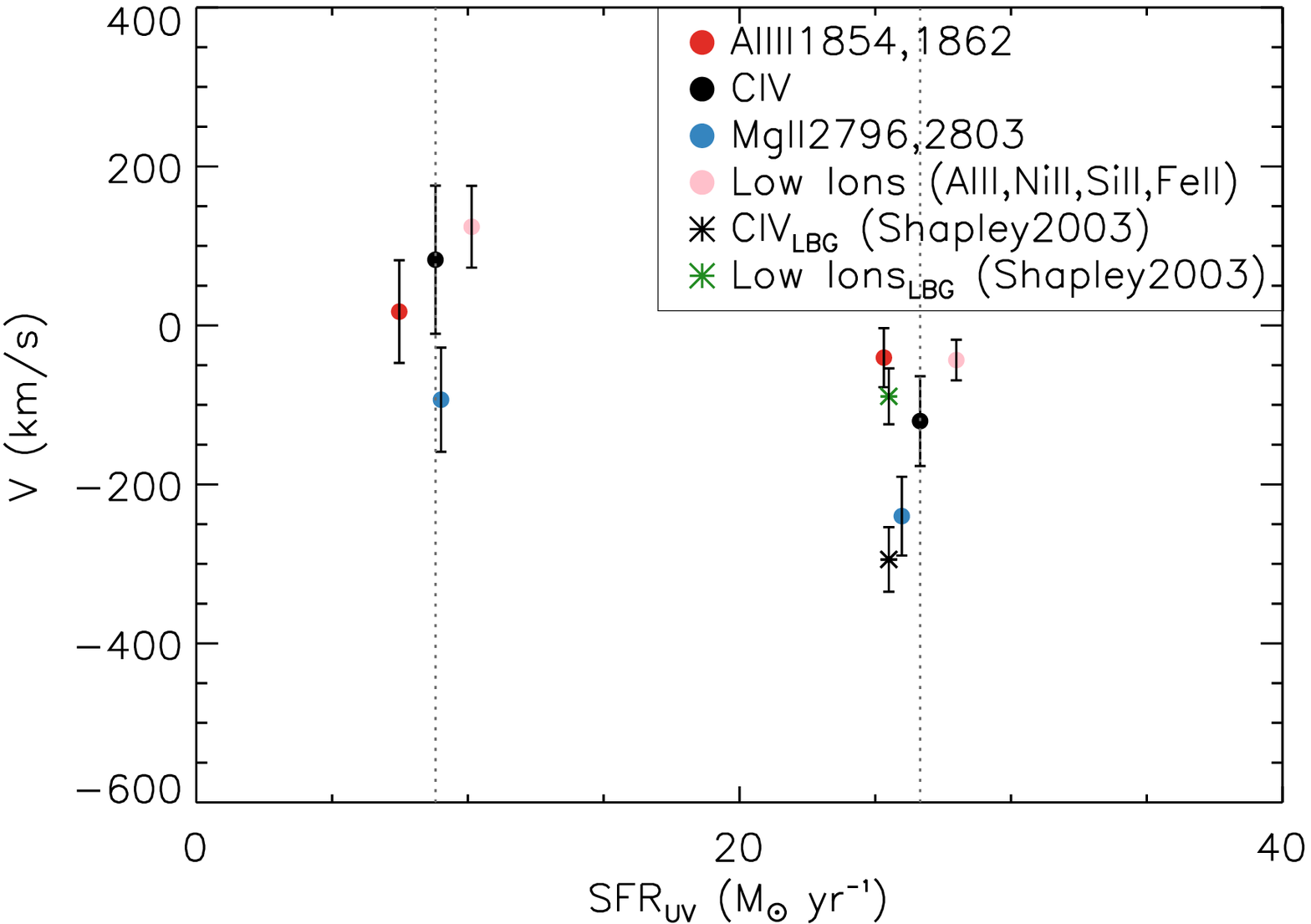}
\includegraphics[width=0.5\linewidth]{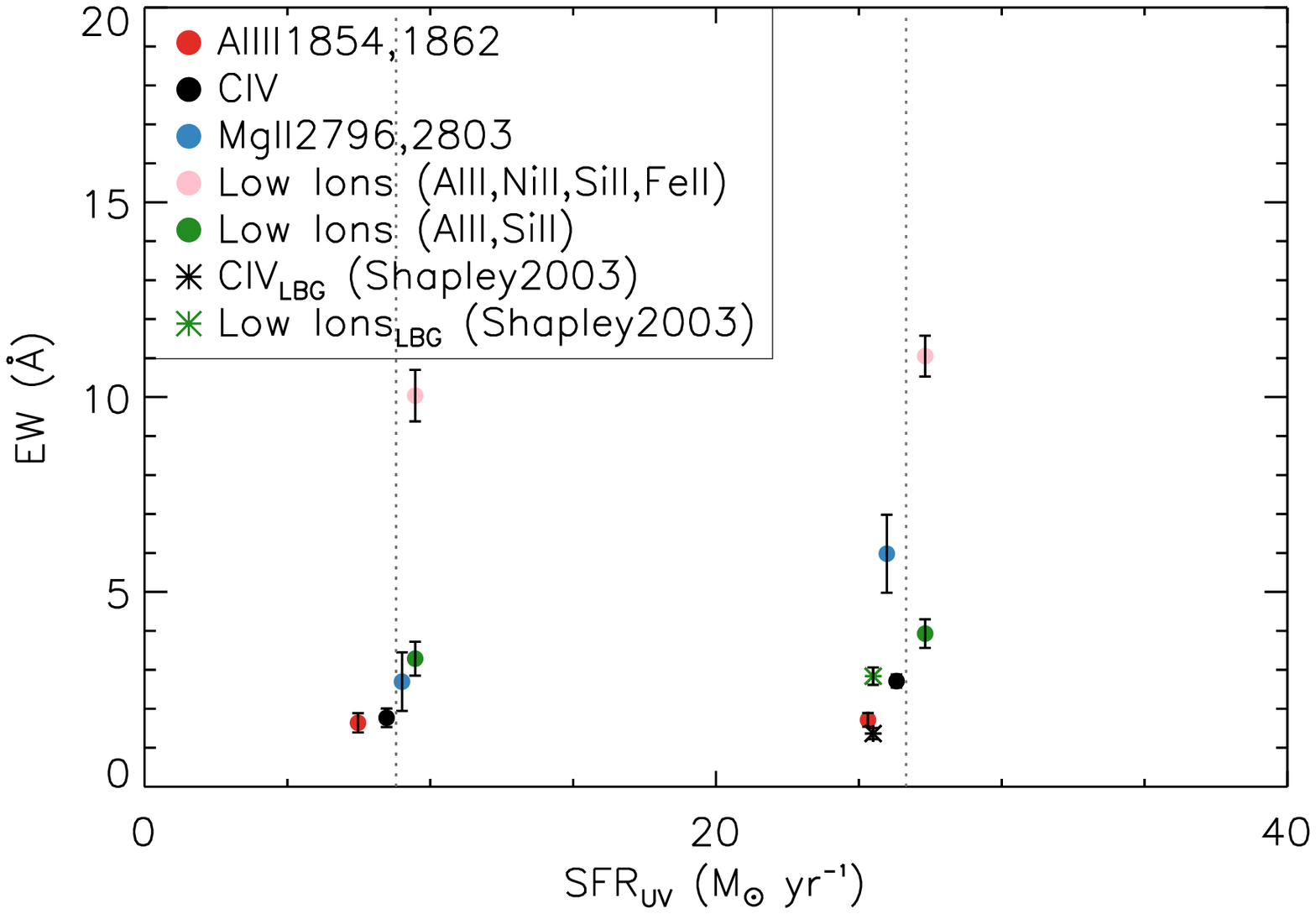}
\caption{Interstellar absorption velocity shift and EW vs. SFR in the composite spectra. Legends are the same as in Figure~\ref{fig:ew-sm}. The green and black stars are measurements from the LBG composite in \citet{Shapley2003}, where only \textrm{Si}~\textsc{ii}$\lambda$1526 and \textrm{Al}~\textsc{ii}$\lambda$1670 are available for the low-ionization line measurements. The errors for the LBG sample was estimated from fitting 500 fake composites, perturbing the returned parameters (centroids, EW) with associated error for each fake composite, and taking the standard deviation from the final distribution of the 500 perturbed centroids or EWs. The green circle in the right panel represents the total EW of \textrm{Si}~\textsc{ii}$\lambda$1526 and \textrm{Al}~\textsc{ii}$\lambda$1670 in the DEEP2 sample, in order to make reasonable comparison with the measurements from \citet{Shapley2003}. }
\label{fig:ctr-ew-SFR}
\end{figure*}

Figure~\ref{fig:ctr-ew-SFR} shows how the measured velocity shift and EW vary with SFR for absorption transitions tracing different ionization states. In the right panel, the green circles represent the total EW of only \textrm{Si}~\textsc{ii}$\lambda$1526 and \textrm{Al}~\textsc{ii}$\lambda$1670 for the comparison with the measurements from Lyman break galaxies (LBGs) from \citet{Shapley2003} (see Section \ref{sec:lbg}).

We find that both the blueshift and EW of \textrm{C}~\textsc{iv} increase with an increasing SFR. The same trends of both blueshift and EW with SFR also hold for the low-ionization lines, \textrm{Mg}~\textsc{ii} included. In addition, it is interesting that the centroids of \textrm{C}~\textsc{iv} and the low-ionization lines (excluding \textrm{Mg}~\textsc{ii}) are $redshifted$ in the low-SFR bin, and the low-ionization lines are barely blueshifted in the high-SFR bin. Figure~\ref{fig:ctr-ew-SFR} shows that the \textrm{C}~\textsc{iv} properties are exclusively related to SFR, suggesting that massive star formation and associated processes drive the properties of the highly-ionized gas traced by \textrm{C}~\textsc{iv}, while for low-ionization lines, the properties correlate with both SFR and stellar mass. Many studies have revealed a relationship between SFR and stellar mass in star-forming galaxies \citep[often referred as the `main sequence';][]{Elbaz2007,Noeske2007,Kashino2013}. Hence, the observed correlation between low-ionization absorption line properties and SFR may simply be a byproduct of the relationship with stellar mass. The intermediate-ionization doublet, \textrm{Al}~\textsc{iii}, appears to have a nearly zero velocity shift (at systemic velocity) in both bins and displays no significant correlation with SFR.

It is worth mentioning that \textrm{C}~\textsc{iv} shows a larger blueshift relative to the low ions in the high-SFR bin than in the low-SFR bin. This larger differential blueshift may result from a larger contribution of shocked, collisionally ionized gas to the \textrm{C}~\textsc{iv} absorption in the high-SFR bin. Alternatively, \textrm{C}~\textsc{iv} may simply be more affected by emission filling at higher SFRs.

The larger blueshift of \textrm{Mg}~\textsc{ii} relative to that of \textrm{C}~\textsc{iv} in the low-SFR bin at first seems to be inconsistent with our results stated in Section \ref{sec:civnuv} that these two transitions display similar velocity shifts on average. However, we find that the objects that contribute to the low-SFR composite are not representative of the parent sample (93 objects with \textrm{C}~\textsc{iv} coverage) in terms of \textrm{C}~\textsc{iv} kinematics, and those with detected \textrm{C}~\textsc{iv} tend to have less blueshifted and even significantly redshifted absorption profiles (measured velocity shifts ranged from -52 km $\mbox{s}^{-1}$ to +191 km $\mbox{s}^{-1}$, with a median of +44 km $\mbox{s}^{-1}$).\footnote{In order to reduce the potential bias introduced by small sample size, we did the same analysis using the Bundy SFR. Although we believe $\mbox{SFR}_{UV}$ is more reliable, they only cover 25 out of 93 objects in our sample. Bundy SFRs, on the other hand, are available for all 93 objects. Given these two SFR measurements are correlated (Figure~\ref{fig:sfrcom}), the composite spectra created using the Bundy SFR should qualitatively reproduce the results derived from $\mbox{SFR}_{UV}$-generated composites. Therefore, we produced three composite spectra (low-, median- and high-SFR bins) according to Bundy SFR and each bin contained 31 objects. Again, \textrm{C}~\textsc{iv} is less blueshifted than \textrm{Mg}~\textsc{ii} in the low-$\mbox{SFR}_{Bundy}$ bin (measured \textrm{C}~\textsc{iv} velocity shifts ranged from -472 km $\mbox{s}^{-1}$ to +191 km $\mbox{s}^{-1}$, with a median of -32 km $\mbox{s}^{-1}$). Note that the median blueshift of \textrm{C}~\textsc{iv} in this case is still not as large compared to that in the sample with robust \textrm{C}~\textsc{iv} measurements (-64 km $\mbox{s}^{-1}$). Hence, a larger sample cannot fully account for the discrepancy we observe.} We are not certain at this point whether this discrepancy is due to the selection effect of a small sample (13 objects in the low-SFR bin), or whether galaxies with low SFRs are intrinsically associated with high-ionization inflows. A larger sample is required to address this question.

\subsection{Specific SFR}
\label{sec:ssfr}

\begin{figure*}
\includegraphics[width=0.5\linewidth]{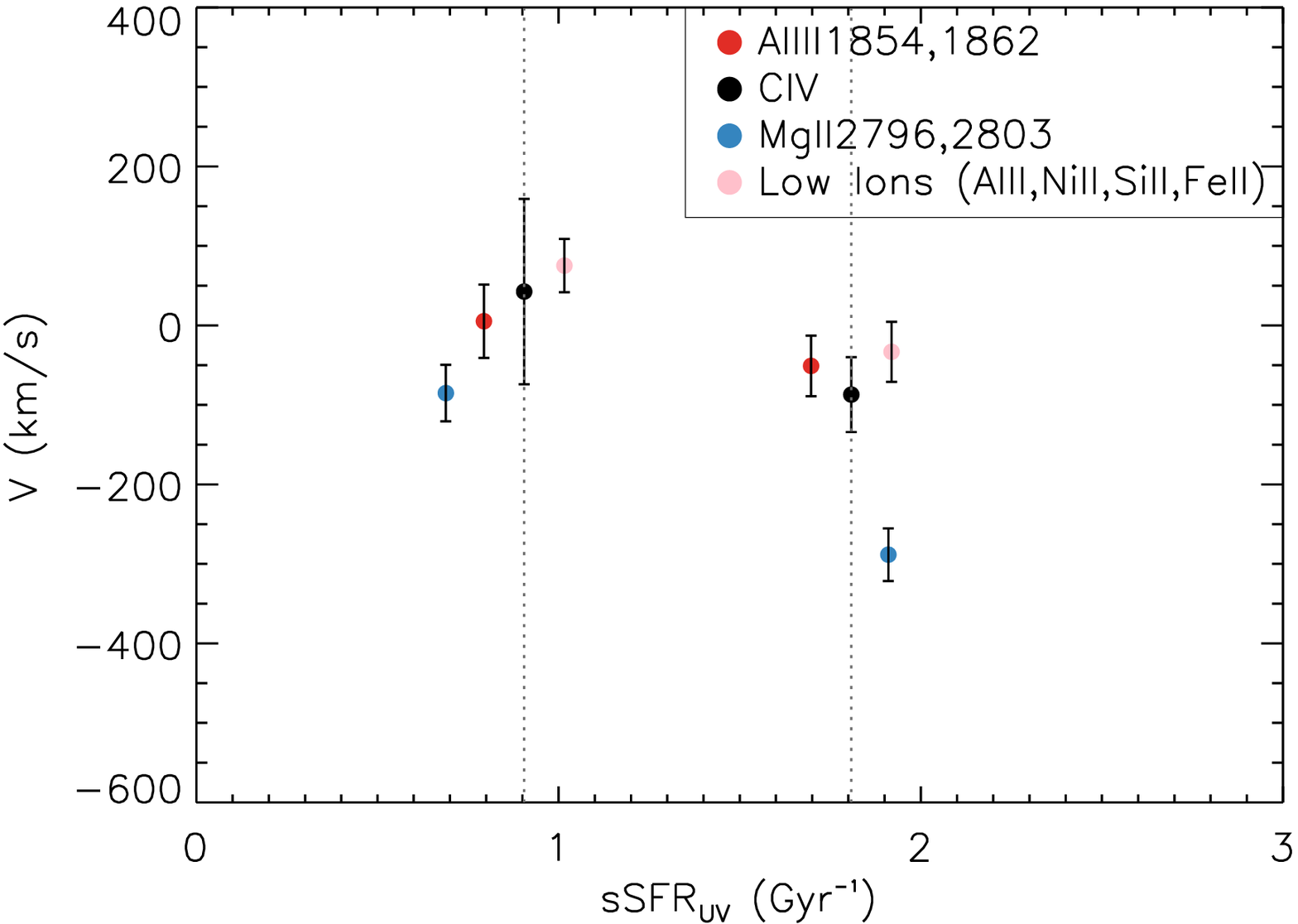}
\includegraphics[width=0.5\linewidth]{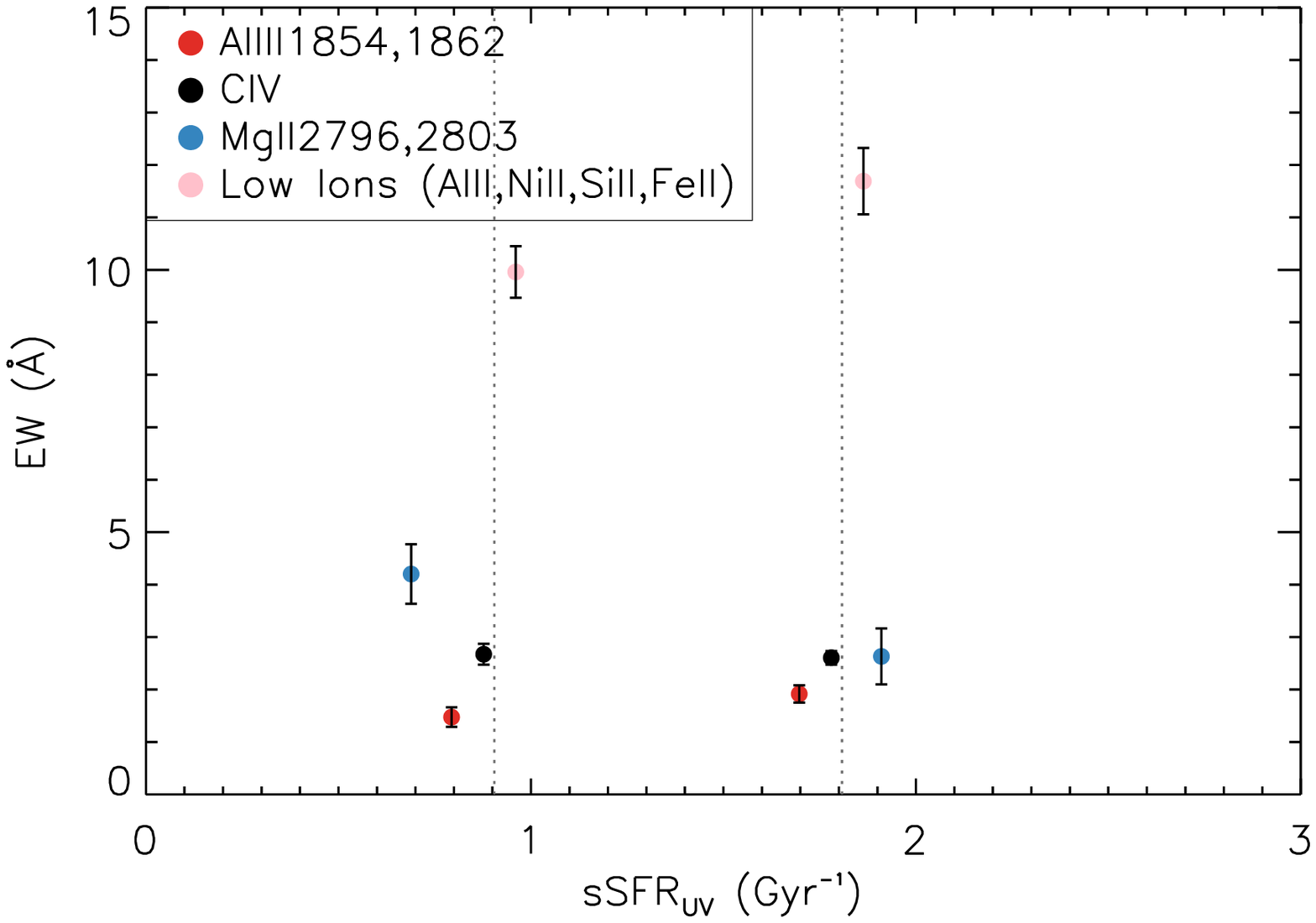}
\caption{Interstellar absorption velocity shift and EW vs. sSFR in the composite spectra. The legends are the same as in Figure~\ref {fig:ew-sm}.}
\label{fig:ctr-ew-sSFR}
\end{figure*}

Figure~\ref{fig:ctr-ew-sSFR} shows the relation between kinematics of low-, intermediate- and high-ionization lines and sSFR. While becoming more blueshifted and stronger at higher sSFR, both \textrm{C}~\textsc{iv} and the low-ionization lines (excluding \textrm{Mg}~\textsc{ii}) are redshifted in the low-sSFR bin and have marginally blueshifted centroids in the high-sSFR bin. Although \textrm{Mg}~\textsc{ii} also shows a larger blueshift at higher sSFR due to the increased effects of emission filling, its strength actually decreases at higher sSFR. The contrasting behavior of the EWs of \textrm{Mg}~\textsc{ii} and the other low-ionization lines can be explained in terms of the interplay between SFR and stellar mass. All low-ionization lines, \textrm{Mg}~\textsc{ii} included, increase in strength with both increasing SFR and stellar mass. The observed trends with sSFR (i.e., SFR/$\mbox{M}_{*}$) are therefore weak, and reflect a stronger dependence of \textrm{Mg}~\textsc{ii} emission filling on stellar mass than on SFR. The intermediate-ionization phase \textrm{Al}~\textsc{iii}, is not sensitive to sSFR and shows no strong correlations.

\section{Summary and Discussion}
\label{sec:sum}

In this paper, we have analyzed the far-UV LRIS spectra of a sample of 93 DEEP2 galaxies at $z\sim$1 with coverage of \textrm{C}~\textsc{iv}$\lambda\lambda$1548,1550. We focused on 46 objects with continuum $S/N$ $>$ 5 and investigated the kinematics of highly-ionized gas in 32 objects with significant \textrm{C}~\textsc{iv} detections. We have also measured low-ionization far-UV lines as well as near-UV \textrm{Mg}~\textsc{ii} and \textrm{Fe}~\textsc{ii}. We found that the typical blueshift of \textrm{C}~\textsc{iv} absorption was slightly greater than that of near-UV \textrm{Fe}~\textsc{ii} resonant absorption lines at the $\sim$ 3-sigma level (intercept of the best-fit linear regression -76 $\pm$ 26 km $\mbox{s}^{-1}$), with a similar detection fraction of blueshifted \textrm{C}~\textsc{iv} absorption. We further compared the \textrm{C}~\textsc{iv} absorption profile with that of  \textrm{Mg}~\textsc{ii}. Given that both the centroid velocity shifts and the maximum velocities of \textrm{C}~\textsc{iv} and \textrm{Mg}~\textsc{ii} are nearly identical, we concluded that the larger blueshift of \textrm{C}~\textsc{iv} absorption with respect to the near-UV \textrm{Fe}~\textsc{ii} may be caused by either faster motion of the highly-ionized gas, or, more likely, filling in on the red side from resonant \textrm{C}~\textsc{iv} emission in analogy to what has been observed in \textrm{Mg}~\textsc{ii} and \textrm{Fe}~\textsc{ii} \citep{Martin2012,Kornei2012}. 

We also investigated the scaling relations between \textrm{C}~\textsc{iv} kinematics and galaxy properties by making composite spectra of galaxies binned according to stellar mass, SFR, sSFR, SFR surface density, $\ub$ color and $B$-band luminosity. We found that both the blueshift and EW of \textrm{C}~\textsc{iv} increase as SFR and sSFR increase, suggesting that outflowing high-ionization gas is associated with processes related to star formation, such as energy and momentum input from massive stars and their supernovae. The strength of the low-ionization lines, on the other hand, is more correlated with the stellar mass, which can be explained by larger covering fractions and velocity dispersions of the cold phase of the ISM in more massive galaxies, due, respectively, to the greater amount of interstellar gas and overall dynamical mass. The properties of \textrm{Al}~\textsc{iii} are not sensitive to any of the galaxy properties we studied.

\subsection{Redshift Evolution}
\label{sec:lbg}

To investigate the possible evolution of the \textrm{C}~\textsc{iv} kinematics, we compared our results with the $z\sim3$ Lyman Break Galaxy (LBG) composite from \citet{Shapley2003}. The composite was constructed from 811 individual LBG spectra, and spans from 900$\mbox{\AA}$ to 2000$\mbox{\AA}$ in wavelength. Since no near-UV spectral features were available, and all the absorption lines beyond 1800$\mbox{\AA}$ were poorly detected in the LBG composite, we only measured \textrm{C}~\textsc{iv}, \textrm{Si}~\textsc{ii}$\lambda$1526 and \textrm{Al}~\textsc{ii}$\lambda$1670. Due to the lack of the corresponding composite error spectrum, we estimated the uncertainties using a method consistent with that of \citet{Shapley2003}: we measured the noise from a relatively featureless range of the continuum (1700 $\mbox{\AA}$ to 1825 $\mbox{\AA}$) and assigned this value as a constant error across the full spectrum to account for pixel variance. To account for sample variation, we fit the far-UV absorption lines in 500 fake composite spectra, which were constructed from bootstrap resampling from the original sample used for the real composite. For each fake composite, the parameters returned by MPFIT were then perturbed by the associated errors. The final error on the parameters for the real composite was derived from the standard deviation of the parameter distributions over the fake composites. SFRs for LBGs were estimated from dust-corrected rest-frame UV luminosities, combining apparent $R$ magnitudes, and extinction based on observed $G-R$ colors. The median SFR from the LBG sample is 26 $\mbox{M}_{\odot}$ $\mbox{yr}^{-1}$ (assuming a Chabrier IMF and $H_{0}=70$ km $\mbox{s}^{-1}$), which is close to that of the high SFR bin in the DEEP2/LRIS sample. Stellar mass and sSFR are not available for the LBGs, thus SFR is the only galaxy property we study here.

As shown in Figure~\ref{fig:ctr-ew-SFR}, both \textrm{C}~\textsc{iv} and far-UV low-ionization lines in LBGs have a higher blueshift yet a lower EW compared to our sample with similar SFRs. This difference may be caused by a different proportion between 
outflowing and interstellar gas in galaxies. Absorption from the systemic and outflowing portions of the ISM jointly contribute to the observed absorption profile. If instead, we model each absorption line (individual doublet member in the \textrm{C}~\textsc{iv} case) with a two-component fit: a dominant systemic component with no Doppler shift representing the absorption from ISM, and a Doppler component characterizing the contribution from outflows \citep{Steidel2010,Coil2011,Martin2012}, the change in either component may alter the shape of the absorption trough. Lack of ISM absorption around systemic velocity results in a more blueshifted centroid as well as a smaller EW, as the LBG sample suggests.

However, this difference can also result from emission filling, especially for resonant lines like \textrm{C}~\textsc{iv}. In order to investigate the real cause, we further compared the maximum velocity of \textrm{C}~\textsc{iv}, $\mbox{V}_{max,\textrm{C}~\textsc{iv}}$, in both the LBG composite and the DEEP2/LRIS sample. The measurement of $\mbox{V}_{max,\textrm{C}~\textsc{iv}}$ in LBGs was performed in an analogous way to that described in Section \ref{sec:civfuv}. For a rough estimate of uncertainties in the LBG composite, we increased the constant error, which was derived from the continuum noise over the wavelength range of 1700 $\mbox{\AA}$ - 1825 $\mbox{\AA}$, by a factor of two. We then measured $\mbox{V}_{max,\textrm{C}~\textsc{iv}}$ of the LBGs in 500 fake composite spectra (the same set as described previously in this section) and derived an average value of -649 km $\mbox{s}^{-1}$. In comparison, $\mbox{V}_{max,\textrm{C}~\textsc{iv}}$ of the high-SFR bin in the DEEP2/LRIS sample is -542 km $\mbox{s}^{-1}$ with a similar SFR (30 $\mbox{M}_{\odot}$ $\mbox{yr}^{-1}$). The larger $\mbox{V}_{max}$, along with a bluer centroid, suggests that the entire absorption profile of \textrm{C}~\textsc{iv} in LBGs is shifted bluewards with less absorption, rather than only having emission filling on the red side of the absorption trough. This observed trend may arise as a consequence of a patchier systemic ISM component and a more significant Doppler component in the LBGs compared to that in the DEEP2/LRIS galaxies.

\subsection{\textrm{C}~\textsc{iv} vs. \textrm{O}~\textsc{vi}}
\label{sec:ovi}

Our results demonstrate that low- and high-ionization interstellar absorption lines only have small differences in terms of kinematics, in the spectra of star-forming galaxies at z $\sim1$. Objects with robust measurements of both \textrm{C}~\textsc{iv} and near-UV \textrm{Fe}~\textsc{ii} show a median velocity difference of $\mbox{V}_{1:1}-\mbox{V}_{\textrm{nuvFe}~\textsc{ii}}$ = -65 km $\mbox{s}^{-1}$, and the best-fit linear regression of $\mbox{V}_{1:1}$ on $\mbox{V}_{\textrm{nuvFe}~\textsc{ii}}$ of this sample is charaterized by an intercept of -76 $\pm$ 26 km $\mbox{s}^{-1}$, assuming a slope of unity. Moreover, given that the velocity shifts of \textrm{C}~\textsc{iv} measured from both line centroids and blue wings agree with those of \textrm{Mg}~\textsc{ii}, we conclude that \textrm{C}~\textsc{iv}, being a resonant transition like \textrm{Mg}~\textsc{ii}, is affected by emission filling to the same extent. Therefore, the small apparent velocity discrepancy between \textrm{C}~\textsc{iv} and near-UV \textrm{Fe}~\textsc{ii} can be explained by filling in on the red side of the \textrm{C}~\textsc{iv} absorption trough by resonant emission, instead of evidence for highly-ionized gas outflowing at a larger speed than that of gas traced by low-ionization lines.

Similar results on the relative velocity shifts between low and high ions have been reported for $z\sim3$ LBGs and low-redshift star-forming galaxies. In the composite spectrum from \citet{Shapley2003}, the high-ionization \textrm{Si}~\textsc{iv} feature displays a similar mean blueshift and velocity FWHM to those of low-ionization lines, including \textrm{Si}~\textsc{ii}, \textrm{O}~\textsc{i}, \textrm{C}~\textsc{ii}, \textrm{Fe}~\textsc{ii}, and \textrm{Al}~\textsc{ii}. In addition, \citet{Pettini2002} studied the kinematics of different phases in ISM using the gravitationally lensed LBG, MS 1512-cB58, with much higher spectral resolution (58 km $\mbox{s}^{-1}$ FWHM). The spectrum of cB58 suggests that all the observed ion stages span the same overall range in velocity, and that the gas with the highest optical depth is blueshifted by roughly the same amount for both low (e.g., \textrm{Si}~\textsc{ii}, \textrm{C}~\textsc{ii},) and high ions (e.g., \textrm{Si}~\textsc{iv}, \textrm{C}~\textsc{iv}). At low redshift, recent work by \citet{Chisholm2016} yields similar results. These authors study 37 nearby star-forming galaxies using tracers at different ionization stages (\textrm{O}~\textsc{i}, \textrm{Si}~\textsc{ii}, \textrm{Si}~\textsc{iii}, and \textrm{Si}~\textsc{iv}), and show that absorption lines with similar strength, regardless of ionization state, have comparable outflow velocities. These findings suggest that gas clouds traced by low- and high-ionization lines may actually be co-moving.

In apparent contrast, \citet{Heckman2001} find that high-ionization lines in general have larger outflow velocities compared to the low ions by studying the rest-far-UV spectrum of the dwarf starburst galaxy NGC 1705. One distinction highlighted by \citet{Heckman2001} is that \textrm{O}~\textsc{vi}, which traces the coronal-phase gas, has broader and more blueshifted absorption profile than the low ions. \citet{Grimes2009} extended the sample to 16 local starbursts, finding similar results. Specifically, \citet{Grimes2009} detect a more significant blueshift in \textrm{O}~\textsc{vi} than in the low-ionization \textrm{C}~\textsc{ii}, \textrm{N}~\textsc{ii} features, and the intermediate-ionization stage \textrm{C}~\textsc{iii} line, suggesting that highly-ionized gas in general has larger outflow velocities.

These seemingly contradictory conclusions are in fact due to different tracers used as `high-ionization lines.' In fact, although both \textrm{C}~\textsc{iv} and \textrm{O}~\textsc{vi} can be produced both from cooler (T $< 10^{5}$ K), photoionized gas, and hotter (T $> 10^{5}$ K), collisionally ionized gas, observations and simulations suggest that the hot, collisionally ionized phase makes a more significant contribution to circumgalactic \textrm{O}~\textsc{vi}. Most observations of circumgalactic \textrm{C}~\textsc{iv} \citep[e.g.,][]{Lehner2011,Liang2014} find that this transition arises in photoionized gas \citep[but see ][]{Borthakur2013}. In simulations of galaxies including feedback \citep[e.g.,][]{Shen2013,Liang2015}, the gas traced by \textrm{O}~\textsc{vi} absorption is hotter on average, and more diffuse and spatially extended, while the gas giving rise to \textrm{C}~\textsc{iv} absorption is clumpier, and traces gas at smaller galactocentric radii. In terms of line strength, although the covering fraction of both \textrm{C}~\textsc{iv} and \textrm{O}~\textsc{vi} decreases less rapidly in the radial direction than that of the low ions \citep{Steidel2010,Shen2013, Liang2014}, \textrm{C}~\textsc{iv} displays a sharper drop in the radial profile of column density compared to that of \textrm{O}~\textsc{vi} \citep{Liang2015}. In short, both observations and simulations show that \textrm{C}~\textsc{iv} and \textrm{O}~\textsc{vi} probe different distributions of circumgalactic gas, and, therefore, may trace distinct outflow kinematics.

Given similar velocity shifts observed in \textrm{C}~\textsc{iv} and the low ions in our sample, it is possible that the cool and warm gas, traced by the low-ionization lines and \textrm{C}~\textsc{iv}, respectively, are largely co-spatial and moving at similar velocities. At the same time, these gas phases are cooler, with lower characteristic outflow velocity compared to the gas typically probed by \textrm{O}~\textsc{vi}. One possible explanation is that \textrm{O}~\textsc{vi}, which is a characteristic feature of starburst galaxies, is produced when hot, X-ray emitting gas interacts with cold, dense clouds in its path \citep{Grimes2009}. Meanwhile, \textrm{C}~\textsc{iv} may arise from the interface layer between the low-ionization clouds and the hot, radiative cooling gas flow where \textrm{O}~\textsc{vi} is seen.

\subsection{Future Outlook}
\label{sec:future}

Investigations comparing the properties of different phases of the outflowing gas are crucial for mapping detailed multi-phase structure and estimating the overall mass loss rate through galactic winds. In this paper, we do not observe large difference in the velocity shifts between the high-ionization \textrm{C}~\textsc{iv} and the low-ionization near-UV \textrm{Fe}~\textsc{ii} features. This result suggests that the gas traced by \textrm{C}~\textsc{iv} and low ions may be co-moving, and, even co-spatial. In addition, we find that the strength of \textrm{C}~\textsc{iv} scales mainly with SFR, while that of the low-ionization lines is more sensitive to stellar mass. These results indicate that the gas probed by \textrm{C}~\textsc{iv} is more associated with outflows, while the low-ionization lines additionally trace the systemic ISM component in the galaxies. Considering the different cases discussed in Section \ref{sec:kine}, we propose two scenarios to explain the origin of the small blueshift interstellar \textrm{C}~\textsc{iv} shows relative to near-UV \textrm{Fe}~\textsc{ii} lines.

If the more highly-ionized gas is indeed traveling faster, our findings suggest that although \textrm{C}~\textsc{iv} and the low ions both have absorption profiles consisting of a Doppler (outflow) component and a systemic ISM component, the ISM component is dominant in the low ions while the outflow component contributes more to the CIV absorption profile. If the outflowing gas traced by \textrm{C}~\textsc{iv} and by the low ions are co-moving, one would expect their outflow components to have roughly the same blueshifts. Therefore, the stronger outflow component relative to that of the low-ionization lines makes \textrm{C}~\textsc{iv} show a slightly more blueshifted overall absorption profile than the low ions, as suggested by our data. Alternatively, if the larger blueshift of \textrm{C}~\textsc{iv} is caused by emission filling -- the scenario we currently favor -- our results indicate that in high-SFR galaxies more \textrm{C}~\textsc{iv} absorption is produced in the ISM due to various reasons (e.g., stellar wind from massive stars, supernova explosions, shocks). In this case, the highly-ionized gas is not necessarily associated with outflows, and the apparent larger blueshift of \textrm{C}~\textsc{iv} is only a result of stronger emission filling due to higher \textrm{C}~\textsc{iv} column density. Future simulations studying the relation between feedback and the circumgalactic medium should be able to reproduce these results.

At the same time, it is essential to have access to spectroscopic data with higher $S/N$ and spectral resolution in order to resolve the \textrm{C}~\textsc{iv} doublet members. Such data would enable us to perform the two-component fit to individual \textrm{C}~\textsc{iv} doublet members, modeling the outflow and ISM components separately. In this way, one would be able to draw more robust conclusions about the low- and high-ionization kinematics. A larger sample is also crucial to investigate the relative blueshifts shown by \textrm{C}~\textsc{iv}, \textrm{Mg}~\textsc{ii} and near-UV \textrm{Fe}~\textsc{ii}. Such improvements will enable us to fully comprehend the links and distinctions among the kinematics of different gas phases in star-forming galaxies.

\acknowledgements We thank Claus Leitherer for help with the identification of stellar absorption lines, Max Pettini for providing the stellar models and Michael Fitzgerald for suggestions about statistical methods. We acknowledge support from the David $\&$ Lucile Packard Foundation (A.E.S. and C.L.M.),
and the National Science Foundation through grants AST-0808161 and AST-1109288 (C.L.M.) and CAREER award AST-1055081 (A.L.C.).
We are grateful to the DEEP2 and AEGIS teams for providing both the galaxy sample and ancillary data on galaxy properties.
We wish to extend special thanks to those of Hawaiian ancestry on
whose sacred mountain we are privileged to be guests. Without their generous hospitality, most
of the observations presented herein would not have been possible.

\clearpage
\newpage
\bibliographystyle{apj}
\bibliography{ms}

\begin{thebibliography}{}
\expandafter\ifx\csname natexlab\endcsname\relax\def\natexlab#1{#1}\fi

\bibitem[{{Bordoloi} {et~al.}(2014){Bordoloi}, {Lilly}, {Hardmeier}, {Contini},
  {Kneib}, {Le Fevre}, {Mainieri}, {Renzini}, {Scodeggio}, {Zamorani},
  {Bardelli}, {Bolzonella}, {Bongiorno}, {Caputi}, {Carollo}, {Cucciati}, {de
  la Torre}, {de Ravel}, {Garilli}, {Iovino}, {Kampczyk}, {Kova{\v c}},
  {Knobel}, {Lamareille}, {Le Borgne}, {Le Brun}, {Maier}, {Mignoli}, {Oesch},
  {Pello}, {Peng}, {Perez Montero}, {Presotto}, {Silverman}, {Tanaka}, {Tasca},
  {Tresse}, {Vergani}, {Zucca}, {Cappi}, {Cimatti}, {Coppa}, {Franzetti},
  {Koekemoer}, {Moresco}, {Nair}, \& {Pozzetti}}]{Bordoloi2014}
{Bordoloi}, R., {Lilly}, S.~J., {Hardmeier}, E., {et~al.} 2014, \apj, 794, 130

\bibitem[{{Borthakur} {et~al.}(2013){Borthakur}, {Heckman}, {Strickland},
  {Wild}, \& {Schiminovich}}]{Borthakur2013}
{Borthakur}, S., {Heckman}, T., {Strickland}, D., {Wild}, V., \&
  {Schiminovich}, D. 2013, \apj, 768, 18

\bibitem[{{Bruzual} \& {Charlot}(2003)}]{Bruzual2003}
{Bruzual}, G., \& {Charlot}, S. 2003, \mnras, 344, 1000

\bibitem[{{Bundy} {et~al.}(2006){Bundy}, {Ellis}, {Conselice}, {Taylor},
  {Cooper}, {Willmer}, {Weiner}, {Coil}, {Noeske}, \& {Eisenhardt}}]{Bundy2006}
{Bundy}, K., {Ellis}, R.~S., {Conselice}, C.~J., {et~al.} 2006, \apj, 651, 120

\bibitem[{{Chabrier}(2003)}]{Chabrier2003}
{Chabrier}, G. 2003, \pasp, 115, 763

\bibitem[{{Chisholm} {et~al.}(2016){Chisholm}, {Tremonti}, {Leitherer}, {Chen},
  \& {Wofford}}]{Chisholm2016}
{Chisholm}, J., {Tremonti}, C.~A., {Leitherer}, C., {Chen}, Y., \& {Wofford},
  A. 2016, ArXiv e-prints, arXiv:1601.05090

\bibitem[{{Coil} {et~al.}(2011){Coil}, {Weiner}, {Holz}, {Cooper}, {Yan}, \&
  {Aird}}]{Coil2011}
{Coil}, A.~L., {Weiner}, B.~J., {Holz}, D.~E., {et~al.} 2011, \apj, 743, 46

\bibitem[{{Dean} \& {Bruhweiler}(1985)}]{Dean1985}
{Dean}, C.~A., \& {Bruhweiler}, F.~C. 1985, \apjs, 57, 133

\bibitem[{{Elbaz} {et~al.}(2007){Elbaz}, {Daddi}, {Le Borgne}, {Dickinson},
  {Alexander}, {Chary}, {Starck}, {Brandt}, {Kitzbichler}, {MacDonald},
  {Nonino}, {Popesso}, {Stern}, \& {Vanzella}}]{Elbaz2007}
{Elbaz}, D., {Daddi}, E., {Le Borgne}, D., {et~al.} 2007, \aap, 468, 33

\bibitem[{{Erb} {et~al.}(2012){Erb}, {Quider}, {Henry}, \& {Martin}}]{Erb2012}
{Erb}, D.~K., {Quider}, A.~M., {Henry}, A.~L., \& {Martin}, C.~L. 2012, \apj,
  759, 26

\bibitem[{{Erb} {et~al.}(2006){Erb}, {Shapley}, {Pettini}, {Steidel}, {Reddy},
  \& {Adelberger}}]{Erb2006}
{Erb}, D.~K., {Shapley}, A.~E., {Pettini}, M., {et~al.} 2006, \apj, 644, 813

\bibitem[{{Faber} {et~al.}(2007){Faber}, {Willmer}, {Wolf}, {Koo}, {Weiner},
  {Newman}, {Im}, {Coil}, {Conroy}, {Cooper}, {Davis}, {Finkbeiner}, {Gerke},
  {Gebhardt}, {Groth}, {Guhathakurta}, {Harker}, {Kaiser}, {Kassin},
  {Kleinheinrich}, {Konidaris}, {Kron}, {Lin}, {Luppino}, {Madgwick},
  {Meisenheimer}, {Noeske}, {Phillips}, {Sarajedini}, {Schiavon}, {Simard},
  {Szalay}, {Vogt}, \& {Yan}}]{Faber2007}
{Faber}, S.~M., {Willmer}, C.~N.~A., {Wolf}, C., {et~al.} 2007, \apj, 665, 265

\bibitem[{{Grimes} {et~al.}(2009){Grimes}, {Heckman}, {Aloisi}, {Calzetti},
  {Leitherer}, {Martin}, {Meurer}, {Sembach}, \& {Strickland}}]{Grimes2009}
{Grimes}, J.~P., {Heckman}, T., {Aloisi}, A., {et~al.} 2009, \apjs, 181, 272

\bibitem[{{Heckman} {et~al.}(1990){Heckman}, {Armus}, \& {Miley}}]{Heckman1990}
{Heckman}, T.~M., {Armus}, L., \& {Miley}, G.~K. 1990, \apjs, 74, 833

\bibitem[{{Heckman} {et~al.}(2000){Heckman}, {Lehnert}, {Strickland}, \&
  {Armus}}]{Heckman2000}
{Heckman}, T.~M., {Lehnert}, M.~D., {Strickland}, D.~K., \& {Armus}, L. 2000,
  \apjs, 129, 493

\bibitem[{{Heckman} {et~al.}(2001){Heckman}, {Sembach}, {Meurer}, {Strickland},
  {Martin}, {Calzetti}, \& {Leitherer}}]{Heckman2001}
{Heckman}, T.~M., {Sembach}, K.~R., {Meurer}, G.~R., {et~al.} 2001, \apj, 554,
  1021

\bibitem[{{Kashino} {et~al.}(2013){Kashino}, {Silverman}, {Rodighiero},
  {Renzini}, {Arimoto}, {Daddi}, {Lilly}, {Sanders}, {Kartaltepe}, {Zahid},
  {Nagao}, {Sugiyama}, {Capak}, {Carollo}, {Chu}, {Hasinger}, {Ilbert},
  {Kajisawa}, {Kewley}, {Koekemoer}, {Kova{\v c}}, {Le F{\`e}vre}, {Masters},
  {McCracken}, {Onodera}, {Scoville}, {Strazzullo}, {Symeonidis}, \&
  {Taniguchi}}]{Kashino2013}
{Kashino}, D., {Silverman}, J.~D., {Rodighiero}, G., {et~al.} 2013, \apjl, 777,
  L8

\bibitem[{{Kornei} {et~al.}(2012){Kornei}, {Shapley}, {Martin}, {Coil}, {Lotz},
  {Schiminovich}, {Bundy}, \& {Noeske}}]{Kornei2012}
{Kornei}, K.~A., {Shapley}, A.~E., {Martin}, C.~L., {et~al.} 2012, \apj, 758,
  135

\bibitem[{{Kornei} {et~al.}(2013){Kornei}, {Shapley}, {Martin}, {Coil}, {Lotz},
  \& {Weiner}}]{Kornei2013}
---. 2013, \apj, 774, 50

\bibitem[{{Lehner} {et~al.}(2011){Lehner}, {Zech}, {Howk}, \&
  {Savage}}]{Lehner2011}
{Lehner}, N., {Zech}, W.~F., {Howk}, J.~C., \& {Savage}, B.~D. 2011, \apj, 727,
  46

\bibitem[{{Leitherer} {et~al.}(2010){Leitherer}, {Ortiz Ot{\'a}lvaro},
  {Bresolin}, {Kudritzki}, {Lo Faro}, {Pauldrach}, {Pettini}, \&
  {Rix}}]{Leitherer2010}
{Leitherer}, C., {Ortiz Ot{\'a}lvaro}, P.~A., {Bresolin}, F., {et~al.} 2010,
  \apjs, 189, 309

\bibitem[{{Liang} \& {Chen}(2014)}]{Liang2014}
{Liang}, C.~J., \& {Chen}, H.-W. 2014, \mnras, 445, 2061

\bibitem[{{Liang} {et~al.}(2015){Liang}, {Kravtsov}, \& {Agertz}}]{Liang2015}
{Liang}, C.~J., {Kravtsov}, A.~V., \& {Agertz}, O. 2015, ArXiv e-prints,
  arXiv:1507.07002

\bibitem[{{Markwardt}(2009)}]{Mark2009}
{Markwardt}, C.~B. 2009, in Astronomical Society of the Pacific Conference
  Series, Vol. 411, Astronomical Data Analysis Software and Systems XVIII, ed.
  D.~A. {Bohlender}, D.~{Durand}, \& P.~{Dowler}, 251

\bibitem[{{Martin}(1999)}]{Martin1999}
{Martin}, C.~L. 1999, \apj, 513, 156

\bibitem[{{Martin}(2005)}]{Martin2005}
---. 2005, \apj, 621, 227

\bibitem[{{Martin} \& {Bouch{\'e}}(2009)}]{Martin2009}
{Martin}, C.~L., \& {Bouch{\'e}}, N. 2009, \apj, 703, 1394

\bibitem[{{Martin} {et~al.}(2012){Martin}, {Shapley}, {Coil}, {Kornei},
  {Bundy}, {Weiner}, {Noeske}, \& {Schiminovich}}]{Martin2012}
{Martin}, C.~L., {Shapley}, A.~E., {Coil}, A.~L., {et~al.} 2012, \apj, 760, 127

\bibitem[{{Murray} {et~al.}(2005){Murray}, {Quataert}, \&
  {Thompson}}]{Murray2005}
{Murray}, N., {Quataert}, E., \& {Thompson}, T.~A. 2005, \apj, 618, 569

\bibitem[{{Newman} {et~al.}(2013){Newman}, {Cooper}, {Davis}, {Faber}, {Coil},
  {Guhathakurta}, {Koo}, {Phillips}, {Conroy}, {Dutton}, {Finkbeiner}, {Gerke},
  {Rosario}, {Weiner}, {Willmer}, {Yan}, {Harker}, {Kassin}, {Konidaris},
  {Lai}, {Madgwick}, {Noeske}, {Wirth}, {Connolly}, {Kaiser}, {Kirby},
  {Lemaux}, {Lin}, {Lotz}, {Luppino}, {Marinoni}, {Matthews}, {Metevier}, \&
  {Schiavon}}]{Newman2013}
{Newman}, J.~A., {Cooper}, M.~C., {Davis}, M., {et~al.} 2013, \apjs, 208, 5

\bibitem[{{Noeske} {et~al.}(2007){Noeske}, {Weiner}, {Faber}, {Papovich},
  {Koo}, {Somerville}, {Bundy}, {Conselice}, {Newman}, {Schiminovich}, {Le
  Floc'h}, {Coil}, {Rieke}, {Lotz}, {Primack}, {Barmby}, {Cooper}, {Davis},
  {Ellis}, {Fazio}, {Guhathakurta}, {Huang}, {Kassin}, {Martin}, {Phillips},
  {Rich}, {Small}, {Willmer}, \& {Wilson}}]{Noeske2007}
{Noeske}, K.~G., {Weiner}, B.~J., {Faber}, S.~M., {et~al.} 2007, \apjl, 660,
  L43

\bibitem[{{Oke} {et~al.}(1995){Oke}, {Cohen}, {Carr}, {Cromer}, {Dingizian},
  {Harris}, {Labrecque}, {Lucinio}, {Schaal}, {Epps}, \& {Miller}}]{Oke1995}
{Oke}, J.~B., {Cohen}, J.~G., {Carr}, M., {et~al.} 1995, \pasp, 107, 375

\bibitem[{{Papovich} {et~al.}(2001){Papovich}, {Dickinson}, \&
  {Ferguson}}]{Papo2001}
{Papovich}, C., {Dickinson}, M., \& {Ferguson}, H.~C. 2001, \apj, 559, 620

\bibitem[{{Pettini} {et~al.}(2002){Pettini}, {Rix}, {Steidel}, {Adelberger},
  {Hunt}, \& {Shapley}}]{Pettini2002}
{Pettini}, M., {Rix}, S.~A., {Steidel}, C.~C., {et~al.} 2002, \apj, 569, 742

\bibitem[{{Pettini} {et~al.}(2001){Pettini}, {Shapley}, {Steidel}, {Cuby},
  {Dickinson}, {Moorwood}, {Adelberger}, \& {Giavalisco}}]{Pettini2001}
{Pettini}, M., {Shapley}, A.~E., {Steidel}, C.~C., {et~al.} 2001, \apj, 554,
  981

\bibitem[{{Prochaska} {et~al.}(2011){Prochaska}, {Kasen}, \&
  {Rubin}}]{Prochaska2011}
{Prochaska}, J.~X., {Kasen}, D., \& {Rubin}, K. 2011, \apj, 734, 24

\bibitem[{{Rix} {et~al.}(2004){Rix}, {Pettini}, {Leitherer}, {Bresolin},
  {Kudritzki}, \& {Steidel}}]{Rix2004}
{Rix}, S.~A., {Pettini}, M., {Leitherer}, C., {et~al.} 2004, \apj, 615, 98

\bibitem[{{Rubin} {et~al.}(2014){Rubin}, {Prochaska}, {Koo}, {Phillips},
  {Martin}, \& {Winstrom}}]{Rubin2014}
{Rubin}, K.~H.~R., {Prochaska}, J.~X., {Koo}, D.~C., {et~al.} 2014, \apj, 794,
  156

\bibitem[{{Rupke} {et~al.}(2005){Rupke}, {Veilleux}, \& {Sanders}}]{Rupke2005}
{Rupke}, D.~S., {Veilleux}, S., \& {Sanders}, D.~B. 2005, \apjs, 160, 115

\bibitem[{{Salim} {et~al.}(2007){Salim}, {Rich}, {Charlot}, {Brinchmann},
  {Johnson}, {Schiminovich}, {Seibert}, {Mallery}, {Heckman}, {Forster},
  {Friedman}, {Martin}, {Morrissey}, {Neff}, {Small}, {Wyder}, {Bianchi},
  {Donas}, {Lee}, {Madore}, {Milliard}, {Szalay}, {Welsh}, \& {Yi}}]{Salim2007}
{Salim}, S., {Rich}, R.~M., {Charlot}, S., {et~al.} 2007, \apjs, 173, 267

\bibitem[{{Salpeter}(1955)}]{Salpeter1955}
{Salpeter}, E.~E. 1955, \apj, 121, 161

\bibitem[{{Schwartz} \& {Martin}(2004)}]{Schwartz2004}
{Schwartz}, C.~M., \& {Martin}, C.~L. 2004, \apj, 610, 201

\bibitem[{{Schwartz} {et~al.}(2006){Schwartz}, {Martin}, {Chandar},
  {Leitherer}, {Heckman}, \& {Oey}}]{Schwartz2006}
{Schwartz}, C.~M., {Martin}, C.~L., {Chandar}, R., {et~al.} 2006, \apj, 646,
  858

\bibitem[{{Seibert} {et~al.}(2005){Seibert}, {Martin}, {Heckman}, {Buat},
  {Hoopes}, {Barlow}, {Bianchi}, {Byun}, {Donas}, {Forster}, {Friedman},
  {Jelinsky}, {Lee}, {Madore}, {Malina}, {Milliard}, {Morrissey}, {Neff},
  {Rich}, {Schiminovich}, {Siegmund}, {Small}, {Szalay}, {Welsh}, \&
  {Wyder}}]{Seibert2005}
{Seibert}, M., {Martin}, D.~C., {Heckman}, T.~M., {et~al.} 2005, \apjl, 619,
  L55

\bibitem[{{Shapley} {et~al.}(2003){Shapley}, {Steidel}, {Pettini}, \&
  {Adelberger}}]{Shapley2003}
{Shapley}, A.~E., {Steidel}, C.~C., {Pettini}, M., \& {Adelberger}, K.~L. 2003,
  \apj, 588, 65

\bibitem[{{Shen} {et~al.}(2013){Shen}, {Madau}, {Guedes}, {Mayer}, {Prochaska},
  \& {Wadsley}}]{Shen2013}
{Shen}, S., {Madau}, P., {Guedes}, J., {et~al.} 2013, \apj, 765, 89

\bibitem[{{Steidel} {et~al.}(2010){Steidel}, {Erb}, {Shapley}, {Pettini},
  {Reddy}, {Bogosavljevi{\'c}}, {Rudie}, \& {Rakic}}]{Steidel2010}
{Steidel}, C.~C., {Erb}, D.~K., {Shapley}, A.~E., {et~al.} 2010, \apj, 717, 289

\bibitem[{{Steidel} {et~al.}(2004){Steidel}, {Shapley}, {Pettini},
  {Adelberger}, {Erb}, {Reddy}, \& {Hunt}}]{Steidel2004}
{Steidel}, C.~C., {Shapley}, A.~E., {Pettini}, M., {et~al.} 2004, \apj, 604,
  534

\bibitem[{{Weiner} {et~al.}(2009){Weiner}, {Coil}, {Prochaska}, {Newman},
  {Cooper}, {Bundy}, {Conselice}, {Dutton}, {Faber}, {Koo}, {Lotz}, {Rieke}, \&
  {Rubin}}]{Weiner2009}
{Weiner}, B.~J., {Coil}, A.~L., {Prochaska}, J.~X., {et~al.} 2009, \apj, 692,
  187

\bibitem[{{Whitaker} {et~al.}(2014){Whitaker}, {Franx}, {Leja}, {van Dokkum},
  {Henry}, {Skelton}, {Fumagalli}, {Momcheva}, {Brammer}, {Labb{\'e}},
  {Nelson}, \& {Rigby}}]{whitaker2014}
{Whitaker}, K.~E., {Franx}, M., {Leja}, J., {et~al.} 2014, \apj, 795, 104

\bibitem[{{Willmer} {et~al.}(2006){Willmer}, {Faber}, {Koo}, {Weiner},
  {Newman}, {Coil}, {Connolly}, {Conroy}, {Cooper}, {Davis}, {Finkbeiner},
  {Gerke}, {Guhathakurta}, {Harker}, {Kaiser}, {Kassin}, {Konidaris}, {Lin},
  {Luppino}, {Madgwick}, {Noeske}, {Phillips}, \& {Yan}}]{Willmer2006}
{Willmer}, C.~N.~A., {Faber}, S.~M., {Koo}, D.~C., {et~al.} 2006, \apj, 647,
  853

\bibitem[{{Wolfe} \& {Prochaska}(2000)}]{Wolfe2000}
{Wolfe}, A.~M., \& {Prochaska}, J.~X. 2000, \apj, 545, 591

\end{thebibliography}

\newpage
\LongTables
\begin{deluxetable*}{ccccccccccc}
\tablewidth{0pt}
  \tablecaption{Galaxy properties and velocity measurements of 93 objects with $\textrm{C}~\textsc{iv}$ coverage.}
\tablecolumns{11}
  \tablewidth{0pc}
  \tablehead{
    \colhead{ID} &
    \colhead{Redshift} &
    \colhead{$\mbox{M}_{B}$} &
    \colhead{$\ub$} &
    \colhead{$\log(\mbox{M}_{*})$} &
    \colhead{$\mbox{A}_{UV}$} &
    \colhead{$\mbox{SFR}_{UV}$} &
    \colhead{$\mbox{V}_{\textrm{C}~\textsc{iv},1:1}$} &
    \colhead{$\mbox{V}_{\textrm{Al}~\textsc{ii}}$} & 
    \colhead{$\mbox{V}_{\textrm{Si}~\textsc{ii}}$} &
    \colhead{$\mbox{V}_{\mbox{nuv} \textrm{Fe}~\textsc{ii}}$}\\
    \colhead{} &
    \colhead{} &
    \colhead{(mag)} &
    \colhead{(mag)} &
    \colhead{($\mbox{M}_{\sun}$)} &
    \colhead{(mag)} &
    \colhead{($\mbox{M}_{\sun} \mbox{yr}^{-1}$)} &
    \colhead{(km $\mbox{s}^{-1}$)} &  
    \colhead{(km $\mbox{s}^{-1}$)} &
    \colhead{(km $\mbox{s}^{-1}$)} &
    \colhead{(km $\mbox{s}^{-1}$)} \\ 
    }
  \startdata
12008166&1.2854&-19.69&0.32&9.78&\nodata&\nodata&\nodata&-86$\pm$54&\nodata&\nodata\\
12008445&1.2774&-20.69&0.52&10.22&\nodata&\nodata&-3$\pm$ 217&67$\pm$58&\nodata&-76$\pm$40\\
12008509&1.2158&-19.80&0.36&9.76&\nodata&\nodata&-362$\pm$ 120&\nodata&\nodata&\nodata\\
12008550&1.3025&-21.23&0.62&10.08&3.65&24&-205$\pm$39&-86$\pm$40&-187$\pm$35&-146$\pm$29\\
12008811&1.2156&-20.80&0.68&10.16&1.97&9&13$\pm$ 121&119$\pm$32&90$\pm$51&31$\pm$22\\
12011428&1.2841&-19.88&0.18&9.66&2.00&9&\nodata&194$\pm$55&\nodata&\nodata\\
12011493&1.2638&-20.74&0.62&10.22&2.49&11&191$\pm$ 148&548$\pm$77&\nodata&-2$\pm$41\\
12011619&1.0745&-19.31&0.44&9.11&0.05&2&\nodata&\nodata&\nodata&\nodata\\
12011742&1.3358&-20.81&0.71&10.23&\nodata&\nodata&-32$\pm$91&123$\pm$63&\nodata&11$\pm$24\\
12011767&1.2817&-22.19&0.75&11.07&6.25&76&-298$\pm$ 161&\nodata&\nodata&-46$\pm$34\\
12012764&1.2353&-20.26&0.62&9.80&\nodata&\nodata&-466$\pm$ 124&\nodata&\nodata&-69$\pm$51\\
12012777&1.2743&-21.07&0.52&10.11&3.37&22&-287$\pm$40&-270$\pm$26&-156$\pm$45&-171$\pm$22\\
12012817&1.2158&-20.92&0.89&10.85&\nodata&\nodata&\nodata&\nodata&\nodata&28$\pm$44\\
12012842&1.3148&-21.68&0.75&11.09&\nodata&\nodata&-398$\pm$87&-24$\pm$56&170$\pm$78&27$\pm$35\\
12012871&1.3443&-20.75&0.56&10.01&3.20&14&-52$\pm$88&-37$\pm$53&-10$\pm$ 142&-5$\pm$38\\
12013002&1.2184&-20.23&0.65&9.56&0.72&4&\nodata&\nodata&\nodata&1$\pm$49\\
12013145&1.3406&-19.67&0.27&9.52&1.67&6&\nodata&\nodata&59$\pm$92&\nodata\\
12013242&1.2868&-21.32&0.45&10.19&2.84&25&-180$\pm$48&-8$\pm$28&34$\pm$85&-79$\pm$19\\
12015563&1.2824&-21.30&0.60&10.35&\nodata&\nodata&\nodata&\nodata&\nodata&\nodata\\
12015682&1.2837&-21.53&0.62&11.20&3.25&27&\nodata&-186$\pm$60&\nodata&\nodata\\
12015775&1.2243&-19.36&0.23&9.65&1.82&4&\nodata&-218$\pm$ 177&\nodata&-70$\pm$90\\
12015792&1.2307&-21.24&0.96&11.07&5.09&12&\nodata&\nodata&\nodata&\nodata\\
12015858&1.2307&-19.87&0.22&9.56&1.98&8&75$\pm$ 178&\nodata&320$\pm$ 155&-30$\pm$59\\
12015914&1.1046&-19.83&0.33&10.11&2.60&8&121$\pm$ 179&148$\pm$62&113$\pm$ 116&68$\pm$25\\
12015933&1.2851&-20.35&0.66&9.70&\nodata&\nodata&\nodata&\nodata&\nodata&\nodata\\
12016019&1.0847&-20.72&0.62&10.03&4.52&18&\nodata&69$\pm$ 108&\nodata&47$\pm$17\\
12016075&1.1174&-19.68&0.54&9.41&2.88&5&\nodata&\nodata&\nodata&18$\pm$53\\
12016903&1.1600&-21.46&0.51&10.21&2.35&25&135$\pm$29&61$\pm$46&\nodata&4$\pm$36\\
12019542&1.2785&-21.68&0.71&10.40&5.41&40&\nodata&\nodata&\nodata&\nodata\\
12019996&1.2813&-21.71&0.63&10.67&4.32&43&316$\pm$ 159&52$\pm$68&\nodata&121$\pm$28\\
12020064&1.3148&-21.12&0.60&10.35&4.71&26&\nodata&-136$\pm$58&\nodata&\nodata\\
12024014&1.2977&-20.51&0.50&9.91&\nodata&\nodata&\nodata&\nodata&\nodata&\nodata\\
12024133&1.1241&-21.14&0.81&10.56&7.01&36&\nodata&\nodata&\nodata&-45$\pm$24\\
12024181&1.0869&-21.00&0.80&10.54&7.63&35&\nodata&\nodata&\nodata&\nodata\\
12100420&1.1995&-20.63&0.60&10.02&2.35&9&-2$\pm$94&127$\pm$69&114$\pm$59&68$\pm$18\\
22004858&1.2687&-21.85&0.62&10.54&\nodata&\nodata&-64$\pm$79&199$\pm$29&112$\pm$42&82$\pm$15\\
22005523&1.2188&-21.68&0.61&10.41&\nodata&\nodata&\nodata&94$\pm$69&\nodata&-26$\pm$26\\
22005715&1.2345&-20.28&0.52&9.75&\nodata&\nodata&\nodata&\nodata&\nodata&\nodata\\
22006207&1.2709&-20.60&0.42&9.89&\nodata&\nodata&-472$\pm$60&-285$\pm$46&\nodata&\nodata\\
22012180&1.2894&-19.91&0.46&9.63&\nodata&\nodata&\nodata&\nodata&\nodata&\nodata\\
22012285&1.1700&-20.05&0.54&9.75&\nodata&\nodata&1162$\pm$68&143$\pm$ 111&\nodata&33$\pm$33\\
22012322&1.1941&-20.42&0.57&9.84&\nodata&\nodata&\nodata&214$\pm$ 114&\nodata&-144$\pm$55\\
22012678&1.1959&-21.86&0.60&10.93&\nodata&\nodata&\nodata&82$\pm$ 109&\nodata&148$\pm$57\\
22013029&1.2710&-21.00&0.69&10.32&\nodata&\nodata&-658$\pm$92&\nodata&\nodata&\nodata\\
22013827&1.2378&-21.25&0.75&10.52&\nodata&\nodata&\nodata&\nodata&\nodata&\nodata\\
22013864&1.2283&-21.00&0.72&10.38&\nodata&\nodata&\nodata&-115$\pm$94&\nodata&-99$\pm$28\\
22020736&1.1236&-21.83&0.99&10.99&\nodata&\nodata&\nodata&\nodata&\nodata&-41$\pm$88\\
22036666&1.3468&-20.96&0.53&10.23&\nodata&\nodata&\nodata&\nodata&\nodata&\nodata\\
22036984&1.0804&-20.79&0.70&10.19&\nodata&\nodata&\nodata&\nodata&\nodata&\nodata\\
22044809&1.1869&-19.59&0.31&9.39&\nodata&\nodata&\nodata&\nodata&\nodata&\nodata\\
22100920&1.2735&-21.12&0.57&10.25&\nodata&\nodata&\nodata&-18$\pm$29&\nodata&-32$\pm$17\\
32016683&1.3006&-21.22&0.58&10.42&\nodata&\nodata&-114$\pm$44&22$\pm$28&14$\pm$69&-6$\pm$20\\
32017188&1.2526&-20.83&0.58&9.98&\nodata&\nodata&\nodata&\nodata&\nodata&\nodata\\
32019848&1.3433&-21.14&0.60&10.46&\nodata&\nodata&\nodata&\nodata&\nodata&\nodata\\
32019861&1.3077&-20.54&0.53&10.01&\nodata&\nodata&\nodata&\nodata&\nodata&\nodata\\
32019900&1.2478&-20.83&0.65&9.75&\nodata&\nodata&\nodata&\nodata&\nodata&\nodata\\
32019980&1.3487&-19.92&0.43&9.73&\nodata&\nodata&\nodata&\nodata&127$\pm$54&\nodata\\
32020062&1.2808&-20.32&0.59&9.86&\nodata&\nodata&-244$\pm$92&\nodata&\nodata&-41$\pm$49\\
32020258&1.2461&-20.58&0.66&10.10&\nodata&\nodata&\nodata&\nodata&\nodata&\nodata\\
32020274&1.2589&-20.85&0.72&10.19&\nodata&\nodata&\nodata&\nodata&\nodata&-78$\pm$69\\
32020317&1.1871&-20.28&0.47&9.72&\nodata&\nodata&-220$\pm$ 142&284$\pm$ 178&\nodata&8$\pm$48\\
32020384&1.2493&-20.82&0.42&9.96&\nodata&\nodata&\nodata&51$\pm$51&157$\pm$ 102&-33$\pm$29\\
32020441&1.0451&-20.51&0.61&10.04&\nodata&\nodata&\nodata&254$\pm$ 100&\nodata&-42$\pm$35\\
32020468&1.2355&-21.00&0.73&10.28&\nodata&\nodata&\nodata&\nodata&\nodata&\nodata\\
32020670&1.2330&-21.39&0.90&10.49&\nodata&\nodata&\nodata&\nodata&\nodata&\nodata\\
32020728&1.0446&-19.95&0.47&9.76&\nodata&\nodata&\nodata&\nodata&\nodata&-27$\pm$39\\
32020769&1.3144&-20.77&0.41&10.01&\nodata&\nodata&\nodata&-11$\pm$ 104&1$\pm$75&15$\pm$42\\
32020873&1.2491&-20.49&0.59&9.77&\nodata&\nodata&\nodata&\nodata&\nodata&-112$\pm$40\\
42006781&1.2860&-20.74&0.55&9.69&\nodata&\nodata&-216$\pm$41&-153$\pm$22&-175$\pm$29&-70$\pm$15\\
42006799&1.2021&-21.74&0.91&11.26&\nodata&\nodata&\nodata&\nodata&\nodata&-7$\pm$30\\
42006898&1.2410&-20.06&0.57&10.01&\nodata&\nodata&\nodata&-300$\pm$ 168&\nodata&38$\pm$38\\
42006904&1.0228&-20.74&0.46&10.07&\nodata&\nodata&\nodata&\nodata&\nodata&-78$\pm$46\\
42014585&1.2706&-20.91&0.59&10.18&\nodata&\nodata&-143$\pm$95&-76$\pm$49&30$\pm$97&-152$\pm$38\\
42014618&1.0131&-20.38&0.59&9.81&\nodata&\nodata&\nodata&-150$\pm$84&\nodata&-71$\pm$15\\
42014718&1.1904&-21.27&0.69&10.86&\nodata&\nodata&4$\pm$ 122&-119$\pm$ 194&\nodata&-141$\pm$27\\
42014758&1.2702&-21.82&0.65&10.67&\nodata&\nodata&-184$\pm$93&-98$\pm$49&-126$\pm$62&-32$\pm$20\\
42018218&1.2253&-20.15&0.50&9.79&\nodata&\nodata&\nodata&\nodata&\nodata&142$\pm$58\\
42021285&1.1082&-20.65&0.71&10.17&\nodata&\nodata&\nodata&\nodata&\nodata&-35$\pm$28\\
42022173&1.3112&-20.25&0.32&9.68&\nodata&\nodata&607$\pm$ 147&-338$\pm$58&176$\pm$ 133&-158$\pm$30\\
42022307&1.2595&-21.44&0.47&10.22&\nodata&\nodata&-292$\pm$63&-78$\pm$28&-32$\pm$36&-136$\pm$13\\
42025506&1.3358&-21.10&0.54&10.17&\nodata&\nodata&-10$\pm$84&-32$\pm$77&1$\pm$66&\nodata\\
42025525&1.2946&-20.12&0.49&9.61&\nodata&\nodata&\nodata&\nodata&\nodata&\nodata\\
42025744&1.2024&-20.18&0.65&9.69&\nodata&\nodata&\nodata&\nodata&\nodata&\nodata\\
42025804&1.2435&-20.35&0.50&9.93&\nodata&\nodata&\nodata&\nodata&\nodata&\nodata\\
42026237&1.3385&-21.50&0.65&10.61&\nodata&\nodata&\nodata&\nodata&\nodata&\nodata\\
42026243&1.3457&-21.66&0.67&10.88&\nodata&\nodata&\nodata&-193$\pm$ 102&\nodata&-309$\pm$52\\
42026308&1.1700&-20.19&0.48&10.08&\nodata&\nodata&\nodata&\nodata&\nodata&\nodata\\
42026327&1.3461&-20.63&0.45&10.20&\nodata&\nodata&\nodata&\nodata&\nodata&\nodata\\
42026601&1.2272&-20.02&0.34&9.64&\nodata&\nodata&\nodata&\nodata&\nodata&\nodata\\
42033338&1.2237&-19.65&0.43&9.74&\nodata&\nodata&\nodata&\nodata&\nodata&\nodata\\
42034098&1.3465&-21.25&0.44&10.51&\nodata&\nodata&\nodata&0$\pm$72&\nodata&0$\pm$55\\
42034156&1.3464&-20.69&0.37&10.10&\nodata&\nodata&179$\pm$99&\nodata&91$\pm$ 120&-35$\pm$33\\
42034223&1.2000&-20.09&0.44&9.45&\nodata&\nodata&\nodata&-57$\pm$91&\nodata&8$\pm$35\\

 \enddata
\label{tab:vplot}
\end{deluxetable*}

{\footnotesize \noindent {\sc Note ---} Redshifts are from the DEEP2 catalog. The $B$-band luminosity and $\ub$ color are values taken from \citet{Willmer2006}. $\mbox{M}_{*}$ was estimated from SED fitting with $BRIK$ photometry, as described in \citet{Bundy2006}. $\mbox{A}_{UV}$ is the extinction in the UV, which was used to derive $\mbox{SFR}_{UV}$, the dust-corrected SFR from GALEX measurements (full description in Section \ref{sec:data}). Column 8-11 are velocity shifts measured from objects with continuum $S/N>5$ and a line detection greater than $3\sigma$ (in terms of EW) using simple Gaussian fits. The listed line $S/N$ was calculated by combining the doublet members for \textrm{C}~\textsc{iv}, and \textrm{Fe}~\textsc{ii}$\lambda2344$, $\lambda$2374 and $\lambda$2587 for the near-UV \textrm{Fe}~\textsc{ii}.}

\end{CJK}
\end{document}